\documentclass[sigconf]{acmart}

\usepackage{enumitem}
\usepackage{multirow, xcolor, array}
\newcolumntype{L}[1]{>{\raggedright\arraybackslash}p{#1}}

\AtBeginDocument{%
  }

\copyrightyear{2026}
\acmYear{2026}
\setcopyright{cc}
\setcctype{by}
\acmConference[CHI '26]{Proceedings of the 2026 CHI Conference on Human Factors in Computing Systems}{April 13--17, 2026}{Barcelona, Spain}
\acmBooktitle{Proceedings of the 2026 CHI Conference on Human Factors in Computing Systems (CHI '26), April 13--17, 2026, Barcelona, Spain}
\acmPrice{}
\acmDOI{10.1145/3772318.3790431}
\acmISBN{979-8-4007-2278-3/2026/04}

\begin{document}

\title{Rethinking External Communication of Autonomous Vehicles: Is the Field Converging, Diverging, or Stalling?}

\author{Tram Thi Minh Tran}
\email{tram.tran@sydney.edu.au}
\orcid{0000-0002-4958-2465}
\affiliation{Design Lab, Sydney School of Architecture, Design and Planning,
  \institution{The University of Sydney}
  \city{Sydney}
  \state{NSW}
  \country{Australia}
}

\author{Debargha Dey}
\email{d.dey@tue.nl}
\orcid{0000-0001-9266-0126}
\affiliation{%
  \institution{Eindhoven University of Technology}
  \city{Eindhoven}
  \country{The Netherlands}
 }

\author{Martin Tomitsch}
\email{Martin.Tomitsch@uts.edu.au}
\orcid{0000-0003-1998-2975}
\affiliation{Transdisciplinary School,
  \institution{University of Technology Sydney}
  \city{Sydney}
  \state{NSW}
  \country{Australia}
}
\renewcommand{\shortauthors}{Tran et al.}

\begin{abstract}
As autonomous vehicles enter public spaces, external human–machine interfaces are proposed to support communication with external road users. A decade of research has produced hundreds of studies and reviews, yet it remains unclear whether the field is converging on shared principles or diverging across approaches. We present a multi-dimensional analysis of 620 publications, complemented by industry deployments and regulatory documents, to track research evolution and identify convergence. The analysis reveals several field-level patterns. First, convergence on a safety-first core: simple visual cues that clarify intent. Second, sustained divergence in necessity and implementation. Third, a progressive filtering funnel: broad exploration in research and concepts narrows in deployment and is codified by regulation into a minimal set of permitted signals. These insights point to a shift in emphasis for future work, from producing new prototypes toward consolidating evidence, clarifying points of contention, and developing frameworks that can adapt across contexts.
\end{abstract}



\maketitle

\section{Introduction}

On March 18, 2018, an autonomous test vehicle struck and killed a pedestrian in Tempe, Arizona, the first known fatality of its kind~\cite{wiki2025elaine}. While the causes were technical and procedural~\cite{indident2018report}, the incident highlighted the stakes of integrating autonomous vehicles (AVs) into public spaces, where interactions with pedestrians are inevitable. These encounters raise a long-standing challenge in Human-Computer Interaction (HCI): designing communication between humans and intelligent systems. From collaborative robots to ambient displays, effective interfaces must work for people with no prior training or shared context. External human–machine interfaces (eHMIs)~\cite{zileli2019towards, dey2020taming} for AVs exemplify this challenge, needing to convey intent and state to diverse road users, often under time pressure and in environments where misinterpretation can have serious consequences.

Since the mid-2010s, researchers and automotive companies have proposed everything from light bars and projected crosswalks to text displays and robotic eyes, an assortment referred to as \textit{`eHMI jungles'}~\cite{dey2020taming}. What began as speculative design has grown into a vibrant area of automotive UX and mobility research, yet the field remains conceptually unsettled. eHMIs increasingly resemble a \textit{wicked problem}, where introducing new communication channels can lead to unintended consequences such as over-reliance~\cite{hollander2019overtrust, kaleefathullah2022misleading} or scalability challenges in complex traffic scenarios~\cite{tran2023scoping, colley2023scalability}. As \citet{de2022external} observes, the core question of whether eHMIs are a necessity or an overcomplication remains open to debate, with their review outlining arguments on both sides and emphasising the lack of consensus about their role, value, and necessity in future AV ecosystems.

This conceptual ambiguity is compounded by methodological inconsistency. As \citet{rouchitsas2019external} note, conclusions about eHMI effectiveness often depend on the evaluation method. Virtual reality (VR) simulations tend to yield clear, positive results regarding their usefulness, whereas physical prototypes and Wizard-of-Oz setups produce more varied and sometimes inconclusive findings. Despite this lack of consensus, deployment has already begun. Waymo has trialled a rooftop `dome' interface during operations in the United States~\cite{GoogleWaymo2020, GoogleWaymo2023, GoogleWaymo2025}, and Mercedes-Benz has received approval for turquoise-coloured automated driving marker lights~\cite{MercedesBenz2023Turquoise}. These choices reflect company-level design decisions rather than coordinated standardisation efforts. In contrast, most publicly documented AV deployments in China appear to operate without dedicated eHMIs.

With more than a decade of research on eHMIs, the literature has expanded rapidly, accompanied by numerous synthesis efforts. In our own data collection, we identified 55 such papers, yet none directly address a simple but timely question: \textbf{Is the eHMI research field converging, diverging, or stalling?} This question matters because research trajectories shape the likelihood of coherent design recommendations, the feasibility of regulatory alignment, and the transferability of findings into industry practice. It is timely because real-world deployments are already underway and regulatory bodies are beginning to consider formal guidance. To answer this question, we conducted a multi-dimensional analysis of 620 eHMI studies (2014–2025), combined with a comparison to industry developments (patents, concepts, deployments) and emerging regulatory frameworks.

This paper makes the following contributions:
\begin{enumerate}
    \item A multi-dimensional analysis of 620 eHMI studies mapping the field’s evolution across key research dimensions.
    \item A synthesis of convergence and divergence trends from prior reviews and taxonomies.
    \item A comparison of academic research with industry developments, and regulatory frameworks.
    \item An evidence-based assessment of whether eHMI research is converging, diverging, or stalling, with implications for future research.
\end{enumerate}

Beyond direct contributions to automotive interfaces, this work advances HCI by analysing the evolution of one of its subfields and informing related emerging technologies in public, mobility, and robotic contexts.

\section{Related Work}

The study of eHMIs has matured to the point where research, industrial development, and early regulatory efforts are influencing one another, creating a body of related work that frames our analysis.

\subsection{Prior Syntheses of eHMI Research} 

In current road traffic, implicit cues from vehicle kinematics, e.g., slowing, stopping, edging forward, remain the primary source of information for pedestrians~\cite{rasouli2019autonomous, dey2017pedestrian}, and early AV trials using `ghost driver'~\cite{rothenbucher2016ghost} Wizard-of-Oz setups found that these cues largely carry over to AV interactions~\cite{moore2019case}. However, removing a visible driver also removes eye contact, gestures, and other social signals that support familiarity, acceptance, and trust~\cite{rasouli2019autonomous}. eHMIs respond by framing AV–pedestrian interaction as an explicit communication problem, with transparency of system state and intent as central concerns~\cite{zileli2019towards}. This framing, together with growing global investment and interest in AVs since the early 2010s (as reflected in Google Trends data\footnote{\url{https://trends.google.com/trends/explore} for the keywords `self-driving cars' and `autonomous vehicles'.}), has driven rapid exploration in both research and industry. The first known eHMI patent, \textit{Pedestrian Notifications}, was filed by Google in 2013~\cite{waymo2015pedestrian}, marking the start of formalised protection of such designs. By 2020, \citet{dey2020taming} identified 70 eHMI concepts spanning academic and industrial sources, requiring an 18-item taxonomy to structure the space.

The aim of most eHMI reviews, surveys, and taxonomies has been to consolidate a rapidly expanding body of work and identify ways to move the field forward. From around 2019--2020, when a critical mass of studies had emerged, more reflective contributions began to appear. \citet{de2022external} and \citet{rouchitsas2019external} surfaced contrasting evidence and arguments in the literature regarding the added value of eHMIs. In parallel, \citet{kaß2020standardised} called for standardised evaluation approaches to address the methodological heterogeneity that limits cross-study comparability. Across these syntheses, recurring debates emerge over whether eHMIs are necessary or redundant, and over parameters of implementation such as colour and message content. While valuable in framing the discussion, these works do not assess whether the field is converging, diverging, or stalling, nor do they connect academic trajectories with industry deployments or regulatory developments.

\subsection{Industry and Regulation}

In many areas of HCI, industry and academic research develop in close relation, often through collaborations that share concepts, prototypes, and evaluation findings. eHMI research exemplifies this exchange, with overlaps between ideas emerging in academic work and those explored in industrial trials, and with multiple studies conducted in partnership with automotive companies~\cite{dey2020taming, block2023road}. At the same time, industry efforts must navigate constraints of technology readiness and manufacturability, robustness, and regulatory compliance, while academic work often prioritises exploration of novel or speculative concepts.

Because eHMIs operate in high-stakes, safety-critical public traffic environments, their design and deployment are inherently subject to extensive regulatory oversight. Regulatory engagement with eHMIs remains in its early stages. Bodies such as the United Nations Economic Commission for Europe (UNECE), the International Organization for Standardization (ISO), and the U.S. National Highway Traffic Safety Administration (NHTSA) have begun discussing external signalling for automated vehicles, but binding regulations or standards are rare. Existing discussions often focus on avoiding conflicts with established traffic signals, ensuring visibility across contexts, and minimising the risk of misinterpretation (e.g.,~\cite{SAE_J3134_2019, SAE_J578_2020, ISO_TR23049_2018, tuv2023china}).

Despite the interconnected roles of academia, industry, and policy in shaping viable designs, systematic comparisons across these trajectories are lacking. This paper addresses that gap by examining how research activity aligns with industrial developments and emerging regulatory frameworks.

\subsection{Field-Level Assessment in HCI} 

Across the broader science-of-science literature, meta-research has increasingly examined how fields evolve. Techniques such as co-citation analysis~\cite{white1998visualizing}, topic modelling~\cite{Geeganage2024topicmodelling}, and longitudinal bibliometrics~\cite{donthu_bibliometrics} have been used to reveal the structural dynamics of research domains. Convergence is often reflected in the repeated use of certain methods, a shared vocabulary, or consensus frameworks. 

Applying this perspective to eHMI research makes it possible to move beyond cataloguing individual studies toward assessing their structural trajectory. Bibliometric reviews of eHMIs include \citet{ma2023bibliometric}, who identified keyword-based clusters from 679 Web of Science (WoS) papers, and \citet{siu2025pedestrians}, who reviewed 234 WoS-indexed papers on pedestrian interaction with eHMI-equipped AVs. These works synthesise research hotspots and influencing factors, but do not address the question posed by our study. Unlike bibliometric reviews, our approach combines expert-coded, multi-dimensional analysis of full texts with cross-comparison to industry and regulatory trajectories, enabling a qualitative assessment of whether the field is converging, diverging, or stalling.

\section{Methods}

\subsection{Research Questions}

Our research is guided by the following research questions (RQs):
\begin{itemize}
    \item RQ1: How has eHMI research evolved over time, and what patterns can be observed in its development?
    \item RQ2: To what extent do the findings and recommendations in eHMI research reflect consensus or fragmentation?
    \item RQ3: How do these findings align with real-world deployments and emerging regulations?
\end{itemize}

To address these questions, we adopted an approach comprising two components. We began with a \textit{structured literature review} of eHMI studies, analysed through multi-dimensional coding. We also examined \textit{industry and regulatory sources} to compare academic research with real-world practices. 

\begin{figure*} [h]
    \centering
    \includegraphics[width=1\linewidth]{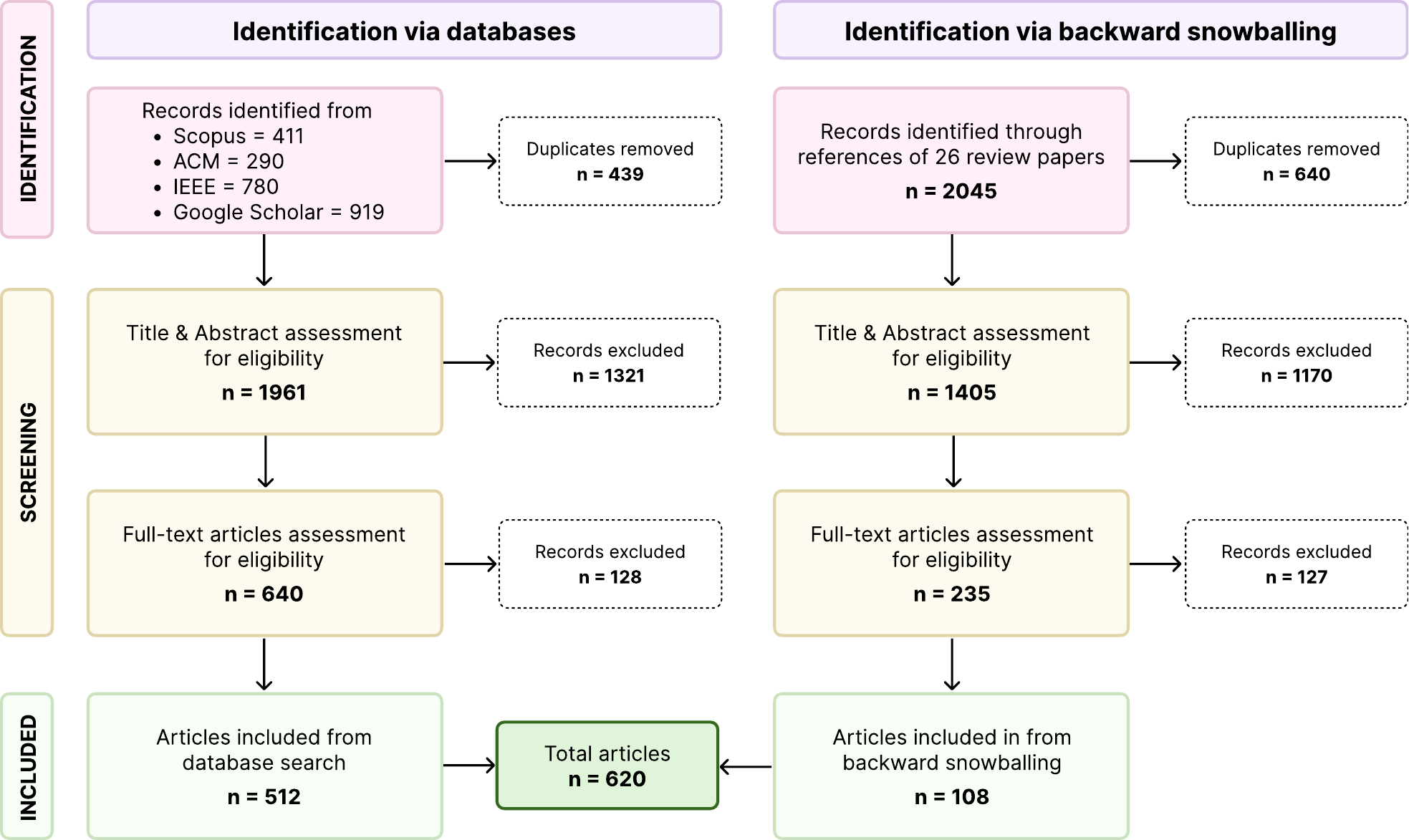}
    \caption{PRISMA flow diagram showing the identification, screening, and inclusion process for eHMI studies. The final dataset comprised 512 articles from database searches and 108 additional articles identified through backward snowballing (total = 620).}
    \Description{The figure presents a PRISMA flow diagram outlining the identification, screening, and inclusion process for eHMI studies. Two pathways are shown: identification via databases and identification via backward snowballing. From database searches (Scopus, ACM, IEEE, Google Scholar), 2,400 records were identified, with 439 duplicates removed. After title and abstract screening (n~=~1,961) and full-text assessment (n~=~640), 515 articles were included. From backward snowballing, 2,045 records were identified, with 640 duplicates removed. Following title and abstract screening (n~=~1,405) and full-text assessment (n~=~239), 118 articles were included. In total, 620 articles were included in the final dataset (515 from database searches and 118 from backward snowballing).}
    \label{fig:prisma}
\end{figure*}

\subsection{Scholarly Sources}

\subsubsection{Databases}

To ensure broad and cross-disciplinary coverage, we searched four complementary sources: Scopus for structured journal and conference indexing, IEEE Xplore for engineering and autonomous systems research, ACM Digital Library for HCI and design-focused research, and Google Scholar to capture grey literature and fill gaps through flexible keyword search.

\subsubsection{Keywords}

To capture eHMI research, we developed a keyword query spanning three categories:

\begin{itemize}
    \item \textbf{Autonomy:} autonomous, automated, self-driving, driverless
    \item \textbf{Vehicle/System:} car, vehicle
    \item \textbf{Interface and Communication:} eHMI, external communication, external human–machine interface, external display
\end{itemize}

\subsubsection{Inclusion and Exclusion Criteria}

To ensure a comprehensive dataset, inclusion criteria were applied loosely in terms of content. Papers were included if they engaged with eHMIs in any form whether as a central focus or as part of a broader discussion on autonomous mobility systems. Specifically, a paper had to:

\begin{itemize}
    \item Address communication between an autonomous mobility system and an external party (e.g., pedestrian, cyclist, manual driver), either as a central or peripheral focus.
    \item Describe, evaluate, or conceptually discuss eHMIs or related external communication mechanisms.
    \item Be published in English and accessible in full text.
    \item Be a peer-reviewed publication or equivalent.
\end{itemize}

Papers were excluded if they met any of the following criteria:

\begin{itemize}
    \item Dealt exclusively with internal HMIs (e.g., in-vehicle displays or driver interfaces) without any mention of external communication.
    \item Described autonomous systems but made no reference to external interfaces, signalling, or communication.
    \item Were unrelated to the topic (e.g., keyword overlap without substantive connection to eHMI).
    \item Were non-peer-reviewed or informal in format (e.g., theses, preprints, or presentation decks).
\end{itemize}

\subsubsection{Search and Screening Results}

We conducted database searches on 9 July 2025. While the core keyword strategy remained consistent, the exact query syntax and search scope (e.g., full-text metadata in IEEE Xplore, title–abstract–keyword in Scopus) were adapted to suit each platform. Filters such as content type were applied where relevant to improve result quality, and no date restrictions were imposed to ensure comprehensive coverage (see \autoref{tab:queries}). The results included a total of \textbf{2,400} documents: 411 from Scopus, 290 from ACM Digital Library, 780 from IEEE Xplore, and 919 from Google Scholar. After merging results from all databases, we removed duplicates using Google Sheets' \textit{Remove Duplicates} function based on the title and URL columns, resulting in \textbf{1,961} unique records. 

In the title and abstract screening, we excluded 1,321 records. In the full-text screening, we excluded an additional 128 records. Across both stages, exclusions were made for the following reasons: papers not about AV communication or interfaces (1003; 2); non-English publications (169; 6); duplicates (116; 6); non-peer-reviewed or informal formats (6; 79); papers not eHMI-related (11; 23); and inaccessible sources (16; 12). This resulted in \textbf{512} documents included for final analysis.



Among the 512 papers included for analysis, we identified 55 review and taxonomy papers, 26 of which contain a large number of eHMI references due to their review focus. These served as secondary sources for backward snowballing~\cite{wohlin2014snowballing}, allowing us to capture additional relevant studies that may not have been indexed or surfaced through keyword search alone. Reference lists were scanned in two rounds (title/abstract and full-text) to identify an additional \textbf{108} research papers aligned with our inclusion criteria.

To ensure screening completeness and transparency, the first author screened all records twice (18 August and 11 November 2025). Although the criteria remained consistent across both rounds, the second pass enabled a more exhaustive application of the criteria and captured papers missed in the initial round. During rescreening, we also clarified exclusion labels to improve consistency in decision recording. A second author independently cross-validated 10\% of randomly selected records to ensure accuracy in screening decisions. We deliberately applied generous relevance criteria at the screening stage, retaining any paper that engaged with AVs and human-facing communication or interaction in a broad sense. This approach allowed us to avoid prematurely excluding borderline cases, which were later distinguished analytically as peripheral eHMIs.

\subsubsection{Data Charting}\hfill

\textit{Metadata}: They were primarily retrieved through DOI queries to Scopus, with records not indexed in Scopus updated manually. The following fields were collected for each paper: document type, DOI, year, title, abstract, author, author affiliations, and author keywords. In addition, we developed Python scripts to extract information from author affiliations. Country information was derived from the first author’s affiliation, while all listed affiliations were used to capture the full set of contributing countries. Industry collaboration was identified by detecting automotive companies in the affiliations.

\textit{Custom Data Fields}: The first and second authors identified and discussed additional data fields relevant to the research questions. Coding was iterative and bottom-up, informed by prior taxonomies (i.e., \cite{dey2020taming}) but adapted to accommodate the diversity of concepts in the dataset. Two groups of fields were charted. The first captures the types of road users and vehicles, providing an overview of the actors and systems represented in the literature. The second comprises a set of high-level orientations—\textit{Design Inquiry}, \textit{Situated Investigation}, and \textit{Field Consolidation}—used to characterise the topical focus of research on external communication. Each orientation contains a set of dimensions and sub-dimensions that define specific aspects of external communication examined in prior work. 

Coding for all papers was conducted manually by the first author. A coding protocol with clear operational definitions was established to support consistency and reduce ambiguity in interpretation. Petal\footnote{\url{https://www.petal.org/}}, a document-analysis platform, was used in parallel as a checking aid to help identify possible omissions or mistakes in the manual coding process. Its outputs were compared against the original PDFs and served only to flag items for review; all final codes were based solely on the manual assessment. As all reviewed papers were publicly accessible, this use of Petal aligns with EU AI Act~\cite{EUAIActWebsite2026} provisions on lawful data use. During the coding process, ambiguous cases were discussed with other authors.

For methodological transparency, the detailed Petal prompts used are provided in \autoref{tab:petal}. A detailed workflow diagram summarising dataset construction, data charting procedures, tools used, verification steps, and links to open-science materials is provided in \autoref{sec:workflow}.

\subsection{Industry and Regulatory Sources}

To complement the academic literature, we included industry and regulatory sources to assess the extent to which eHMI research findings are reflected in real-world deployments and standardisation efforts.

\subsubsection{Industry Concepts and Deployments}  

We began with the taxonomy paper by \citet{dey2020taming}, which contains the largest collection of industry eHMI concepts to date. Covering work up to 2020, it provided a baseline for identifying early conceptual directions. We extended this baseline through a manual search using a curated list of 47 companies (see \autoref{tab:companies}), spanning major vehicle OEMs and AV service operators (e.g., Waymo, Cruise), delivery robot and sidewalk AV companies (e.g., Nuro, Starship), automotive OEMs with AV/eHMI concept work (e.g., Volvo, Nissan), and tech companies with AV platforms (e.g., Apple, Google, NVIDIA). For each company, we conducted targeted keyword searches (e.g., `external display', `pedestrian interaction', `vehicle yield signal') across official domains and publicly available media sources. In recording the data, we differentiated between patents, conceptual designs, and those that had progressed to real-world deployment or testing.

\subsubsection{Regulatory and Standards Documents}  
We identified regulations and standards within our scholarly dataset by searching with the keywords `standards' and `regulations'. Relevant frameworks were then extracted from sources including UNECE, ISO, and various national transport authorities. 



\subsection{Data Analysis}

\subsubsection{Evolution of eHMI Research}

For \textbf{RQ1}, we analysed all publication types within the core dataset (n~=~563, see Section \ref{sec:focus}) using the developed coding scheme. 


\subsubsection{Consensus vs. Fragmentation}

For \textbf{RQ2}, we conducted a meta-synthesis~\cite{jensen1996meta, Thorne2022Qualitative_meta} of review papers identified in our dataset. A meta-synthesis was appropriate because it enables comparison across diverse secondary sources to surface recurring themes of consensus as well as points of divergence that remain unsettled. This approach builds upon existing synthesis efforts by integrating them into a higher-level comparative analysis.  

We initially considered a meta-analysis but realised it was not feasible due to substantial heterogeneity: measures with the same label were often operationalised differently across studies, experimental contexts ranged from VR simulations to on-road trials, and reporting practices varied considerably (e.g., omission of effect sizes or reliance on qualitative findings). Such variation precludes meaningful aggregation of effect sizes without introducing bias.

\subsubsection{Alignment with Industry and Regulation}
For \textbf{RQ3}, we first examined developments in industry deployments and regulatory documents to trace how eHMI concepts have progressed in practice and policy. We then compared these trajectories with scholarly findings. 

\section{Results}

\subsection{Dataset Overview}

This section provides a descriptive overview of the publications in our dataset to contextualise the subsequent analyses.

\subsubsection{Publication Volume}

\begin{figure}[h]
    \centering
    \includegraphics[width=1\linewidth]{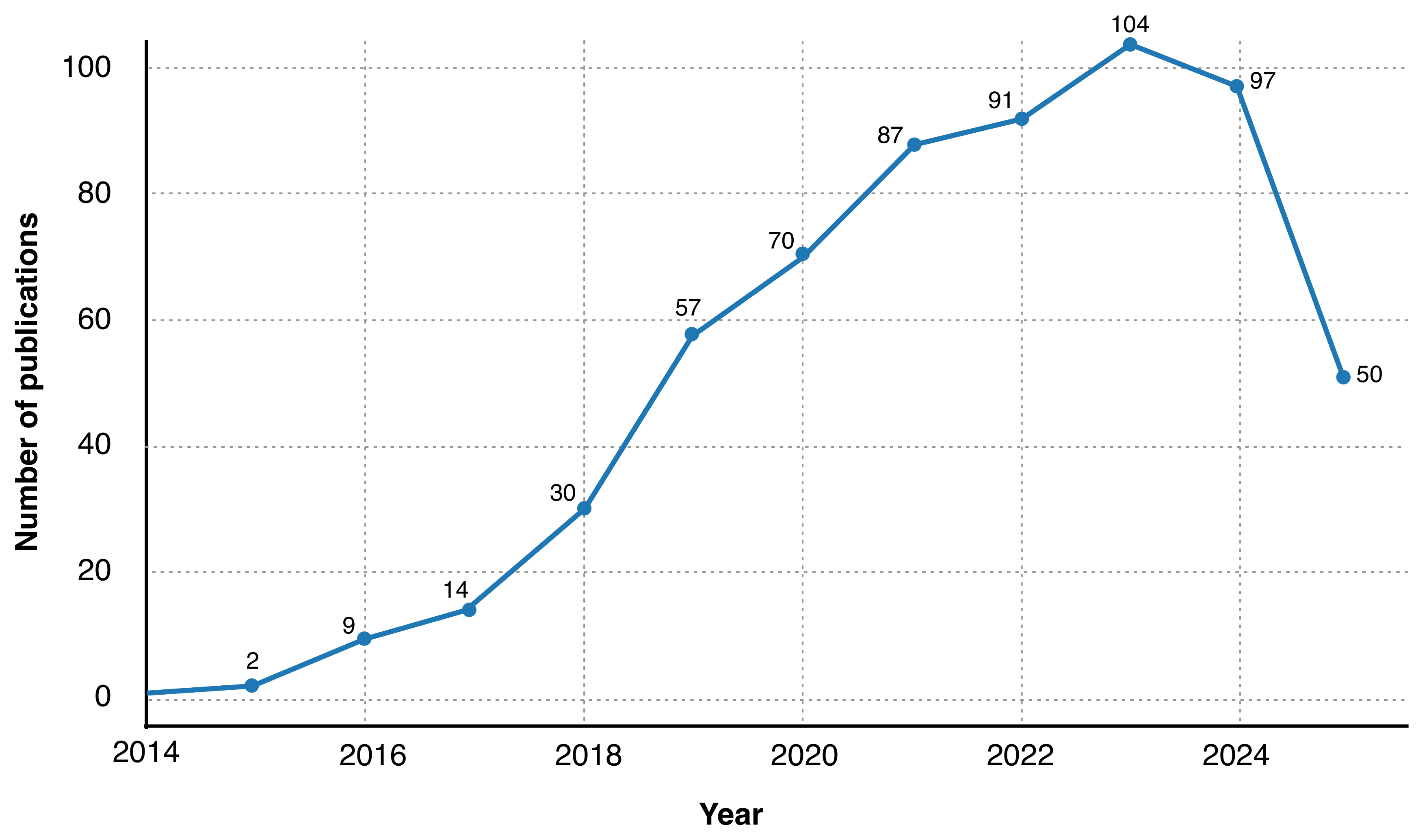}
    \caption{Annual distribution of 620 eHMI publications from 2014–2025.}
    \Description{The figure is a line chart showing the number of eHMI-related publications per year from 2014 to 2025. The x-axis represents the year, and the y-axis represents the number of publications, ranging from 0 to 100. The trend shows a sharp increase from 1 publication in 2014 to 99 in 2023, followed by a slight dip to 98 in 2024 and a noticeable drop to 48 in 2025. Key points are labelled: 1 (2014), 2 (2015), 9 (2016), 14 (2017), 30 (2018), 57 (2019), 70 (2020), 87 (2021), 91 (2022), 104 (2023), 97 (2024), and 50 (2025).}
    \label{fig:annual_distribution}
\end{figure}

Indicated by \autoref{fig:annual_distribution}, the annual distribution indicates that eHMI research was negligible prior to 2017, followed by sharp growth beginning in 2018 and sustained high output between 2019 and 2024, with annual counts consistently around 90–100 papers. The 2025 figure (n~=~50) reflects partial-year data rather than a decline, suggesting that the field remains in a phase of sustained activity. This trajectory closely mirrors previous bibliometric analyses~\cite{ma2023bibliometric, siu2025pedestrians}, which also identify 2017–2018 as the inflection point for rapid growth in eHMI research and show similarly stable levels of publication thereafter.

\begin{figure}[h]
    \centering
    \includegraphics[width=1\linewidth]{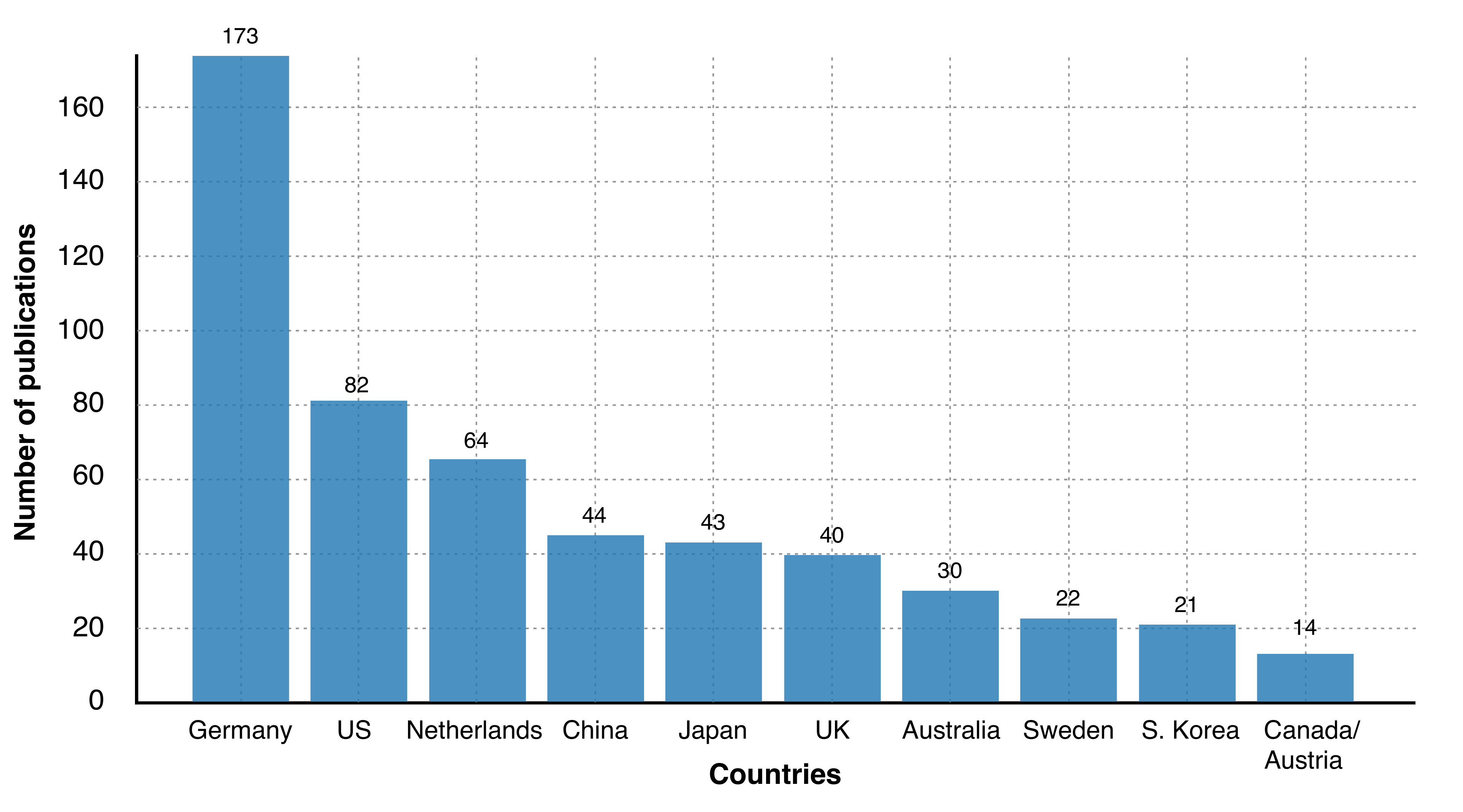}
    \caption{Top 10 countries by first-author affiliation in the eHMI dataset.}
    \Description{The bar chart shows the top 10 countries by first-author affiliation in the eHMI dataset. Germany leads with 173 publications, followed by the US (82) and the Netherlands (64). China (44) and Japan (43) contribute similar volumes, while the UK accounts for 40. Australia (30), South Korea (22), Sweden (21), and Austria (14) and Canada (14) make smaller but notable contributions.}
    \label{fig:countries}
\end{figure}

\begin{figure*}[h]
    \centering
    \includegraphics[width=0.95\linewidth]{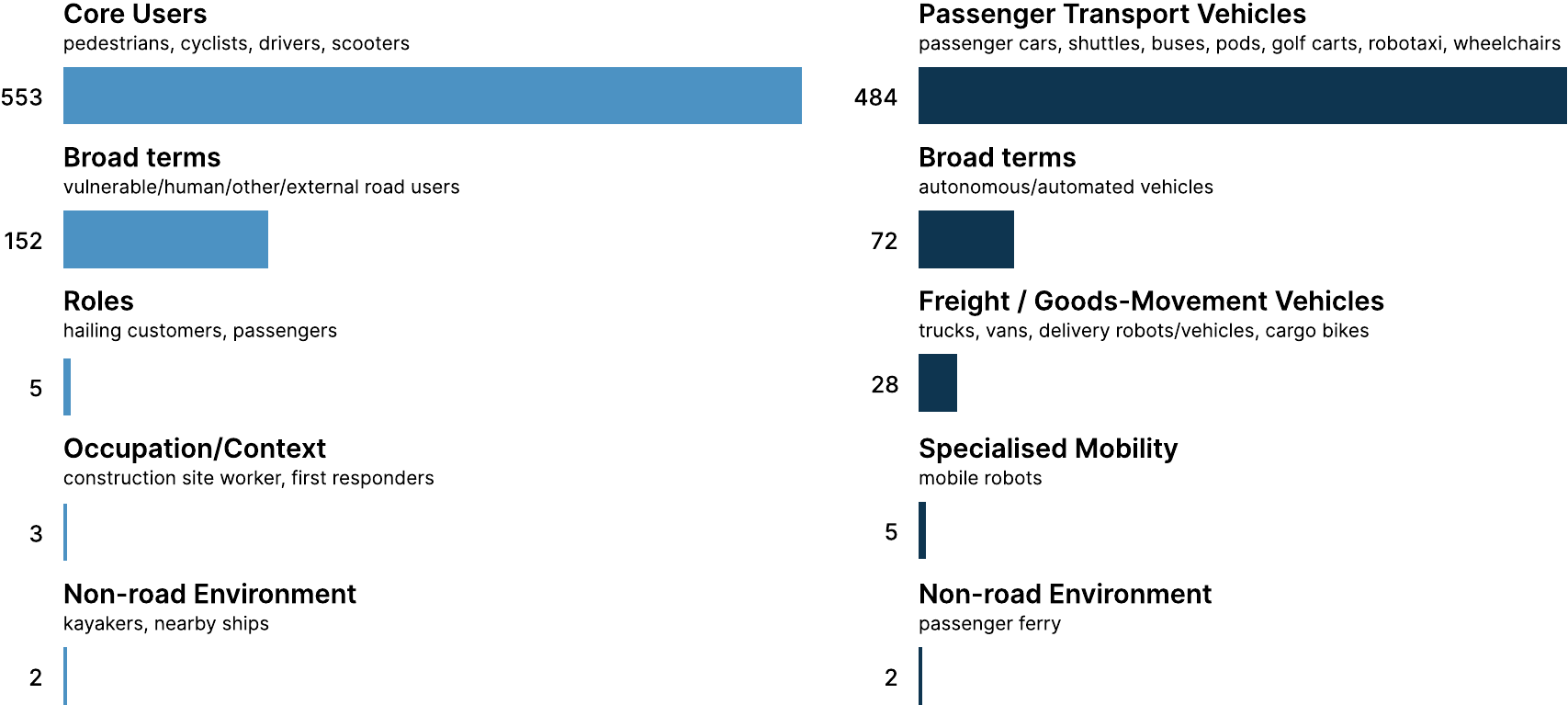}
    \caption{
    Distribution of road-user and vehicle-type categories in the reviewed studies. Counts exceed the number of studies because a single paper may include multiple mentions.}
    \Description{.}
    \label{fig:user_vehicle}
\end{figure*}

\subsubsection{Publication Formats}

The dataset is primarily composed of conference-associated outputs: full papers (n~=~285), extended abstracts (n~=~42), and workshop papers (n~=~11), which together represent over half of all publications. Journal articles (n~=~254) constitute the main alternative outlet, while other formats such as technical reports and book chapters are comparatively rare.

\subsubsection{Countries and Industry Collaborations} 

Based on the first author’s affiliation, publications in the core dataset represent contributions from 35 countries (see \autoref{fig:countries}). Germany leads with 173 papers, followed by the United States (82), the Netherlands (64), China (44), Japan (43), and the United Kingdom (40). Other countries with notable contributions include Australia (30), Sweden (22), and South Korea (21). The remaining countries each account for fewer than 20 publications, indicating a highly centralised concentration of research activity in a small number of nations.

Industry affiliations in the dataset highlight the influence of major automotive and technology companies across several regions. In Germany, 26 papers are linked to companies such as Volkswagen, Porsche, Mercedes-Benz, BMW, Continental, and Bosch. The United States contributes 13 papers through affiliations with firms including General Motors, Toyota, Nissan, Bosch, and Volkswagen. The United Kingdom is represented in seven papers with links to Nissan, Jaguar Land Rover, Fiat, BMW, and Ibex Automation. South Korea (six papers) is associated with Hyundai and Samsung, while Japan (three papers) is linked to Toyota, Honda, and Mitsubishi. China (two papers) includes contributions from SAIC Motor and Changan. Additional links include Nissan-affiliated papers in Denmark and the Netherlands, and a General Motors-affiliated paper in Israel. One technical report from SAE includes authors from Ford, Amazon, and Westat. These connections suggest that industry collaborations are concentrated among a small set of multinational automotive manufacturers with global R\&D footprints.

\subsubsection{Focus}
\label{sec:focus}

The papers were categorised into:
\begin{itemize}
    \item Core eHMI papers (n~=~563): studies that explicitly addressed eHMI as a primary focus.
    \item Peripheral eHMI papers (n~=~57): publications in adjacent domains where eHMIs were discussed tangentially.
\end{itemize}

These 57 papers reference eHMIs within broader contexts. Their contributions range from brief acknowledgements of eHMIs as one component in a larger system, to discussions of their role alongside other interfaces or technologies. Most commonly, these papers offered a brief mention in broader AV or interaction contexts, where eHMIs were noted as one of many design, safety, or communication considerations in automated mobility systems (e.g., \cite{fu2023adopting, Brown2023designing, lim2021ux, deshmukh2023systematic, hu2021review}). A smaller subset featured eHMIs alongside topics such as trust calibration, simulator sickness, and perceived comfort (e.g., \cite{gadermann2024increasing, schrauth2021acceptance}).

\begin{figure*}[h]
    \centering
    \includegraphics[width=1\linewidth]{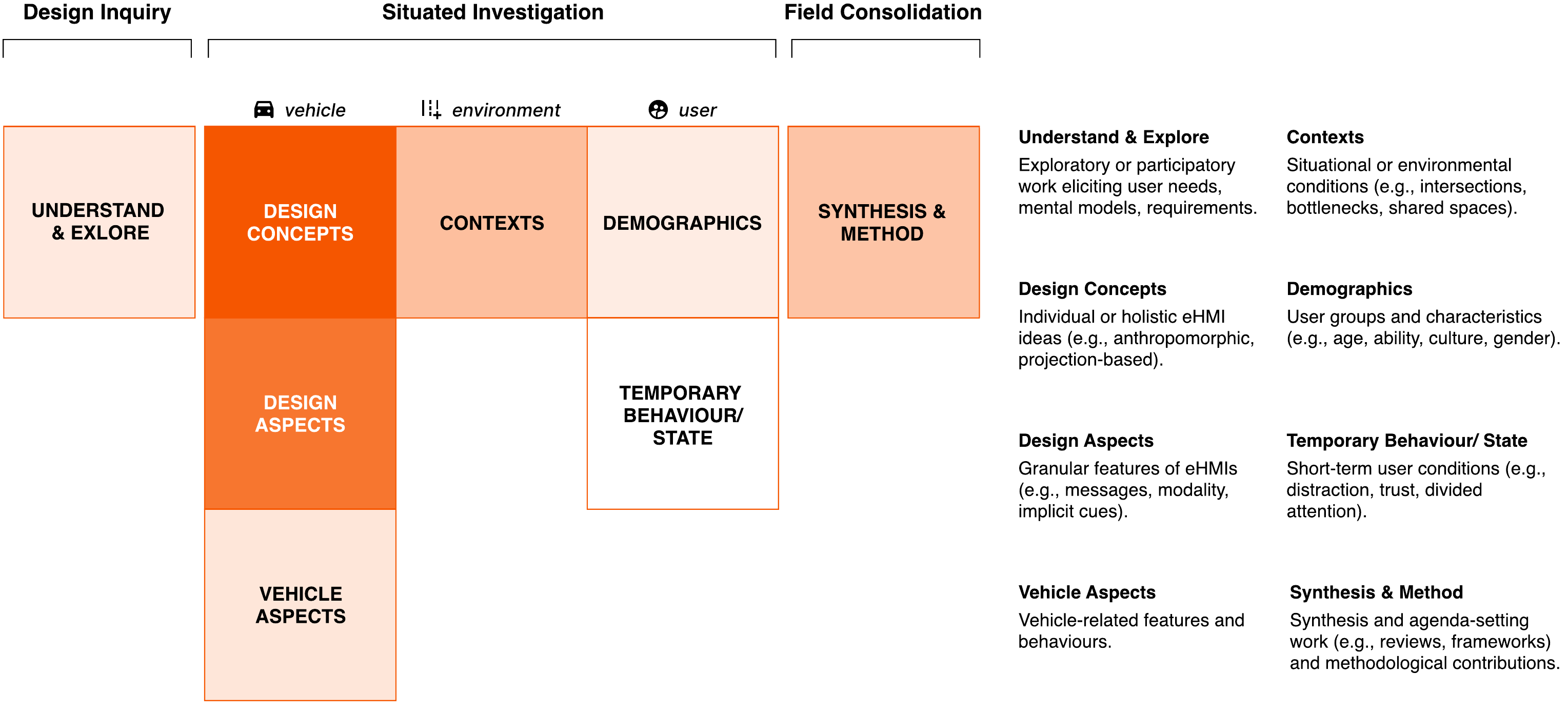}
    \caption{Heatmap overview of the high-level dimensions used to characterise AV external communication research. The framework distinguishes three orientations: Design Inquiry, Situated Investigation, and Field Consolidation. Values range from 11 (smallest) to 506 (largest).}
    \Description{Heatmap matrix showing nine dimensions mapped across three orientations: Design Inquiry, Situated Investigation, and Field Consolidation. Colour intensity encodes frequency, with Design Concepts forming the darkest and therefore most dominant block; mid-range tones appear in Design Aspects, Context, Synthesis \& Method, and Vehicle Aspects; light tones represent Understand \& Explore, Demographics, and Temporary Behaviour/State. The arrangement illustrates how exploratory needs-finding sits at the inquiry stage, concept and contextual work cluster in situated investigation, and methodological synthesis appears in consolidation.}
    \label{fig:orientations}
\end{figure*}

\subsection{Evolution of eHMI Research (RQ1)}

\paragraph{Road Users}

The literature clusters around a clear core user group. Pedestrians are by far the most frequently studied group (n~=~424), highlighting their centrality in AV–human interaction research. Drivers (n =73) and cyclists (n~=~51) follow as the next largest categories. Broader generic categories (n~=~152) such as Vulnerable Road Users, Human Road Users, External Road Users, and Other Road Users also appear in the literature, indicating that many studies adopt an umbrella framing rather than focusing on specific actors. A small number of studies examine specific interaction roles, such as hailing customers or passengers (n~=~5). These cases foreground service-oriented communication demands that differ from typical pedestrian–AV encounters.

Occupation- or context-specific actors also appear, though in far smaller numbers. These include construction-site workers and first responders, where AV interaction is shaped by specialised tasks and constrained, high-risk settings (n~=~3). Their inclusion highlights how eHMI research is beginning to account for specific communicative needs, where signalling is not only about negotiating right-of-way but also about ensuring worker safety, coordinating emergency responses, or supporting road management tasks. A further subset focuses on non-road environments such as waterways, involving kayakers and nearby ships (n~=~2). These cases sit at the periphery of the AV–human interaction literature but signal emerging interest in extending AV communication challenges beyond conventional road spaces.

\paragraph{Vehicle Types}

Passenger transport vehicles dominate the literature (n~=~481), confirming that eHMI research remains anchored in passenger-centric scenarios such as cars, shuttles, and buses. Within this group, passenger cars remain the primary platform studied (n~=~407). A second cluster adopts broad labels such as `autonomous/automated vehicles' (n~=~72), indicating work that abstracts away from platform-specific constraints. Freight and goods-movement vehicles form a smaller but distinct category (n~=~28), covering trucks, vans, delivery robots and vehicles, and cargo bikes. Specialised mobility platforms appear infrequently (n~=~5), generally involving mobile robots in constrained environments. Non-road environments are almost absent (n~=~2), with isolated cases such as passenger ferries.

\begin{figure*}[htbp]
    \centering
    \includegraphics[width=0.96\linewidth]{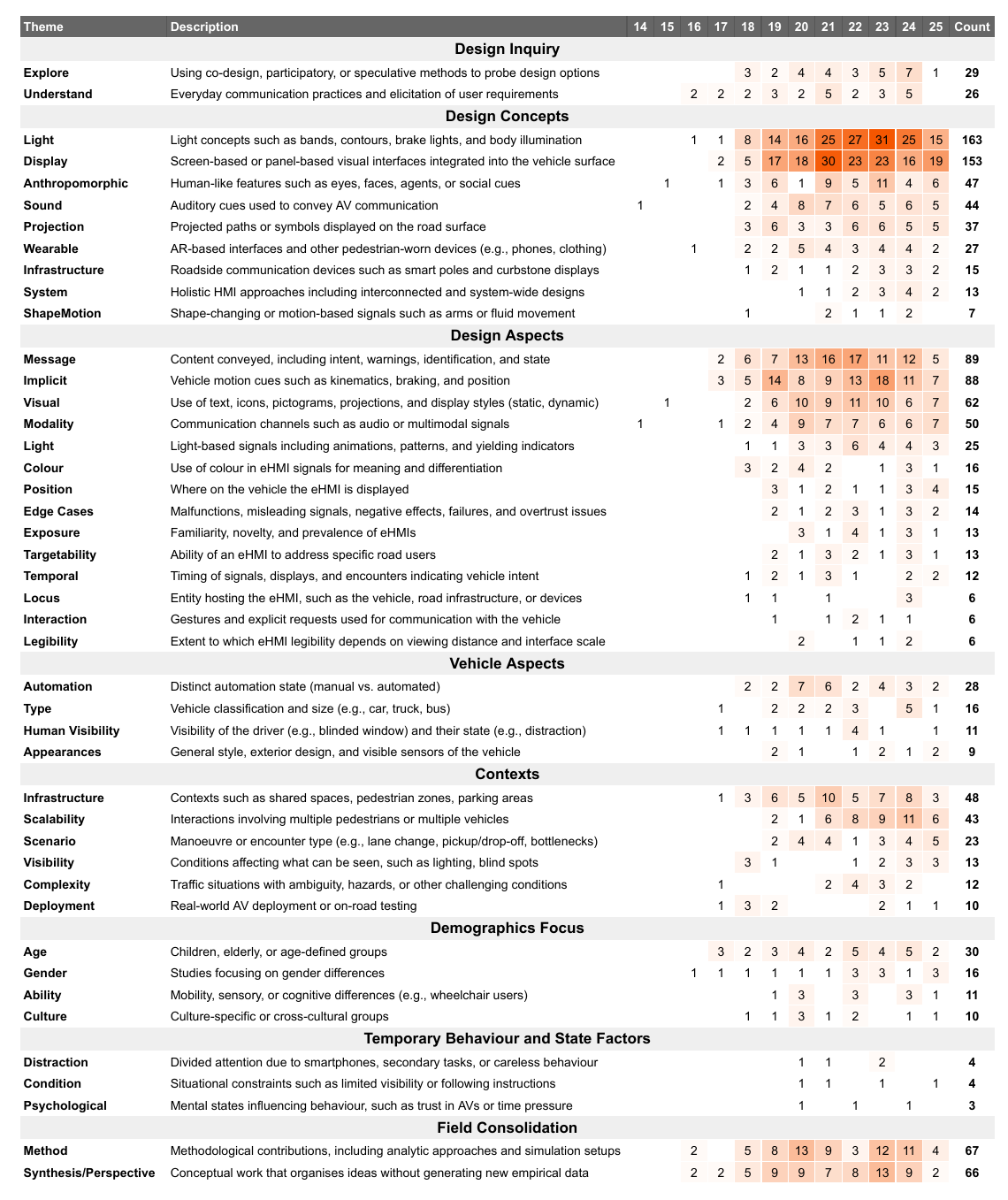}
    \caption{Heatmap illustrating the topical evolution of eHMI research (2014–2025). Each row represents a theme or sub-theme with intensity indicating the distribution of publications over time.}
    \label{fig:dimension_mapping}
    \Description{The heatmap comprises eight coded groupings across 2014–2025: Design Inquiry (understanding, exploration), capturing elicitation and framing; Design Concepts (light, display, anthropomorphic, sound, projection, wearable, infrastructure, system, shape–motion), representing singular or holistic signalling ideas; Design Aspects (message, implicit, visual, modality, light, colour, position, edge cases, exposure, targetability, temporal, locus, interaction, legibility), forming the most granular cluster; Vehicle Aspects (automation state, classification, driver visibility, exterior signalling), outlining vehicle-dependent features; Context (infrastructure, scalability, scenarios, complexity, visibility), defining situational conditions; Temporary Behaviour/State (distraction, divided attention, trust), describing short-term user states; Demographics (age, gender, ability, culture), capturing population variability; and Field Consolidation (method, synthesis), indicating review, integration, and methodological contributions.}
\end{figure*}

Having established the range of road users and vehicle types represented in the dataset, we next examine \textbf{the topical focus of research on external communication}. The three orientations introduced in the Method section guide this analysis. \autoref{fig:orientations} provides an overview heatmap of these orientations and their associated dimensions. \autoref{fig:dimension_mapping} expands this view by detailing the underlying sub-dimensions, together with their development over time.

\subsubsection{Design Inquiry}

\paragraph{Understand}

A body of work examines how road users currently communicate and allocate attention in everyday traffic. Observational studies, video analysis, and digital ethnography capture communication practices such as gestures~\cite{tran2024mapping}, interaction patterns~\cite{altaie2023keep}, and the role of implicit motion cues relative to explicit signals~\cite{lee2021road}. Eye-tracking studies~\cite{dey2019gaze, deWinter2021visualattention} complement this work by examining visual attention and perceptual priorities, providing insight into which cues road users notice or ignore. Other studies focus on early AV deployments or Wizard-of-Oz field studies, examining how interaction practices shift around driverless vehicles operating in public spaces, including emergent behaviours such as honking directed at AVs~\cite{passero2024honkable} and griefing of AVs~\cite{moore2020defense}.

A subset of studies relies on qualitative elicitation methods, including interviews, focus groups~\cite{alhawiti2024exploring}, expert surveys~\cite{ammar2024identifying}, workshops~\cite{colley2022truck}, and travel-along studies~\cite{vinkhuyzen2016developing}. These methods are primarily used to surface user needs, expectations, concerns, and interpretive strategies. This line of research extends beyond pedestrians to include cyclists, drivers, e-scooter riders, older adults, people with disabilities, and first responders~\cite{lee2023safe, harkin2024vulnerable, berge2022cyclist}. Several studies compare perspectives across these groups~\cite{harkin2024vulnerable} or focus on context-specific stakeholders such as construction-site drivers~\cite{colley2022truck} or public transport users~\cite{riener2021shuttle}.

\paragraph{Explore}

Studies in this strand commonly employ design-led and generative methods, including co-design workshops~\cite{asha2021wheelchairs, asha2022towards}, drawing sessions~\cite{alhawiti2024exploring}, creative workshops~\cite{riener2021shuttle}, scenario-based design activities~\cite{lee2021discovering}, and expert panels~\cite{tabone2021towards}. These studies treat participants as contributors to design thinking, using design interventions to open up the design space and surface preferences, values, and concerns around how eHMI concepts might be approached. Outcomes are typically framed as design implications, conceptual frameworks, mapped design spaces, or low- to mid-fidelity design concepts, supporting ideation and informing subsequent research.

A notable subset of Exploration work explicitly targets accessibility and inclusion, examining communication needs for specific groups such as wheelchair users~\cite{asha2021wheelchairs} or exploring alternative modalities beyond vision~\cite{asha2022towards}. Other studies probe novel or unconventional roles for eHMIs, including emotional expression~\cite{dong2024insideout} and help-seeking behaviours~\cite{yu2024agent}, extending the design space beyond signalling intent alone.

\subsubsection{Situated Investigation}

\paragraph{Design Concepts}

Light is the dominant concept by a wide margin (n~=~163), encompassing light bands or strips, vehicle contour illumination, frontal brake lights, headlights, marker lamps, and other light patterns. It accounts for the highest number of instances and appears across the widest range of studies, indicating its role as the de facto baseline for eHMI design. Display forms a strong secondary cluster (n~=~153). It is the most substantial visual alternative to light-based signalling and reflects continued interest in richer, information-dense interfaces.

\begin{figure*}[h]
    \centering
    \includegraphics[width=0.75\linewidth]{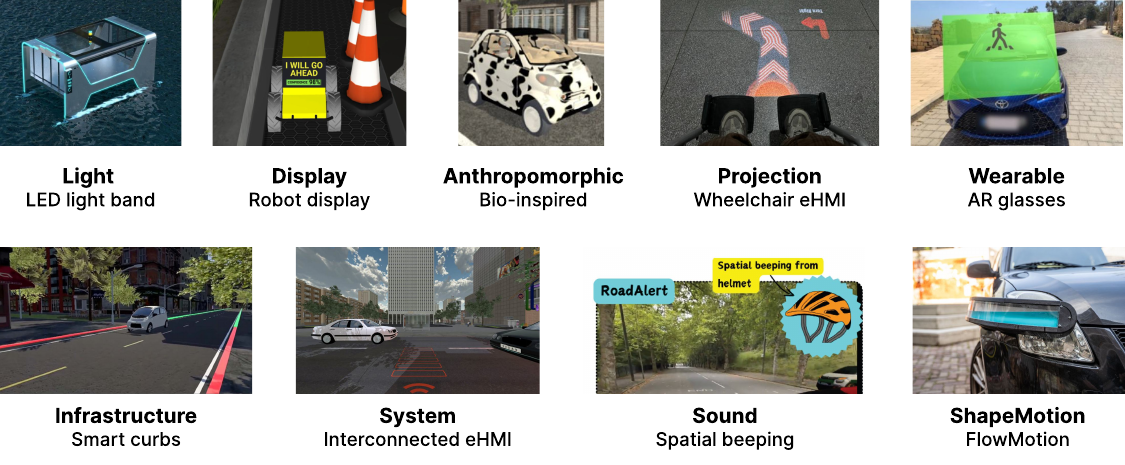}
    \caption{Illustrative examples of the nine design concept categories identified in the dataset. Each image represents one concept type corresponding to the coding scheme used in the review: Light~\cite{simic2023automation}, Display~\cite{cumbal2025visualising}, Anthropomorphic~\cite{oudshoorn2021bio}, Projection~\cite{zhang2024shared}, Wearable~\cite{tabone2021towards}, Infrastructure~\cite{hollander2022take}, System~\cite{tran2024interconnected}, Sound~\cite{altaie2025scalable}, ShapeMotion~\cite{dey2021flow}.}
    \Description{The figure presents nine example images, each illustrating one design concept category used in the review. From left to right, the top row shows: a vehicle-mounted LED light band (Light), a screen-based robot display (Display), a vehicle with bio-inspired facial features (Anthropomorphic), a ground projection showing a wheelchair icon (Projection), and AR glasses used for communication (Wearable). The bottom row shows: smart curb infrastructure emitting signals (Infrastructure), an interconnected vehicle–road communication system (System), a cyclist helmet emitting spatial beeps (Sound), and a vehicle with a dynamic light element that changes shape (ShapeMotion). Each example corresponds to one of the nine coded concept types and is included to illustrate the range of eHMI design approaches.}
    \label{fig:design-concepts}
\end{figure*}

Anthropomorphic (n~=~47) and Projection (n~=~37) occupy a mid-tier. Anthropomorphic concepts most commonly involve animated eyes or emotional expression, while projection is used mainly for path-related or zebra-crossing cues. Sound (n~=~44) covers electronically generated auditory cues, either as standalone auditory HMIs or as redundancy accompanying a visual channel. Wearable (n~=~27) and Infrastructure-integrated designs (n~=~15) remain peripheral relative to vehicle-mounted concepts. System-level concepts (n~=~13) position eHMI within the broader vehicle HMI ecosystem (e.g. encompassing both internal and external interfaces or networked vehicles) and within wider user-experience contexts such as ride-sharing. ShapeMotion is the least represented category (n~=~7), using mechanical arms, gestures, or physical transformations. Overall, design concepts have broadened over time, but the field continues to be anchored by light- and display-based approaches. Illustrative examples of each design concept are shown in \autoref{fig:design-concepts}.

\paragraph{Design Aspects} 

Design aspects are the smallest units of eHMI communication. Attention at this granular level is foundational for explaining how eHMIs function, for accumulating shared knowledge about effective communication strategies, and for informing future standardisation efforts. Among them, studies of Message (communication purpose) form the single largest body of work (n~=~89). This work examines a range of message-level considerations, including which information to communicate~\cite{merat2018what, faas2020external}, communication style (e.g. advisory, informative, polite, or dominant)~\cite{weiss2022external}, perspective (e.g. egocentric or allocentric)~\cite{eisma2021perspective}, and whether messages convey intent, perception, or a combination of both~\cite{loew2022goahead}. Messages are frequently tailored to different road users, vehicle platforms, and contexts. For pedestrians, designs have ranged from situational awareness cues and intent displays to emotional expressions and passenger pick-up–related information, tested across passenger cars, shuttles, buses, delivery robots, and micro-mobility platforms. Cyclist-focused studies have explored awareness messages and HUD-style warnings, while driver-directed work has examined intervene status, minimum-risk manoeuvres, and scenario-based decisions.

Implicit cues conveyed through vehicle kinematics and braking (n~=~88) form the second largest body of work. This area is particularly significant given ongoing arguments that implicit motion cues may be more reliable or preferable than explicit eHMIs. Studies in this cluster examine implicit cues in isolation~\cite{ackermann2019deceleration, moore2019case}, as well as situations in which they align with or conflict with eHMI signals~\cite{lau2021investigating}, providing insight into how pedestrians interpret combined or competing sources of information. In comparison, research on Visual aspects (n~=~62) and Modality (n~=~50) is moderately represented. Modality-focused studies primarily investigate sound-based cues, with a smaller number exploring haptic feedback, such as vibrations delivered through watches or pedestrians’ mobile phones~\cite{altaie2025cultures}. Light-based signalling patterns (n~=~25) constitute a more focused subset, examining variations in animation and rhythm.

Other granular design aspects, including Colour, Position on the vehicle, Timing, and Targetability (i.e., communication directed at specific road users), appear less frequently but contribute important nuance to message interpretation. Similarly, Locus, Interaction, and Legibility remain relatively small but meaningful lines of inquiry. Notably, a subset of studies explicitly investigates Edge Cases, such as communication failures or conflicts~\cite{faas2021calibrating}, as well as the effects of repeated exposure to eHMIs over time~\cite{faas2020longitudinal, colley2022time}.

\paragraph{Vehicle Aspects} 

Vehicle aspects form a comparatively small body of work. Most studies examine Automation distinctions (n~=~28), typically comparing manual and autonomous operation to assess effects on pedestrian behaviour and trust. Type (n~=~16) explores vehicle classification and size, testing how cars, trucks, buses, or smaller platforms shape user judgements. Human Visibility (n~=~11) investigates whether the presence, absence, or obscuring of a driver influences interpretation. Appearances (n~=~9) focuses on exterior styling, visible sensors, and perceived `friendly' or `aggressive' characteristics. Several studies combining multiple vehicle-related factors within the same design to capture interaction effects. 

\paragraph{Contexts} 

Infrastructure and Scalability dominate contextual investigations (n~=~48 and 43 studies respectively). This concentration shows that eHMI research is primarily concerned with how spatial layouts and traffic density shape communication demands. Scenario-focused work (n~=~23) forms a clear secondary cluster, indicating interest in manoeuvre-specific interactions (bottlenecks, turning manoeuvres, and pickup or drop-off events) but far less systematic attention. Complexity (n~=~12) addressed ambiguous or high-risk encounters, while visibility (n~=~13) focused on low-light, blind spot, and approach-angle conditions. They both appear infrequently, suggesting that challenging or degraded conditions remain underexplored despite their relevance to safety-critical design. Current Deployment (n~=~10) reflects studies situated in environments where AVs are already operating.

\paragraph{Temporary Behaviour and State Factors} 
These were the least represented, with 11 studies. Psychological states included specific attitudes towards AVs \cite{ackermans2020effects}, being under time pressure \cite{colley2022time}, and responding to varying crossing opportunities categorised as safe, risky, or unsafe \cite{hochman2024pedestrians}. Distraction-related conditions were common, with studies introducing smartphone use \cite{lanzer2023interaction,hollander2020save}, secondary tasks \cite{zhao2023invisible}, less careful crossing behaviour \cite{lee2025mitigating}, or general carelessness \cite{daimon2021pedestrian} to assess their influence on pedestrian responses. Other work tested situational conditions such as limited visibility \cite{lim2023visibility}, receiving instructions \cite{liu2021instruction,liu2025preinstruction}, or being primed with particular descriptions of AVs before interaction \cite{hagenzieker2020interactions}.

\paragraph{Demographics Focus}
Demographic focus spanned ability, age, culture, and gender. Ability-related research investigated hearing-enhanced pedestrians, people with disabilities, and subgroups such as wheelchair users, vision-impaired pedestrians, and those with reduced mobility (e.g., \cite{asha2022towards,asha2020wheelchair,colley2020inclusive}). Age-focused studies examined differences across children, adults, and elderly participants, with some targeting specific age groups such as children \cite{bluhm2023frontal} or elderly \cite{hensch2022effects}. Cultural context was addressed in cross-country or intercultural comparisons to explore variations in eHMI preferences or interpretations (e.g., \cite{wesseling2020exploring,weber2019crossing}). Gender appeared in the analysis of multiple studies, with one explicitly centring its investigation on gender differences~\cite{chang2020gender}.

\subsubsection{Field Consolidation}

\paragraph{Synthesising and Perspective Contributions}

Synthesising works take diverse forms, including frameworks and conceptual models~\cite{rouchitsas2025framework, enam2024external, Bengler2020}, state-of-the-art reviews and taxonomies~\cite{dey2020taming, tran2023scoping, Leveque2020where}, and theory-oriented papers linking eHMI research to broader domains such as communication, robotics, and traffic engineering~\cite{domeyer2020vehicle}. Other works comprise conference workshops~\cite{locken2022accessible, sahin2021prosocial, dong2023holistic, dong2024insideout, dey2018methodology, bazilinskyy2024multiagent, locken2020wecare, moore2019wizards}, which provide spaces for collective reflection and discussion among researchers.

Perspective papers include vision pieces and position papers presenting expert views~\cite{milford2024integrating, hesse2021holistic, robert2019future, tabone2021vulnerable, schaudt2019judging}. Unlike syntheses, these works do not primarily consolidate evidence but instead articulate conceptual outlooks, challenge assumptions, and connect eHMI debates to broader system-level concerns. For instance, one contribution gathered perspectives from sixteen human factors researchers on AVs and the use of AR for addressing eHMI scalability issues~\cite{tabone2021vulnerable}, illustrating how perspective papers surface expert opinion and diverse viewpoints.

\paragraph{Methodological Contributions}

A total of 67 papers focused on methodological development in the study of eHMIs. These contributions can be grouped into several themes. First, many works introduce new simulation and testbed environments, including VR frameworks for pedestrian and cyclist interactions~\cite{kooijman2019ehmis, camara2019examining, qi2023investigating, lee2019investigating, dalipi2020benchmark, lee2023IntVRsection, yeo2023cycling, kaß2020methodological}, mixed-reality setups~\cite{drechsler2021mixed}, digital twins~\cite{serrano2023digital}, Wizard of Oz vehicles~\cite{rothenbucher2016ghost, bindschadel2022studying, diederichs2021wizard, moore2019wizards}, coupled and multi-vehicle simulators~\cite{bazilinskyy2020coupled, feierle2020multi}, and microsimulation models~\cite{orlicky2021microsimulation, orlicky2021assessment}.

Second, several papers propose standardised protocols, including test procedures~\cite{kaß2020standardised, yan2025comparing}, frameworks for disentangling efficiency and effectiveness~\cite{rouchitsas2025framework}, and reviews of evaluation practices~\cite{zheng2023literature}. Third, some contributions develop new measurement approaches, such as physiological indicators~\cite{wang2024physio, iwamoto2024skin, guo2022virtual}, behavioural measures like onset time and role-switching~\cite{faas2020efficient, gao2024role, almeida2024reward}, and perceptual measures including eye-tracking and alternatives~\cite{Leveque2020where, gadermann2025codecharts}. Finally, a number of works focus on design and prototyping methods, from context-based prototyping~\cite{hoggenmueller2021context, dongas2023virtual} and tangible toolkits~\cite{hoggenmuller2020tangible} to optimisation techniques~\cite{colley2025optimisation} and low-code integration of machine learning models~\cite{winkelmann2023node}.

\subsection{Consensus vs. Fragmentation in Findings (RQ2)}

\begin{table*}[h]
\centering
\caption{Review and taxonomy papers on eHMIs, grouped by synthesis focus.}
\label{tab:review_papers}
\footnotesize
\begin{tabular}{@{}L{3.2cm}L{4.5cm}p{5.5cm}@{}}
\toprule
\textbf{Synthesis Focus} & \textbf{Typical Contributions} & \textbf{Representative Papers} \\
\midrule
Bibliometric / Systematic & 
Map publication growth, research topics, and citation patterns & 
\cite{ma2023bibliometric} 679 articles, Web of Science \newline 
\cite{siu2025pedestrians} 234 articles, Web of Science \\
\addlinespace

General / Integrative Reviews & 
Provide integrative overviews of AV–road user interaction, including issues of necessity, scalability, social roles, and user acceptance & 
\cite{bied2024autonomous} AVs as Social Agents \newline 
\cite{rasouli2019autonomous} Factors Influencing Interactions \newline 
\cite{rouchitsas2019external} Empirical Work \newline 
\cite{wang2021can} Emotional Expressions \newline 
\cite{tran2023scoping} Scalability Issues \newline
\cite{colley2020unveiling} Scalability Issues \newline 
\cite{yan2023user} User Acceptance \newline
\cite{deb2018pedestrians} Pedestrian Receptivity \newline
\cite{de2022external} For and Against Arguments \newline 
\cite{ezzati2021interaction} Challenges and Recommendations \newline 
\cite{block2023road} Challenges and Opportunities \newline 
\cite{joshi2022advancing} Challenges and Opportunities \\
\addlinespace

Evaluation Methods & 
Synthesise evaluation approaches (VR, usability, physiological, eye-tracking) and identify methodological biases & 
\cite{zheng2023literature} Usability Evaluation Practice \newline 
\cite{tran2021review} VR Simulation \newline 
\cite{le2020automotive} VR Simulation \newline 
\cite{colley2019better} VR Simulation \newline 
\cite{wang2024physio} Physiological Measurements \newline 
\cite{leveque2020pedestrians} Eye-tracking Studies \newline
\cite{souman2021overview} Standards and Evaluation Criteria \newline
\cite{joshi2023mapping} Standards Gaps \\
\addlinespace

User Groups and Contexts & 
Review studies on specific road users (e.g., people with disabilities, older adults, VRUs) and interaction contexts (e.g., shared spaces) & 
\cite{enam2024external} People with Disabilities \newline 
\cite{colley2019including} People with Impairments \newline 
\cite{fabricius2022interactions} Highly Automated Trucks \newline 
\cite{predhumeau2021pedestrian} Shared Spaces \newline 
\cite{brill2023external} Shared Spaces \\
\addlinespace

Guidelines / Taxonomies & 
Propose design principles, classification schemes, and taxonomies for eHMIs & 
\cite{schieben2019designing} Design Considerations \newline 
\cite{carmona2021ehmi} Guidelines for Deployment \newline 
\cite{gao2022external} All Road Users Perspective \newline 
\cite{wilbrink2023principles} Principles for eHMIs \newline 
\cite{dey2022shape} Shape-Changing Interfaces \newline 
\cite{dey2020taming} Taxonomy of eHMIs \newline
\cite{hollander2021taxonomy} Taxonomy of VRUs \newline 
\cite{colley2020designspace} Design Space for eHMIs \\
\addlinespace

Peripheral Reviews & 
Survey broader AV/AR/HCI domains where eHMIs are discussed tangentially (e.g., intelligent vehicles, AR applications, trust, accessibility) & 
\cite{hu2021review} Key Challenges in Intelligent Vehicles \newline 
\cite{sankeerthana2022strategic} Adoption of AVs and Risk Perception \newline 
\cite{riegler2019systematic} AR for Automated Driving \newline 
\cite{deshmukh2023systematic} Challenging Scenarios with VRUs \newline 
\cite{riegler2021systematic} VR for Automated Driving \newline 
\cite{riegler2021augmented} AR for Future Mobility \newline 
\cite{stefanidi2024augmented} AR on the Move (VRUs) \newline 
\cite{yuan2024driving} AV Interaction Design \newline 
\cite{bastola2025driving} Accessibility and Disability \newline 
\cite{yan2022inclusive} Inclusive Shared AVs \newline
\cite{omeiza2021explanations} Explanations in Autonomous Driving \newline 
\cite{tekkesinoglu2025advancing} Explainable Autonomous Vehicle Systems \newline
\cite{zhou2021factors} Factors Affecting Pedestrian Trust \newline 
\cite{kumar2024human} AV Interaction Technologies \newline 
\cite{zhang2021human} Human–Machine Interaction \newline 
\cite{gadermann2024increasing} Evaluation Methods for Interaction Design \newline 
\cite{reyes2022vulnerable} VRUs and Connected AVs \newline 
\cite{Bengler2020} HMI Framework \newline
\cite{ayoub2019manual} Manual to Automated Driving \newline
\cite{stanciu2018pedestrian} Driver–cyclist–pedestrian Communication \\
\bottomrule
\end{tabular}
\end{table*}


55 were review or taxonomy papers (see \autoref{tab:review_papers}). These secondary sources cover a wide range of topics, but several trends can be observed. Bibliometric and systematic reviews provide high-level mapping of publication growth and topic clusters. Over nearly a decade, research has consistently centred on safety, trust, and effective communication as key dimensions of pedestrian--AV interaction~\cite{ma2023bibliometric}. Pedestrian crossing behaviour and evaluations of eHMIs are influenced by: human factors (e.g., age, nationality), vehicle factors (e.g., type, colour, position), and environmental factors (e.g., signalisation, distractions)~\cite{siu2025pedestrians}. Designs should therefore be context-sensitive and inclusive.  

A substantial number of reviews concentrate on evaluation methods, particularly VR simulation, usability testing, physiological measures, and eye-tracking, representing attempts to improve and refine how eHMIs are assessed. Another set of reviews focuses on specific user groups and contexts, such as VRUs, people with disabilities, and interactions in shared spaces. A further theme proposes design guidelines and taxonomies, aiming to structure the design space and support standardisation. Finally, integrative reviews discuss broader themes such as the social role of AVs, emotional communication, scalability, and user acceptance. 

The review and taxonomy papers show that secondary synthesis efforts are numerous but typically narrow in scope, with each focusing on a particular method, user group, or design perspective. They also indicate which topics in eHMI research have accumulated sufficient density to justify a dedicated review, providing a proxy for areas that the community has considered especially important. While these review and taxonomy papers vary in focus, they also surface areas of agreement, disagreement, and forward-looking recommendations. Synthesising across them, three key patterns can be identified: (i) themes where the literature shows broad \textbf{\textit{consensus}}, (ii) areas marked by \textbf{\textit{fragmentation}} and debate, and (iii) \textbf{\textit{suggested directions}} for future research and practice. We report these below.


\subsubsection{Consensus} \hfill

\textbf{Necessity of Interaction (Baseline Agreement).}  
There is broad agreement that some form of interaction between pedestrians and AVs is indispensable~\cite{rasouli2019autonomous}. 
Studies consistently show that vehicles equipped with communication interfaces lead to more effective, efficient, safer, and more satisfactory interactions compared to vehicles without them~\cite{rouchitsas2019external}.  
At a minimum, communication should provide a replacement mechanism for \textit{eye contact}, ensuring pedestrians feel acknowledged and reducing uncertainty especially in situations where ambiguity is not resolved by vehicle kinematics alone~\cite{rasouli2019autonomous}.  
Implicit communication remains highly effective, but `just motion is not enough' in all situations; additional cues may be needed to resolve ambiguity~\cite{rasouli2019autonomous}.  

\textbf{Message Qualities and User Experience.}  
Messages should be stress-free and easy to understand~\cite{rasouli2019autonomous}.  
Communication must balance clarity with flexibility, avoiding over-prescription that could mislead pedestrians in dynamic environments.  

\textbf{Communication Style: Allocentric vs. Egocentric.}  
Although empirical findings sometimes show pedestrian preference for \textit{egocentric} messages (i.e., direct instructions), the consensus is that communication should be \textit{allocentric} (i.e., conveying vehicle state and intent). Allocentric cues are less prone to misunderstanding~\cite{bied2024autonomous}.  
Messages should be \textit{informative rather than advisory}, offering information about the vehicle’s behaviour rather than telling pedestrians what to do~\cite{rasouli2019autonomous}.  
Explicit advice or instructions should be avoided, since they may not suit situations with multiple pedestrians at once~\cite{rouchitsas2019external}.  

\subsubsection{Fragmentation} \hfill

\textbf{Explicitness and the Necessity of eHMIs (Ongoing Debate).}  
While there is consensus that interaction of some form is necessary, disagreement persists over whether explicit eHMIs are required in addition to vehicle motion. Some argue that explicit eHMIs are essential to mimic the social interaction pedestrians expect from human drivers, while others contend that motion cues are sufficient and explicit signals are only useful in specific contexts~\cite{bied2024autonomous}.  
\citet{de2022external} reinforce this divide, outlining arguments both against and in favour of eHMIs. Critics argue that vehicle motion already dominates pedestrian interpretation and leaves no interaction void to be filled, that the diversity of eHMI concepts creates unresolvable dilemmas, and that competing for pedestrians’ limited visual attention risks confusion and overreliance. Supporters counter that eHMIs can complement implicit communication to achieve `superhuman performance,' are wanted and accepted by road users, can communicate more than simple stop/go instructions (e.g., automation status), substitute for eye contact, and mitigate misperceptions of motion cues.  

\textbf{Modality and Cue Type.}  
Fragmentation is also pronounced in debates over which modality or cue type is most effective. Visual eHMIs are vulnerable to poor weather or lighting; acoustic cues are difficult in noisy environments and risk confusion if widely adopted; and haptic concepts assume pedestrians carry dedicated devices, which is unrealistic~\cite{bied2024autonomous}.  
Light-based eHMIs, while common, are abstract and require pedestrians to learn their meanings in advance. Text and icon-based systems offer clearer information but face cross-cultural challenges~\cite{siu2025pedestrians}.  
Evidence on content and coding is equally inconsistent: some studies report advice is preferred over vehicle state information, while others find no effect; intention cues are often prioritised over detection cues, yet not universally; and preferences for coding vary across lights, sounds, text, and pictorial designs~\cite{rouchitsas2019external}.  
Beyond individual modalities, the field suffers from a lack of robust multimodal evaluations. Most studies focus on single-modality systems, with the few cross-modality comparisons relying on small sample sizes~\cite{rasouli2019autonomous}, leaving insufficient evidence to converge on best practices~\cite{siu2025pedestrians}.  

\subsubsection{Influences on Fragmentation and Suggestions} \hfill

\textbf{Human Factors and Context.}  
No unified framework exists because expectations and interpretations vary across cultures, social norms, and road contexts (e.g., urban versus rural)~\cite{rasouli2019autonomous}. While factors such as age, gender, and traffic culture are acknowledged as influential, their specific effects remain underexplored~\cite{siu2025pedestrians}. The broader driving context also matters: the type of road (segregated versus complex mixed environments) may determine the communication demands placed on eHMIs~\cite{de2022external}.

\textbf{Ecological Validity.}  
Much of the most convincing evidence stems from laboratory and VR-based studies using objective measures such as reaction time and accuracy, while studies with physical prototypes in real traffic often rely on subjective ratings. This discrepancy raises doubts about how well experimental findings translate to real-world interactions~\cite{rouchitsas2019external, siu2025pedestrians}. The field also lacks standardised methods for evaluating AV–pedestrian communication. Reviews note that many studies adopt a `ceteris paribus' approach that fails to isolate the effects of specific parameters (e.g., technology, location, content type, modality), making comparisons inconclusive~\cite{rouchitsas2019external}. Longitudinal work is needed to understand how perceptions and practices evolve over time. This includes moving beyond narrow laboratory contexts towards broader ecological validity~\cite{rasouli2019autonomous}.  

User-centric design should guide the development of eHMIs, ensuring that standards are grounded in behavioural evidence and diverse user studies~\cite{siu2025pedestrians}. Cross-disciplinary collaboration involving ergonomics, psychology, and transportation safety experts is seen as vital for developing robust standards and avoiding siloed approaches~\cite{bied2024autonomous, siu2025pedestrians}.  

\textbf{Standardisation.}  
Standardisation is considered particularly urgent for light-based eHMIs, whose abstract nature requires prior learning to ensure comprehension~\cite{siu2025pedestrians}.  

Several reviews call for the standardisation of communication modalities and protocols. Recommendations~\cite{de2022external, rasouli2019autonomous, siu2025pedestrians} include:  
(i) Research should address how the recognisability of AVs shapes pedestrian interaction. Factors include the visibility of external sensors (e.g., roof-mounted lidar), the use of `self-driving' stickers, and whether a safety driver is present. These features interact with eHMI cues and may influence trust and behaviour.  
(ii) developing consistent, internationally recognised colours and symbols (e.g., cyan and white) for visibility and neutrality;  
(iii) prioritising multimodal systems that combine visual, text, icon, and auditory signals rather than relying on a single channel;  
(iv) defining specifications for eHMI elements such as colours, sizes, and placements within AV designs.  

Consistency across AV manufacturers and regions is also emphasised, as fragmented practices risk confusing pedestrians and undermining trust~\cite{siu2025pedestrians}. Moreover, liability considerations are largely absent from current research; manufacturers may view eHMIs as a mechanism for transparency and legal accountability~\cite{de2022external}. 

\subsection{Alignment with Industry and Regulation (RQ3)}

\subsubsection{Industry Patents, Concepts, and Deployments} \hfill


\textbf{Patents}: Identified patents (n~=~7) collectively show that major companies are converging on external communication as a core part of AV design. Lyft~\cite{lyft2020notification} and Uber~\cite{uber2018lightoutput} present the most context-aware, multi-channel signalling concepts, using dynamic displays that adapt to user position, state, and environmental conditions. Google/Waymo~\cite{waymo2015pedestrian} covers a more foundational intent-notification approach, namely, detect, decide, notify. Nissan~\cite{nissan2024notification} goes beyond pedestrian-only signalling to explicitly mirror information for internal occupants and external road users. Apple~\cite{apple2023onewayfilter} focuses narrowly on display visibility technology (a one-way filter over LEDs) rather than the messaging logic itself, making it a hardware enhancement patent that could underpin other companies’ systems. GM Cruise~\cite{gmc2024identification} addresses fleet-level passenger–vehicle identification, aiming to prevent misboarding and streamline rideshare operations, a point also reflected in Lyft~\cite{lyft2020notification}. GM Global~\cite{gm2023responses} addresses a different challenge: how to respond to adversarial behaviour from vulnerable road users. Depending on assessed risk, the system could escalate from visual and auditory warnings to evasive rerouting, or even notifying authorities.   

In terms of design patterns, most rely on visual channels supplemented by auditory cues; projection appears in Lyft, Uber, and Waymo; tactile feedback is unique to Nissan. There is a growing emphasis on adaptive directionality, ensuring the right message reaches the right audience from the right part of the vehicle. Regarding timeline, the analysis shows a first wave of patents in 2013–2015 establishing baseline AV–pedestrian communication (Waymo), a 2017–2018 cluster introducing richer multi-modal and context-aware systems (Uber, Lyft, Nissan), and a 2021–2023 wave addressing operational fleet needs (GM Cruise), hardware performance in real-world conditions (Apple), and adversarial interaction scenarios (GM Global).


\textbf{Concepts}: 
Concept designs for delivery robot eHMI (n~=~2) emphasise branding and commercial use. Kiwibot~\cite{Kiwibot2024} explored transforming robots into mobile billboards after acquiring an advertising startup, while Ottobots~\cite{Ottobots2023} proposed multi-screen displays for interactive experiences and customer engagement. Both examples position the delivery robot as a moving platform for media exposure and brand interaction.

Concept designs for passenger car eHMI (n~=~29) reveal the most diverse and experimental space. A strong cluster of work explores road projection to guide pedestrians or signal vehicle intent; examples include Rinspeed’s Oasis~\cite{Rinspeed2017} projecting status or entertainment, Jaguar Land Rover’s Projection Pod~\cite{JaguarLandRover2019} projecting lines and turns, Audi’s Matrix Laser~\cite{Audi2015} spelling `Stop!' at pedestrians, and Hyundai’s Lighting System~\cite{Hyundai2023} as well as Mercedes-Benz's F015 Concept~\cite{Daimler2015} projecting text and symbols like illuminated crosswalks. Parallel to this, several manufacturers focused on 360-degree LED or light bar signalling: Mercedes’ turquoise light systems~\cite{MercedesBenz2019-2}, BMW’s Vision Next 100~\cite{BMW2016}, and Nissan’s IDS Concept~\cite{NissanMotorCorporation2015}, each designed to make autonomous mode and pedestrian recognition visible from all angles. Anthropomorphism was another recurring theme: Jaguar’s Virtual Eyes~\cite{JaguarLandRover2018} and Semcon’s Smiling Car~\cite{Semcon2016} simulated eye contact or smiles to reassure pedestrians, while Toyota’s e-Palette~\cite{ToyotaMotorCorporation2018} used eye-like headlamps. Others pushed multi-functional or personalised displays, from Intel’s window-mounted passenger name displays~\cite{Intel2018} to Sony-Honda’s media bar~\cite{SonyHonda2023} and Daimler’s smart EQ fortwo~\cite{Daimler2017}, which combined personalised panels with event or weather information.

Auditory innovations were rarer but notable: Kia’s Niro~\cite{Kia2018} proposed directed sounds to warn pedestrians, while Volvo’s 360c~\cite{VolvoCars2018} suggested directional ultrasonics combined with visual cues. Across these concepts, message types ranged from warnings and instructions (e.g., Audi, Mercedes Digital Light) to intent and awareness signalling (e.g., Nissan, Daimler cooperative car) and even entertainment and advertising (e.g., Rinspeed, Sony Honda). Overall, concept-stage designs reflect a highly experimental period where carmakers proposed a broad palette of modalities and messages, often blending safety, reassurance, personalisation, and branding. 

\textbf{Deployments}:  
Deployed eHMI in delivery robots (n~=~4) reveals a dual trajectory: some designs emphasise safety and accessibility, using lights, flags, and operating sounds to ensure presence is noticed (e.g., Coco~\cite{Coco2021}), while others lean into anthropomorphic expression, adding eyes, emojis, and displays to humanise intent and evoke emotion (e.g., Kiwibot~\cite{Kiwibot2022}, Serve Robotics~\cite{ServeRobotics2024}). 

Passenger car deployments of eHMI (n~=~10) range from minimal to highly expressive. In China, Baidu Apollo~\cite{BaidusApollo2021} and Pony.ai~\cite{PonyAI2019} rely mainly on implicit motion cues, with no publicly documented, widely deployed eHMIs, while Zoox~\cite{Zoox2021} pursued similarly minimal signalling through coloured lights and familiar blinkers. In contrast, Navya~\cite{Navya2017} and Drive.ai~\cite{Driveai2016} experimented with text, symbols, and emoji displays, sometimes paired with auditory feedback. Waymo’s iterations reveal a stepwise expansion of external signalling~\cite{GoogleWaymo2020, GoogleWaymo2022, GoogleWaymo2023, GoogleWaymo2025}: in 2020, the company introduced roof-mounted LED domes to display identifiers that helped riders locate their vehicle; by 2023, these domes evolved to include symbolic light patterns (e.g., yellow and green icons warning of pedestrian crossings or ingress/egress), messages to cyclists about upcoming door openings, and external audio alerts to communicate with emergency responders; and in 2025, additional legally mandated back-up beeps were integrated. A key theme across several deployments is the use of honks as an auditory channel: WeRide~\cite{WeRideChina2020} employed a conventional horn in line with local traffic culture, while Drive.ai~\cite{Driveai2016} worked on re-designing honks to be more socially appropriate and context-sensitive. Separately, Waymo~\cite{GoogleWaymo2022} introduced a melodic, high-pitch auditory cue, not a horn but a distinctive vehicle identification signal, to help low-vision riders locate their car. Overall, car-based eHMI reflects a split between minimalism (implicit only) and layered multimodality, with auditory cues serving as a particularly contested design space.

\subsubsection{Regulations, Standards, and Working groups}


To provide clarity, the extracted items were grouped according to their relevance to AVs (see \autoref{tab:standards_regs}).

The first group consists of AV-specific regulations and standards, which explicitly address external AV communication. National efforts in this area remain limited and focus primarily on signalling automated-driving status. Singapore is the only country in the dataset with an enacted, AV-targeted legal instrument governing trial operations, under which external-signalling requirements can be imposed~\cite{Singapore_AV}. China has issued a draft national standard proposing ADS marker-lamp specifications, although it has not yet been formalised~\cite{tuv2023china}. Beyond these national efforts, most AV-specific guidance comes from international standards organisations such as ISO and SAE, which provide ergonomic communication principles, evaluation methods, and recommended practices for ADS marker lamps~\cite{SAE_J3134_2019, ISO_TR23049_2018, ISO_TR23720, ISO_TR23735, ISO_TR12204}. Another emerging theme relates to specific application areas; for example, the ISO 4448 project~\cite{ISO_TR4448} focuses on sidewalk delivery and service robots, extending standards work beyond traditional road vehicles. A number of governance and development structures underpin these initiatives. International bodies such as the UNECE Working Party on Automated/Autonomous and Connected Vehicles (GRVA)~\cite{UNECE_GRVA}, UNECE GRE~\cite{UNECE_GRE}, and the GTB Expert Group~\cite{GTB_GlobalLightingGroup} provide platforms for harmonising technical requirements and coordinating cross-national collaboration. 

The second group consists of long-standing vehicle and ITS regulations that, while not written for AVs, nonetheless impose hard constraints on the design of eHMIs. Core lighting regulations such as FMVSS 108 in the United States and CMVSS 108 in Canada restrict permissible colours, positions, intensities, and activation patterns for exterior lamps. UNECE Regulation No. 149~\cite{UNECE_R149} further restricts projection lighting by allowing only white symbols, leaving shapes and other visual elements unspecified. National traffic-control documents, including the US MUTCD, define colour and symbol conventions for roadside signage, influencing how new AV signals must avoid visual interference.

Broader standards and ITS documents also shape the eHMI design space. SAE J578 defines chromaticity limits for automotive signal colours~\cite{SAE_J578_2020}. ISO 9241-303~\cite{ISO_9241_303}, originally designed for electronic displays, has been cited as a baseline for text size legibility, even though its applicability to eHMIs viewed at longer distances remains contested. ISO 18682~\cite{ISO_18682} introduces hazard detection and notification systems for intelligent transport systems (ITS), with potential relevance for protecting vulnerable road users such as cyclists. ETSI TR 102 638~\cite{ETSI_TR102638}, while centred on V2X communication, establishes a broader framework that can be adapted for AV signalling. 

\begin{table*}[htbp]
\centering
\caption{Regulations, standards, and working groups relevant to eHMI development, as cited in the reviewed literature.}
\label{tab:standards_regs}
\footnotesize
\renewcommand{\arraystretch}{1.2}
\begin{tabular}{@{}L{3.5cm}L{6.3cm}L{2.3cm}L{2cm}@{}}
\toprule
\textbf{Name} & \textbf{Description} & \textbf{Version / Date} & \textbf{Mentioned In} \\
\midrule
\multicolumn{4}{@{}l}{\textbf{AV-SPECIFIC REGULATIONS AND STANDARDS}} \\
\midrule
Singapore Road Traffic Rules (Autonomous Motor Vehicles)
& Requires AVs to display distinctive lighting to indicate automated mode status. 
& G.N. S 89/2024 (2024)
& \cite{block2023road, joshi2023mapping} \\

Chinese National Standard 
& Specifications for optical ADS marker lamps. 
& Draft GB/T (2023) 
& \cite{tuv2023china} \\

SAE J3134-201905 
& Recommended practice for ADS marker lamps signalling automated status. 
& Draft (2019) 
& \cite{block2023road, joshi2023mapping, mirnig2021stop} \\

ISO/TR 4448 series 
& Ground-based automated mobility (sidewalk robots). 
& Draft series (2022–) 
& \cite{block2023road} \\

ISO/TR 23049:2018 
& Ergonomic aspects of AV external visual communication. 
& Edition 1 (2018)
& \cite{block2023road, joshi2023mapping, souman2021overview, shmueli2023ehmi} \\

ISO/PAS 23735:2025
& Ergonomic guidance for AV external visual communication. 
& Edition 1 (2025) 
& \cite{souman2021overview} \\

ISO/AWI TR 23720 
& Evaluation methods for road-user behaviour with AV communication. 
& AWI registered (2021) 
& \cite{souman2021overview} \\

UNECE GRVA 
& Working party on AVs; eHMI-related regulation work. 
& Established (2018)
& \cite{block2023road, joshi2023mapping, werner2018new, shmueli2023ehmi} \\

UNECE GRE / AVSR Task Force 
& Lighting and signalling requirements for AVs. 
& AVSR mandate (2019)
& \cite{souman2021overview} \\

GTB (Lighting Expert Group) 
& Global lighting and signalling expert group. 
& Established (1952)
& \cite{souman2021overview} \\

\midrule
\multicolumn{4}{@{}l}{\textbf{GENERAL VEHICLE / ITS REGULATIONS CONSTRAINING eHMI}} \\
\midrule
FMVSS 108 (US)
& Governs allowable colours/activation patterns for vehicle lamps. 
& 49 CFR §571.108 (2025)
& \cite{yi2024impact, shmueli2023ehmi, singer2022display} \\

CMVSS 108 (Canada)
& Governs allowable colours/activation patterns for vehicle lamps. 
& SOR/2018-43 (2018) 
& \cite{yi2024impact, shmueli2023ehmi, singer2022display} \\

UNECE Regulation No.\ 149 
& Restricts road-projection symbols to white 
& Revised (2019)
& \cite{sun2023onroad} \\

MUTCD (US DOT) 
& Specifies colours and design conventions for US road signs. 
& Revised (2022) 
& \cite{yu2023way} \\

SAE J578 
& Chromaticity limits for automotive signal colours. 
& Revised (2020)
& \cite{werner2018new, shmueli2023ehmi} \\

ISO 9241-303:2011 
& General standard for electronic visual displays. 
& Edition 1 (2011)
& \cite{souman2021overview} \\

ISO 18682 
& Hazard detection and notification systems for ITSs. 
& Edition 1 (2016)
& \cite{vonsawitzky2022hazard} \\

ISO/TR 12204:2012 
& In-vehicle warning signal integration. 
& Edition 1 (2012)
& \cite{vonsawitzky2022hazard} \\

ETSI TR 102 638  
& Intelligent Transport Systems report (V2X). 
& V2.1.1 (2024)
& \cite{block2023road, joshi2023mapping} \\

\bottomrule
\end{tabular}
\end{table*}

\subsubsection{Alignment}

\begin{figure*}[htbp]
    \centering
    \includegraphics[width=0.8\linewidth]{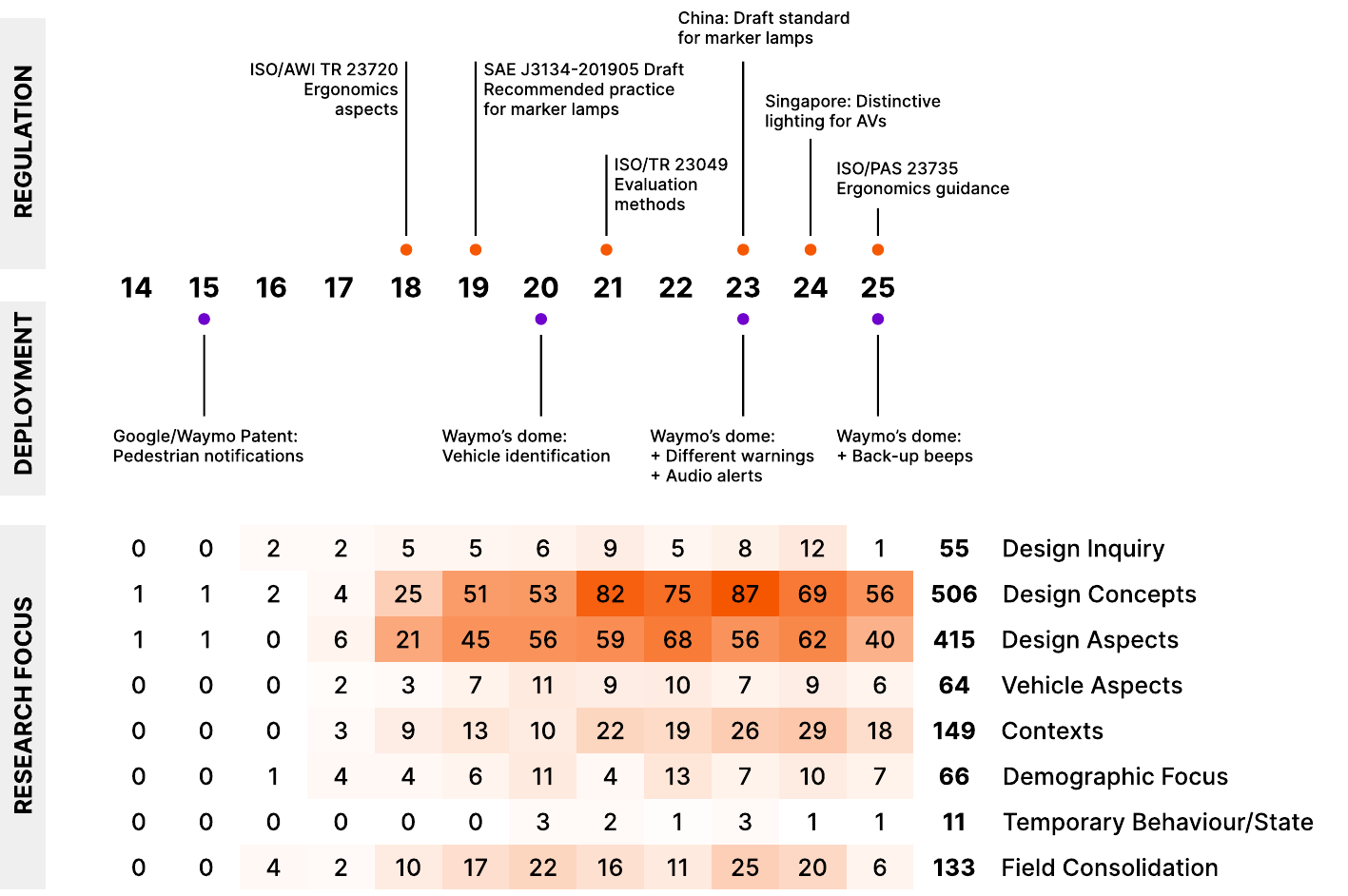}
 \caption{Time-aligned overview of eHMI research focus, regulatory activity, and real-world AV deployment events (2014–2025). Google/Waymo is used as a representative, high-visibility case to anchor the timeline, not as an exhaustive account of AV deployments.}
    \label{fig:alignment_time}
    \Description{The figure presents a three-part, vertically aligned timeline covering the years 2014–2025. At the bottom, a heatmap shows the annual distribution of eHMI research across eight categories: Design Inquiry, Design Concepts, Design Aspects, Vehicle Aspects, Contexts, Demographic Focus, Temporary Behaviour/State, and Field Consolidation. Above the heatmap, a middle band marks real-world AV deployment examples. These include the Google/Waymo notification dome (pedestrian alerts, vehicle identification, audio alerts, and back-up beeps), each positioned at the year of introduction. At the top, a third band lists major regulatory milestones aligned to their publication years, including ISO TR 23049 (evaluation methods), ISO/AWI TR 23720 (ergonomics aspects), ISO/PAS 23735 (ergonomics guidance), SAE J3134 (draft recommended practice for marker lamps), and national activities such as Singapore’s distinctive AV lighting requirement and China’s draft marker-lamp standard.}
\end{figure*}

\autoref{fig:alignment_time} provides a temporal view of how industry activity, academic research, and regulation have developed in relation to one another. Industry developments appear earliest, with the Google/Waymo patent filed in 2013 and approved in 2025~\cite{waymo2015pedestrian}, preceding the rise of published eHMI research. Industry patents serve a different role than research or concepts. Rather than exploring new forms, they cover a broad range of potential communication needs, ensuring companies secure legal ground across possible eHMI directions. Many of these ideas remain undeployed, but patents illustrate how industry positions itself broadly before deployment and regulation narrow the space. Academic activity expands next and covers a wide range of communication approaches. Regulatory milestones emerge later once research activity has become more established.

\begin{figure*}[htbp]
    \centering
    \includegraphics[width=1\linewidth]{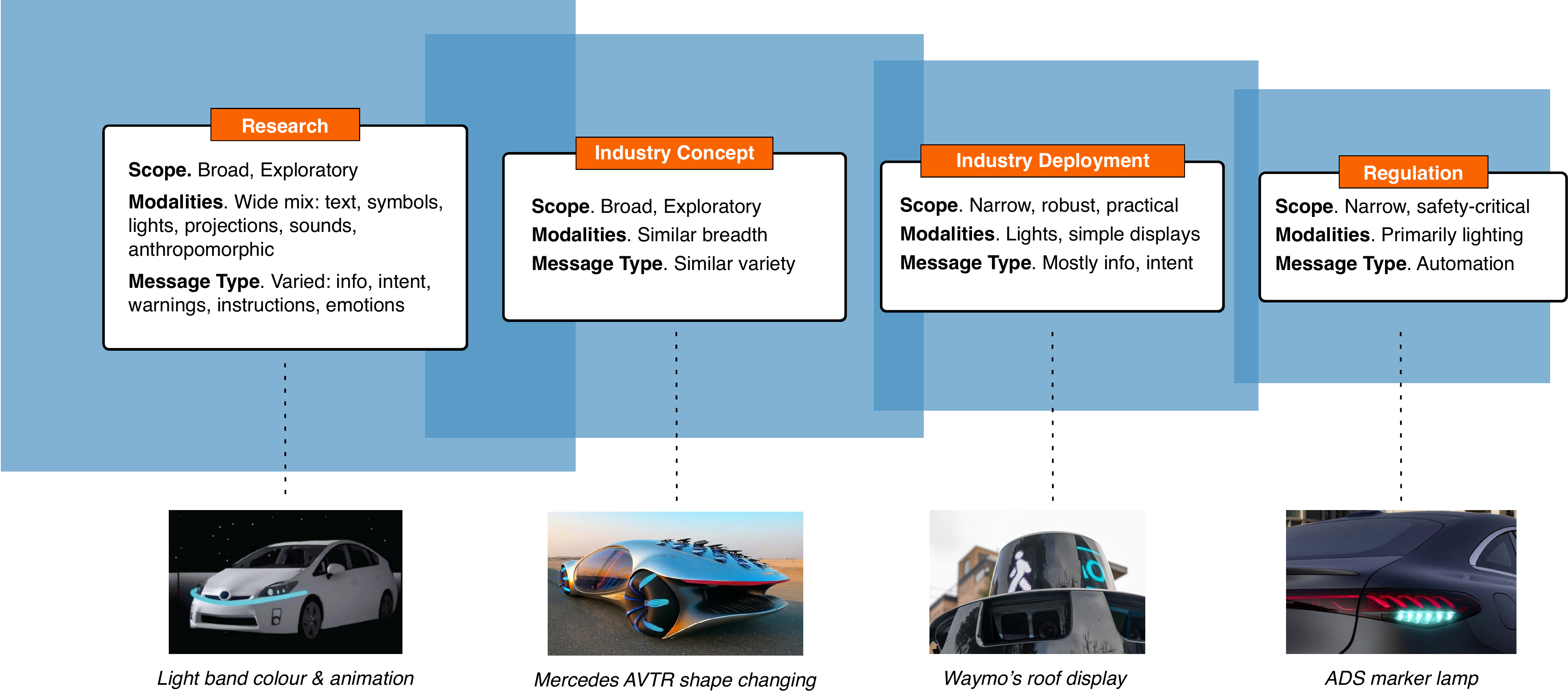}
 \caption{Filtering funnel showing how eHMI design options progress from broad exploration in research and industry concepts, to narrower sets in deployment, and finally to the most constrained forms under regulation. Illustrative examples from AV prototypes~\cite{dey2020color}, concept vehicles \cite{mercedes2025visionavtr}, deployed systems~\cite{GoogleWaymo2023}, and regulatory indicators~\cite{SAE_J3134_2019} are included.}
    \label{fig:alignment_funnel}
    \Description{The figure presents an alignment funnel illustrating how eHMI design options become progressively constrained. In the Research stage, the scope is broad and exploratory, with diverse modalities including text, symbols, lights, projections, sounds, and anthropomorphic cues, and varied message types spanning information, intent, warnings, instructions, and emotions. In Industry Concepts, the scope remains broad with similar breadth in modalities and message types. By the Industry Deployment stage, the scope narrows to robust and practical applications, with modalities limited to lights and simple displays, and message types focusing mostly on information and intent. Under Regulation, the scope is safety-critical, modalities are primarily restricted to lighting, and message types are reduced to automation.}
\end{figure*}

In terms of focus, comparing the research, industry, and regulatory developments reported in earlier sections reveals a process of progressive filtering rather than outright misalignment. Both academic research and industry concepts span a broad spectrum of approaches, from minimalist light-based cues to expressive, anthropomorphic, and multimodal designs. At the stage of deployment, this space narrows considerably, with companies favouring simpler, more robust solutions such as LED domes, light bars, and basic auditory cues that can withstand everyday use and public scrutiny. Regulation and standards activity further constrains the design space by codifying only a small subset of modalities, primarily lighting, colour, and safety-critical indicators. This progression can be understood as a funnel-shaped trajectory (see \autoref{fig:alignment_funnel}). This alignment reflects how practicalities and safety requirements act as filters on the design space, shaping which eHMI options transition from speculation to standardisation. This filtering is not a loss but a convergence process, ensuring safety and usability while retaining room for expressive exploration in research and concept design.


\section{Discussion}

eHMI research has expanded rapidly in the past decade, yet beneath this growth lie unresolved tensions that continue to shape the field. In this section, we trace the evolution of eHMI research, revisit the debate on necessity, examine the challenges of standardisation, and reflect on where the field should go next.

\subsection{The Evolution of eHMI Research}

The results show sustained growth in empirical work, but also clear divergence in both design philosophy and necessity. On one side, minimalist approaches treat eHMIs as supplementary, offering only the bare minimum needed to clarify vehicle intent. On the other, expressive approaches envision eHMIs as essential, providing rich, directive, or even socially communicative signals. This split echoes two dimensions in the taxonomy by \citet{dey2020taming}: \textit{Complexity in Implementation}, where 64\% of concepts could be realised with widely available technology (e.g., light strips), 7\% would require new but already available technology dependent on large-scale deployment or infrastructure change,; and \textit{New Vehicle Design}, where 77\% of concepts were shown on existing vehicle models, while 23\% required entirely new designs that could influence pedestrian acceptance, especially in early mixed-traffic phases. The persistence of such polarisation (despite the prevalence of readily implementable solutions) raises the question of whether the field is approaching a form of premature stabilisation, in which familiar design tensions are revisited without resolving foundational disagreements. Alternatively, this diversity may be necessary to accommodate the wide range of contexts in which AVs will operate, from highly regulated urban intersections to informal shared spaces.

Patterns in the dataset suggest that the field is beginning to consolidate around certain focal points. The heatmaps and taxonomy reveal concentrations of work on message design, implicit cues, and comparative modality studies. These clusters could serve as starting points for consensus-building, especially if future research shifts from proposing new prototypes to synthesising existing evidence. At the same time, other areas remain sparse, including large-scale evaluations, multimodal frameworks, and integration with industry deployments.  

Seen in this light, the evolution of eHMI research is marked by both growth and fragmentation. The challenge ahead is to channel this dichotomy into constructive frameworks rather than letting it dissipate into parallel, disconnected lines of inquiry. Whether the current trajectory reflects healthy exploration or risks entrenching unresolved debates will depend on the field’s ability to build cumulative knowledge from the foundations already established.

\subsection{Rethinking the Necessity of eHMIs}

One of the most persistent debates in the literature concerns the necessity of explicit eHMIs. Reviews often frame this as a binary question: some argue that explicit interfaces are essential to replace the social interaction traditionally provided by human drivers, while others contend that implicit communication through vehicle motion is sufficient, with eHMIs adding little beyond specific edge cases~\cite{bied2024autonomous, de2022external}. However, such framing risks oversimplifying a multifaceted design problem. The question is not whether eHMIs are necessary \textit{in general}, but rather \textit{when}, \textit{for whom}, and \textit{which implementations} are most appropriate.  

\textbf{When?} Explicit eHMIs are likely to play a greater role in situations of ambiguity, complexity, or shared space, where vehicle motion alone does not provide sufficient clarity. In contrast, in straightforward traffic contexts with clear rules and signals, implicit cues may suffice. The necessity of explicit eHMIs may also depend on geographic regions and the maturity of AV deployment: in areas where AVs are novel, pedestrians may require more explicit reassurance, whereas over time reliance on such cues may diminish as familiarity grows.  

\textbf{For whom?} Different pedestrian groups may require different forms of support. Children, older adults, and people with disabilities may benefit more from explicit communication, while experienced road users or those familiar with AVs may rely primarily on motion cues. Beyond human road users, animals represent an additional and largely unaddressed interaction case, for which conventional eHMI assumptions may not apply~\cite{tran2025animal}. 

\textbf{Which implementations?} Even when explicit eHMIs are needed, their form remains an open design question. Text, icons, light patterns, and auditory signals each have strengths and limitations, and their effectiveness depends on context, inclusivity, and standardisation. Multimodal~\cite{dey2024multimodal} or adaptive approaches~\cite{tran2025adaptive} may therefore be promising, providing redundancy and flexibility to suit different road users and environments.

Historical analogues show that such debates are not new. The introduction of \textit{Accessible Pedestrian Signals} (APS) in the mid-twentieth century was initially contested~\cite{harkey2007accessible}: many argued that conventional visual signals were sufficient for most pedestrians, and even major organisations representing blind people in the United States held opposing views on whether APS were necessary. Concerns centred on inconsistent or poor implementations that could increase risk, create overreliance, or offer only marginal benefit---issues that echo many of the reservations currently raised about eHMIs. Over time, the framing shifted away from a binary question toward more conditional considerations of \textit{when} APS should be installed (e.g., at complex intersections), \textit{for whom} (e.g., people with visual impairments, older adults), and \textit{which implementations} (auditory, vibrotactile, or spoken-message cues). A similar trajectory can be traced in the history of \textit{turn signals and brake lights}, which were once regarded as optional additions to hand signals~\cite{abingrose2022blinkers} but gradually became standardised in terms of colour, placement, and redundancy.  
Reframing the necessity debate for eHMIs in these terms highlights the need to move beyond a false dichotomy of `eHMIs versus no eHMIs.' Instead, necessity should be treated as contingent, context-dependent, and sensitive to user diversity. As with APS and turn signals, eHMIs may follow a path from contested adoption to eventual standardisation, with debates shifting toward context, inclusivity, and implementation details. This perspective also aligns with the research directions identified in review papers, which emphasise the importance of context-sensitive, multimodal, and standardised designs grounded in user studies and ecological validity. However, if necessity is conditional rather than absolute, then standardisation must also grapple with flexibility rather than uniformity.

\subsection{The Challenge of Standardising eHMIs}

Although reviews converge in emphasising the need for standardisation~\cite{rasouli2019autonomous, siu2025pedestrians}, what `standardisation' should mean in practice remains contested. Standardisation is often assumed to imply convergence on a small number of fixed designs. Yet the necessity of eHMIs is conditional, depending on when they are used, for whom they are designed, and which implementations are most appropriate. This makes the prospect of complete convergence both unrealistic and undesirable.  

A more productive way forward is to distinguish between \textit{universal parameters} and \textit{contextual extensions}. Universal parameters would define a \textit{basic eHMI}, a minimum layer of communication that can be enforced globally to guarantee baseline safety and minimise negative side-effects. Such a baseline should convey only the essential information needed to prevent confusion or misinterpretation, for example through highly visible, culturally neutral colours, simple forms, or redundant modalities. By reducing communication to a bare minimum, this layer would address many of the concerns raised in the literature, including distraction, overreliance, or conflicting interpretations across regions.  

Beyond this basic layer, contextual extensions should allow design flexibility. These might adapt modalities, richness of information, or degree of explicitness according to the road context (e.g., shared spaces versus signalised intersections), user group (e.g., children, older adults, people with disabilities), or level of AV maturity in a given region. In areas where AVs are novel, pedestrians may initially require more explicit reassurance, while in more mature contexts minimal signals may suffice.  

One more time, the history of turn signals provides a useful analogue. Despite being highly standardised today, turn signals still vary in form: some cars use simple flashing bulbs, others adopt LED strips or dynamic `wiping' animations. Universal parameters such as colour, blink frequency, and visibility are tightly regulated, ensuring recognisability and safety, while manufacturers retain scope for stylistic differentiation. A similar layered model could guide eHMI development: enforce a basic standard to ensure minimum safety and consistency, while permitting adaptation and innovation above that baseline.  

Standardising eHMIs is therefore less about enforcing a single global solution than about establishing a roadmap that separates what must be harmonised universally from what can remain open to contextual adaptation. This approach balances the practicality of implementation with the inclusivity and flexibility required to serve diverse road users worldwide.  

\subsection{Where eHMI Research Should Go Next}

Despite the rapid growth of eHMI studies, the discourse risks becoming saturated with incremental work. Each month new papers appear, often testing yet another light bar, icon, or auditory cue in a small-scale lab setting. While these studies add isolated data points, they seldom extend the conversation in meaningful ways. What is needed now is not additional prototypes in isolation, but a shift in orientation toward frameworks, validity, and integration.

The design space for eHMIs is already crowded, and small variations in colour, form, or modality do little to advance cumulative knowledge. The key question is no longer whether a specific eHMI `works,' but how to structure the design space so that multiple eHMIs can coexist without confusion. Progress requires organising principles that can clarify what belongs to a universal minimum of communication (i.e., the `basic eHMI') and where contextual adaptations are appropriate. Such roadmaps would enable both convergence on shared parameters and flexibility in implementation, rather than a continued search for a single best solution, as nearly a decade of research has shown that there is no `silver bullet'.

Another challenge concerns interoperability. Academic research often treats designs in isolation, overlooking that manufacturers are already deploying eHMIs on public roads. Without cross-manufacturer coordination, there is a real risk of fragmented communication systems that erode rather than build trust. Future research must therefore engage more directly with industry practices and regulatory processes, recognising that deployment is already underway and that experimental findings need to inform and not lag behind these developments.

Finally, the scope of eHMI research is beginning to expand beyond pedestrian crossings. Future mobility ecosystems will require communication not only with pedestrians but also with cyclists, delivery robots, conventional drivers, and even animals in shared spaces. This expansion is significant because it shifts the very notion of what an eHMI is. Whereas early work treated the eHMI as a discrete interface, typically a light bar or icon signalling intent to pedestrians, its meaning increasingly extends toward a broader communication framework that supports diverse autonomous agents within complex environments. The field must therefore confront whether this diversification represents healthy growth toward an inclusive, ecological perspective on mobility, or whether it risks diluting the concept of eHMI and fragmenting the discourse further.  


\subsection{Practical Implications}

Our analysis is deliberately field-level, tracing how hundreds of studies, concepts, deployments, and regulatory efforts collectively shape the design space for external communication of AVs. Yet these patterns also have immediate implications for practitioners and policymakers who must make concrete decisions about what to build, approve, or study next. In this section, we translate our findings into practical guidance for eHMI design and deployment, for regulation and standardisation, and for future research practice.

\paragraph{\textit{Implications for eHMI Design \& Deployment}}

The alignment funnel highlights how a broad design `jungle' at the level of research and concept exploration narrows into a small set of robust solutions at deployment and regulation stages. For design and engineering teams, this suggests treating speculative interfaces and deployment-ready systems as distinct layers of work. Concept cars, speculative prototypes, and VR explorations are best used to probe unfamiliar modalities, social expectations, and edge cases. In contrast, deployed eHMIs should converge on a safety-first core: simple, allocentric visual cues that clarify automation status and intent, and are compatible with existing lighting and signalling conventions. One practical way to operationalise this is to specify eHMIs as layered assemblies, with (i) a minimal baseline set of signals that is consistently available across vehicles and markets, and (ii) contextual extensions that may be selectively enabled for particular environments (e.g., shared spaces, logistics yards) or user groups (e.g., occupational road users, people with sensory impairments). Designers should also be cautious in transplanting interaction patterns that work for pedestrian crossings with passenger cars into freight, public transport, or mixed-traffic scenarios, where our heatmaps show much thinner empirical coverage.

\paragraph{\textit{Implications for Regulation \& Standardisation}}
For regulators and standards bodies, our findings argue against codifying a single canonical eHMI design, and in favour of standardising parameters and constraints that can accommodate multiple implementations. Regulatory activity already concentrates on lighting, colour spaces, and projection constraints, reflecting a move toward codifying a narrow subset of the wider research space. The layered model suggested above can serve as a roadmap: standards can focus on defining a minimal set of universal parameters (e.g., permitted colour ranges, luminance levels, temporal patterns, mounting locations, and activation logic for automation markers), while explicitly allowing contextual extensions and manufacturer differentiation above this baseline. The alignment funnel can be used as a diagnostic tool when drafting or revising standards: it makes visible where regulations already align with convergent industrial practice (e.g., marker lights for automated mode) and where gaps remain (e.g., for multimodal interfaces, non-passenger platforms, or occupational roles). This perspective encourages regulators to support safe convergence without prematurely closing off design options that may be needed in specific local or emerging contexts.

\paragraph{\textit{Implications for Research Practice}}
For researchers, the most immediate implication is not to produce yet another isolated prototype, but to use existing maps of the field to select topics and methods that contribute to cumulative knowledge. Our dataset and heatmaps show saturation in certain combinations of context, modality, and task (e.g., small-scale VR studies of crossing scenarios with passenger cars and pedestrians), alongside clear gaps for mixed traffic, non-pedestrian road users, non-standard vehicles, and longitudinal, in-the-wild deployments. Future work can have greater practical impact by (i) prioritising comparative and multimodal studies over single-concept evaluations, (ii) increasing ecological validity through on-road trials, test beds, and longer-term studies, and (iii) deliberately targeting under-represented contexts and user groups, including occupational roles and people with diverse abilities. At the methodological level, our synthesis emphasises the value of more transparent reporting of scenarios, vehicle types, and user characteristics, as well as clearer separation of implicit and explicit cues, to support replicability and meta-analysis.

\paragraph{\textit{Summary}}
These implications posit the need for a more coordinated division of labour across research, industry, and regulation. Research retains a broad exploratory mandate, but should increasingly be guided by field-level maps and designed to feed into cumulative evidence. Industry design should focus on integrating a small set of robust, allocentric, and regulation-ready signals into vehicle platforms, treating more expressive or speculative concepts as optional layers. Regulators can concentrate on defining flexible but firm baselines that align with emerging practice, rather than prescribing entire interfaces. By making these roles explicit, the field can move from a loosely coupled `jungle' of proposals toward an ecosystem where external communication for AVs is both practically deployable and adaptable across diverse contexts.

\subsection{Limitations and Future Work}

Our dataset was restricted to peer-reviewed publications in English, which likely excludes relevant evidence from non-English sources and grey literature. This limitation is especially pronounced for industry deployments in China, where details of real-world AV practices are sparsely documented. As a result, the analysis reflects the published record and may not capture ongoing or undocumented industry activities.

Although coding was conducted systematically using an iteratively developed framework, alternative classifications are possible. Additionally, several coding dimensions, such as vehicle types, target road users, and demographic focus, were extracted from titles and abstracts only. This approach ensured consistency across the large corpus, as titles and abstracts typically provide the primary conceptual framing of each study. However, this constraint means that details reported only in the full text may be under-coded or omitted. A small number of cases may therefore be missed, which should be considered when interpreting the coverage of specific categories.

Given the sheer number of papers included in our analysis (620 in total), it was not feasible to cite each individual study within the manuscript. To support transparency, reproducibility, and future research, the full coded dataset is published on the Open Science Framework (OSF). Building on examples such as the Locomotion Vault~\cite{luca2021locomotion}, we are planning to develop an interactive platform that enables filtering by parameters such as year, modality, study type, and user group. 

A natural extension of this work is to complement the literature-driven synthesis with expert perspectives. A Delphi study~\cite{linstone1975delphi, rowe1999delphi} could be designed directly around the areas of consensus, divergence, and uncertainty identified here, enabling structured engagement with experts across the board of researchers, practitioners, and policymakers. Such an approach would help prioritise open questions, validate emerging points of agreement, and guide the development of standards and design recommendations for eHMIs.

\section{Conclusion}

This review set out to ask whether eHMI research is converging, diverging, or stalling. Our analysis shows that eHMI research is marked by all three dynamics at once. Convergence is visible in emerging agreement that interaction is indispensable, that allocentric rather than egocentric communication reduces ambiguity, and that some form of baseline standardisation will be required. At the same time, divergence persists in design philosophy (minimalist versus expressive), in the perceived necessity of explicit signals (supplementary versus essential), and in competing preferences across modalities. Signs of stalling are also evident: many studies revisit the same unresolved debates with small-scale VR experiments or incremental prototypes, raising the risk of premature stabilisation. The field’s future trajectory depends on whether this diversity can be channelled into constructive frameworks. Progress will require building on areas of consensus, addressing fragmentation through flexible roadmaps, and moving beyond incremental novelty toward cumulative, ecologically valid knowledge that can support deployment.

\begin{acks}
This research was supported by the Australian Research Council (ARC) Discovery Project DP220102019, Shared-space interactions between people and autonomous vehicles. The authors thank the anonymous reviewers for their valuable comments and suggestions, which have been incorporated into the final version.
\end{acks}

\bibliographystyle{ACM-Reference-Format}
\bibliography{references}

@inproceedings{hollander2019overtrust,
author = {Holl\"{a}nder, Kai and Wintersberger, Philipp and Butz, Andreas},
title = {Overtrust in External Cues of Automated Vehicles: An Experimental Investigation},
year = {2019},
isbn = {9781450368841},
publisher = {Association for Computing Machinery},
address = {New York, NY, USA},
url = {https://doi.org/10.1145/3342197.3344528},
doi = {10.1145/3342197.3344528},
booktitle = {Proceedings of the 11th International Conference on Automotive User Interfaces and Interactive Vehicular Applications},
pages = {211–221},
numpages = {11},
location = {Utrecht, Netherlands},
series = {AutomotiveUI '19}
}

@misc{abingrose2022blinkers,
author = {Abingrose, Ann It Girl},
title = {The Evolution of Car Blinkers: From Hand Signals to High-Tech Indicators},
howpublished = {\url{https://medium.com/@abingroseannitgirl/the-evolution-of-car-blinkers-from-hand-signals-to-high-tech-indicators-97330fa126d4}}
,
year = {2022}
}

@misc{harkey2007accessible,
  title={Accessible pedestrian signals: A guide to best practices},
  author={Harkey, David L and Carter, Daniel L and Barlow, Janet M and Bentzen, Billie Louise},
  journal={National Cooperative Highway Research Program, Contractor’s Guide for NCHRP Project},
  year={2007},
  publisher={Transportation Research Board Washington, WA, USA}
}

@inproceedings{luca2021locomotion,
author = {Di Luca, Massimiliano and Seifi, Hasti and Egan, Simon and Gonzalez-Franco, Mar},
title = {Locomotion Vault: the Extra Mile in Analyzing VR Locomotion Techniques},
year = {2021},
isbn = {9781450380966},
publisher = {Association for Computing Machinery},
address = {New York, NY, USA},
url = {https://doi.org/10.1145/3411764.3445319},
doi = {10.1145/3411764.3445319},
booktitle = {Proceedings of the 2021 CHI Conference on Human Factors in Computing Systems},
articleno = {128},
numpages = {10},
location = {Yokohama, Japan},
series = {CHI '21}
}

@article{white1998visualizing,
  title={Visualizing a discipline: An author co-citation analysis of information science, 1972--1995},
  author={White, Howard D and McCain, Katherine W},
  journal={Journal of the American society for information science},
  volume={49},
  number={4},
  pages={327--355},
  year={1998},
  publisher={Wiley Online Library}
}

@article{Geeganage2024topicmodelling,
author = {Geeganage, Dakshi Kapugama and Xu, Yue and Li, Yuefeng},
title = {A Semantics-enhanced Topic Modelling Technique: Semantic-LDA},
year = {2024},
issue_date = {May 2024},
publisher = {Association for Computing Machinery},
address = {New York, NY, USA},
volume = {18},
number = {4},
issn = {1556-4681},
url = {https://doi.org/10.1145/3639409},
doi = {10.1145/3639409},
journal = {ACM Trans. Knowl. Discov. Data},
month = feb,
articleno = {93},
numpages = {27}
}

@article{donthu_bibliometrics,
  author       = {Donthu, Naveen and Kumar, Satish and Mukherjee, Debmalya and Pandey, Nitesh and Lim, W. M.},
  title        = {How to conduct a bibliometric analysis: An overview and guidelines},
  journal      = {Journal of Business Research},
  volume       = {133},
  pages        = {285--296},
  year         = {2021},
  doi          = {10.1016/j.jbusres.2021.04.070},
  url          = {https://doi.org/10.1016/j.jbusres.2021.04.070}
}

@misc{MercedesBenz2023Turquoise,
  author       = {{Mercedes-Benz}},
  title        = {{Mercedes-Benz receives approvals for turquoise-coloured automated driving marker lights in California and Nevada}},
  year         = {2023},
  howpublished = {\url{https://media.mercedes-benz.be/fr/mercedes-benz-receives-approvals-for-turquoise-coloured-automated-driving-marker-lights-in-california-and-nevada/}},
}

@misc{wiki2025elaine,
  author       = {{Wikipedia contributors}},
  title        = {{Death of Elaine Herzberg}},
  year         = {2018},
  howpublished = {\url{https://en.wikipedia.org/wiki/Death_of_Elaine_Herzberg}},
}

@techreport{indident2018report,
  author       = {{National Transportation Safety Board}},
  title        = {{Preliminary Report: Highway Incident HWY18MH010}},
  institution  = {National Transportation Safety Board},
  year         = {2018},
  type         = {Preliminary report},
  number       = {HWY18MH010},
  url          = {https://www.ntsb.gov/investigations/Pages/HWY18MH010.aspx},
}

@inproceedings{zileli2019towards,
  author    = {Zileli, Selin and Boyd Davis, Stephen and Wu, Jiayu},
  title     = {Towards Transparency between the Autonomous Vehicle and the Pedestrian},
  booktitle = {Design and Semantics of Form and Movement (DesForm 2019): Beyond Intelligence},
  year      = {2019},
  pages     = {96--104},
  publisher = {MIT Press},
  address   = {Cambridge, MA, USA}
}

@inproceedings{asha2022towards,
  author    = {Asha, Ashratuz Zavin and Brierley, Owen and Somanath, Sowmya and Finn, Patrick and Sharlin, Ehud},
  title     = {Towards Designing Audio Interactions with Autonomous Vehicles: A Hearing-Enhanced Pedestrian Story},
  booktitle = {AutomationXP@CHI: Engaging with Automation},
  year      = {2022},
  numpages  = {8},
  publisher = {CEUR Workshop Proceedings},
  address   = {Aachen, Germany}
}

@article{hesse2021holistic,
  title={Holistic, context-sensitive human-machine interaction for automated vehicles},
  author={Hesse, Tobias and Oehl, Michael and Drewitz, Uwe and Jipp, Meike},
  journal={ATZ worldwide},
  volume={123},
  number={3},
  pages={46--49},
  year={2021},
  publisher={Springer}
}

@article{robert2019future,
  title={The future of pedestrian-automated vehicle interactions},
  author={Robert Jr, Lionel P},
  journal={XRDS: Crossroads, The ACM Magazine for Students},
  volume={25},
  number={3},
  pages={30--33},
  year={2019},
  publisher={ACM New York, NY, USA}
}

@inproceedings{bindschadel2022studying,
author = {Bindsch\"{a}del, Janina and Kiesel, Andrea},
title = {Studying pedestrians´ crossing behavior during automated vehicle interactions: A Wizard of Oz study},
year = {2023},
isbn = {9781450396240},
publisher = {Association for Computing Machinery},
address = {New York, NY, USA},
url = {https://doi.org/10.1145/3558884.3558885},
doi = {10.1145/3558884.3558885},
booktitle = {Proceedings of the 7th International Workshop on Sensor-Based Activity Recognition and Artificial Intelligence},
articleno = {1},
numpages = {7},
location = {Rostock, Germany},
series = {iWOAR '22}
}

@article{daimon2021pedestrian,
  title={Pedestrian Carelessness toward Traffic Environment Due to External Human--Machine Interfaces of Automated Vehicles},
  author={Daimon, Tatsuru and Taima, Masahiro and Kitazaki, Satoshi},
  journal={traffic},
  volume={10},
  pages={11},
  year={2021}
}

@article{werner2018new,
  author  = {Werner, Annette},
  title   = {New Colours for Autonomous Driving: An Evaluation of Chromaticities for the External Lighting Equipment of Autonomous Vehicles},
  journal = {Colour Turn},
  number  = {1},
  volume    = {1},
  year    = {2019},
numpages = {14},
  doi     = {10.25538/tct.v0i1.692},
  url     = {https://doi.org/10.25538/tct.v0i1.692},
}

@article{Thorne2022Qualitative_meta,
author = {Thorne, Sally},
title = {Qualitative meta-synthesis},
journal = {Nurse Author \& Editor},
volume = {32},
number = {1},
pages = {15-18},
doi = {https://doi.org/10.1111/nae2.12036},
url = {https://onlinelibrary.wiley.com/doi/abs/10.1111/nae2.12036},
year = {2022}
}

@inproceedings{lee2019investigating,
  title={Investigating pedestrians' crossing behaviour during car deceleration using wireless head mounted display: an application towards the evaluation of eHMI of automated vehicles},
  author={Lee, Yee Mun and Uttley, Jim and Solernou, Albert and Giles, Oscar and Romano, Richard and Markkula, Gustav and Merat, Natasha},
  booktitle={Proceedings of the Tenth International Driving Symposium on Human Factors in Driving Assessment, Training and Vehicle Design},
  pages={252--258},
  year={2019},
  publisher = {University of Iowa Public Policy Center},
address   = {Iowa City, IA, USA}
}

@article{qi2023investigating,
  title={Investigating pedestrian crossing decision with autonomous cars in virtual reality},
  author={Qi, Shuaixin and Menozzi, Marino},
  journal={Zeitschrift f{\"u}r Arbeitswissenschaft},
  volume={77},
  number={2},
  pages={218--229},
  year={2023},
  publisher={Springer}
}

@article{weiss2022external,
  title={External Human-Machine-Interfaces on Automated Vehicles: Which message and perspective do pedestrians in crossing situations understand best?},
  author={Wei{\ss}, Sebastian Ludwig and Eisele, Daniel and Petzoldt, Tibor},
  journal={Intelligent Human Systems Integration (IHSI 2022): Integrating People and Intelligent Systems},
  volume={22},
  numpages = {10},
  number={22},
  year={2022},
  publisher={AHFE Open Acces}
}

@inproceedings{alhawiti2024exploring,
  author    = {Alhawiti, A. and Kwigizile, V. and Oh, J. and Asher, Z. and Hakimi, O.},
  title     = {Exploring External Human--Machine Interface Design for Autonomous Vehicle to Pedestrian Communication: Insights from Discussions and Drawing Sessions},
  booktitle = {Intelligent Human Systems Integration (IHSI 2024): Integrating People and Intelligent Systems},
  editor    = {Ahram, Tareq and Karwowski, Waldemar and Russo, Dario and Di Bucchianico, Giuseppe},
  series    = {AHFE Open Access},
  volume    = {119},
  numpages = {10},
  year      = {2024},
  publisher = {AHFE International},
  address   = {USA},
  doi       = {10.54941/ahfe1004469}
}

@inproceedings{camara2019examining,
  title={Examining pedestrian-autonomous vehicle interactions in virtual reality},
  author={Camara, Fanta and Dickinson, Patrick and Merat, Natasha and Fox, Charles W},
  booktitle={Proceedings of 8th Transport Research Arena TRA 2020},
  year={2019},
numpages={5},
  publisher={Transport Research Arena},
address={Helsinki, Finland}
}

@article{vinkhuyzen2016developing,
author = {Vinkhuyzen, Erik and Cefkin, Melissa},
title = {Developing Socially Acceptable Autonomous Vehicles},
journal = {Ethnographic Praxis in Industry Conference Proceedings},
volume = {2016},
number = {1},
pages = {522-534},
doi = {https://doi.org/10.1111/1559-8918.2016.01108},
url = {https://anthrosource.onlinelibrary.wiley.com/doi/abs/10.1111/1559-8918.2016.01108},
year = {2016}
}

@inproceedings{bazilinskyy2020coupled,
  author    = {Bazilinskyy, Pavlo and Kooijman, Lars and Dodou, Dimitra and de Winter, Joost C. F.},
  title     = {Coupled Simulator for Research on the Interaction Between Pedestrians and (Automated) Vehicles},
  booktitle = {Driving Simulation Conference Europe},
  year      = {2020},
  numpages  = {2},
  publisher = {Driving Simulation Conference Europe},
  address   = {Antibes, France}
}

@misc{mercedes2025visionavtr,
  author       = {{Mercedes-Benz Canada}},
  title        = {VISION AVTR | Future Vehicles},
  howpublished = {\url{https://www.mercedes-benz.ca/en/future-vehicles/vision-avtr}},
  year         = {2025},
  organization = {Mercedes-Benz Canada}
}

@misc{guo2022virtual,
  title={A virtual reality framework to measure psychological and physiological responses of the self-driving car passengers},
  author={Guo, Xiaolei and Wan, Dayu and Liu, Dongfang and Mousas, Christos and Chen, Yingjie},
  year={2022}
}

@inproceedings{simic2023automation,
author = {Simic, Vedran and Alsos, Ole Andreas},
title = {Automation transparency: Designing an external HMI for autonomous passenger ferries in urban waterways},
year = {2023},
isbn = {9781450398930},
publisher = {Association for Computing Machinery},
address = {New York, NY, USA},
url = {https://doi.org/10.1145/3563657.3596130},
doi = {10.1145/3563657.3596130},
booktitle = {Proceedings of the 2023 ACM Designing Interactive Systems Conference},
pages = {1145–1158},
numpages = {14},
location = {Pittsburgh, PA, USA},
series = {DIS '23}
}

@inproceedings{diederichs2021wizard,
  title={A Wizard-of-Oz vehicle to investigate human interaction with AI-driven automated cars},
  author={Diederichs, Frederik and Mathis, Lesley-Ann and Bopp-Bertenbreiter, Valeria and Bednorz, Benjamin and Widlroither, Harald and Flemisch, Frank},
  booktitle={Proceedings of the DSC},
  year={2021},
numpages={6},
publisher = {Driving Simulation Association},
address   = {Paris, France}
}

@article{jensen1996meta,
  title={Meta-synthesis of qualitative findings},
  author={Jensen, Louise A and Allen, Marion N},
  journal={Qualitative health research},
  volume={6},
  number={4},
  pages={553--560},
  year={1996},
  publisher={Sage Publications Sage CA: Thousand Oaks, CA}
}

@article{rouchitsas2019external,
  title={External human--machine interfaces for autonomous vehicle-to-pedestrian communication: A review of empirical work},
  author={Rouchitsas, Alexandros and Alm, H{\aa}kan},
  journal={Frontiers in psychology},
  volume={10},
  pages={2757},
  year={2019},
  publisher={Frontiers Media SA}
}

@article{de2022external,
  title={External human--machine interfaces: Gimmick or necessity?},
  author={de Winter, Joost and Dodou, Dimitra},
  journal={Transportation research interdisciplinary perspectives},
  volume={15},
  pages={100643},
  year={2022},
  publisher={Elsevier}
}

@article{brill2023external,
  title={External human--machine interfaces for automated vehicles in shared spaces: A review of the human--computer interaction literature},
  author={Brill, Sarah and Payre, William and Debnath, Ashim and Horan, Ben and Birrell, Stewart},
  journal={Sensors},
  volume={23},
  number={9},
  pages={4454},
  year={2023},
  publisher={MDPI}
}

@misc{souman2021overview,
  title        = {An overview of standards and evaluation criteria to assess the interaction of drivers and other road users with automated vehicles},
  author       = {Souman, Jan and van Weperen, Marijke and van Dam, Esra and Hogema, Marika Hoedemaeker},
  year         = {2021},
  howpublished = {Report TNO-2021-R12633},
  institution  = {TNO},
  address      = {The Hague, Netherlands},
  url          = {https://publications.tno.nl/publication/34639061/DWcJIS/TNO-2021-R12633.pdf}
}

@inproceedings{colley2020unveiling,
author = {Colley, Mark and Walch, Marcel and Rukzio, Enrico},
title = {Unveiling the Lack of Scalability in Research on External Communication of Autonomous Vehicles},
year = {2020},
isbn = {9781450368193},
publisher = {Association for Computing Machinery},
address = {New York, NY, USA},
url = {https://doi.org/10.1145/3334480.3382865},
doi = {10.1145/3334480.3382865},
booktitle = {Extended Abstracts of the 2020 CHI Conference on Human Factors in Computing Systems},
pages = {1–9},
numpages = {9},
location = {Honolulu, HI, USA},
series = {CHI EA '20}
}

@inproceedings{block2023road,
  title={The road ahead: Advancing interactions between autonomous vehicles, pedestrians, and other road users},
  author={Block, Avram and Joshi, Swapna and Tabone, Wilbert and Pandya, Aryaman and Lee, Seonghee and Patil, Vaidehi and Britten, Nicholas and Schmitt, Paul},
  booktitle={2023 32nd IEEE International Conference on Robot and Human Interactive Communication (RO-MAN)},
  pages={16--23},
  year={2023},
  publisher = {IEEE},
address   = {Piscataway, NJ, USA}
}

@article{yan2023user,
  title={User acceptance of autonomous vehicles: Review and perspectives on the role of the human-machine interfaces},
  author={Yan, Ming and Rampino, Lucia and Caruso, Giandomenico and others},
  journal={Computer-Aided Design and Applications},
  volume={20},
  number={5},
  pages={987--1004},
  year={2023}
}

@article{rowe1999delphi,
  title={The Delphi technique as a forecasting tool: issues and analysis},
  author={Rowe, Gene and Wright, George},
  journal={International journal of forecasting},
  volume={15},
  number={4},
  pages={353--375},
  year={1999},
  publisher={Elsevier}
}

@book{linstone1975delphi,
  title={The delphi method},
  author={Linstone, Harold A and Turoff, Murray and others},
  volume={1975},
  year={1975},
  publisher={Addison-Wesley Reading},
address = {Reading, MA, USA}
}

@inproceedings{wohlin2014snowballing,
author = {Wohlin, Claes},
title = {Guidelines for snowballing in systematic literature studies and a replication in software engineering},
year = {2014},
isbn = {9781450324762},
publisher = {Association for Computing Machinery},
address = {New York, NY, USA},
url = {https://doi.org/10.1145/2601248.2601268},
doi = {10.1145/2601248.2601268},
booktitle = {Proceedings of the 18th International Conference on Evaluation and Assessment in Software Engineering},
articleno = {38},
numpages = {10},
location = {London, England, United Kingdom},
series = {EASE '14}
}

@inproceedings{ma2023bibliometric,
  title={A Bibliometric and Visual Analysis of autonomous vehicles-pedestrians interaction},
  author={Ma, Chaomin and Zhang, Wanjia},
  booktitle={AHFE International Conference on Human Factors and Systems Interaction},
  year={2023},
numpages={6},
  doi={10.54941/ahfe1003745},
  url={https://doi.org/10.54941/ahfe1003745},
publisher = {AHFE International},
address   = {USA},
}

@article{siu2025pedestrians,
title = {Pedestrians’ Interaction with eHMI-equipped Autonomous Vehicles: A Bibliometric Analysis and Systematic Review},
journal = {Accident Analysis \& Prevention}, 
volume = {209},
pages = {107826},
year = {2025},
issn = {0001-4575},
doi = {https://doi.org/10.1016/j.aap.2024.107826},
url = {https://www.sciencedirect.com/science/article/pii/S0001457524003713},
author = {Siu Shing Man and Chuyu Huang and Qing Ye and Fangrong Chang and Alan Hoi Shou Chan},
}

@ARTICLE{tran2021review,
  author={Tran, Tram Thi Minh and Parker, Callum and Tomitsch, Martin},
  journal={IEEE Transactions on Human-Machine Systems}, 
  title={A Review of Virtual Reality Studies on Autonomous Vehicle–Pedestrian Interaction}, 
  year={2021},
  volume={51},
  number={6},
  pages={641-652},
  doi={10.1109/THMS.2021.3107517}}

@InProceedings{le2020automotive,
author="Le, Duc Hai
and Temme, Gerald
and Oehl, Michael",
editor="Stephanidis, Constantine
and Antona, Margherita
and Ntoa, Stavroula",
title="Automotive eHMI Development in Virtual Reality: Lessons Learned from Current Studies",
booktitle="HCI International 2020 -- Late Breaking Posters",
year="2020",
publisher="Springer International Publishing",
address="Cham",
pages="593--600",
isbn="978-3-030-60703-6"
}

@article{leveque2020pedestrians,
  title={Where do pedestrians look when crossing? A state of the art of the eye-tracking studies},
  author={L{\'e}v{\^e}que, Lucie and Ranchet, Maud and Deniel, Jonathan and Bornard, Jean-Charles and Bellet, Thierry},
  journal={IEEE Access},
  volume={8},
  pages={164833--164843},
  year={2020},
  publisher={IEEE}
}

@inproceedings{colley2019better,
author = {Colley, Mark and Walch, Marcel and Rukzio, Enrico},
title = {For a better (simulated) world: considerations for VR in external communication research},
year = {2019},
isbn = {9781450369206},
publisher = {Association for Computing Machinery},
address = {New York, NY, USA},
url = {https://doi.org/10.1145/3349263.3351523},
doi = {10.1145/3349263.3351523},
booktitle = {Proceedings of the 11th International Conference on Automotive User Interfaces and Interactive Vehicular Applications: Adjunct Proceedings},
pages = {442–449},
numpages = {8},
location = {Utrecht, Netherlands},
series = {AutomotiveUI '19}
}

@inproceedings{joshi2022advancing,
  title={Advancing the State of AV--Vulnerable Road User Interaction: Challenges and Opportunities},
  author={Joshi, Swapna and Block, Avram and Tabone, Wilbert and Pandya, Aryaman and Schmitt, Paul and MassRobotics and Motional},
  booktitle={Proceedings of the AAAI 2023 Spring Symposium},
  year={2023},
  numpages = {10},
  publisher={AAAI},
  address={Palo Alto, CA, USA}
}

@INPROCEEDINGS{enam2024external,
  author={Enam, Md Atik and Bastola, Ananta and Brinkley, Julian},
  booktitle={2024 IEEE 4th International Conference on Human-Machine Systems (ICHMS)}, 
  title={Are the External Human-Machine Interfaces (eHMI) Accessible for People with Disabilities? A Systematic Review}, 
  year={2024},
  volume={},
  number={},
  pages={1-6},
  doi={10.1109/ICHMS59971.2024.10555703},
publisher = {IEEE},
address   = {Piscataway, NJ, USA}
}

@inproceedings{colley2019including,
author = {Colley, Mark and Walch, Marcel and Gugenheimer, Jan and Rukzio, Enrico},
title = {Including people with impairments from the start: external communication of autonomous vehicles},
year = {2019},
isbn = {9781450369206},
publisher = {Association for Computing Machinery},
address = {New York, NY, USA},
url = {https://doi.org/10.1145/3349263.3351521},
doi = {10.1145/3349263.3351521},
booktitle = {Proceedings of the 11th International Conference on Automotive User Interfaces and Interactive Vehicular Applications: Adjunct Proceedings},
pages = {307–314},
numpages = {8},
location = {Utrecht, Netherlands},
series = {AutomotiveUI '19}
}

@article{ezzati2021interaction,
  title={Interaction of automated driving systems with pedestrians: Challenges, current solutions, and recommendations for eHMIs},
  author={Ezzati Amini, Roja and Katrakazas, Christos and Riener, Andreas and Antoniou, Constantinos},
  journal={Transport Reviews},
  volume={41},
  number={6},
  pages={788--813},
  year={2021},
  publisher={Taylor \& Francis}
}

@article{fabricius2022interactions,
  title={Interactions between heavy trucks and vulnerable road users—A systematic review to inform the interactive capabilities of highly automated trucks},
  author={Fabricius, Victor and Habibovic, Azra and Rizgary, Daban and Andersson, Jonas and W{\"a}rnest{\aa}l, Pontus},
  journal={Frontiers in Robotics and AI},
  volume={9},
  pages={818019},
  year={2022},
  publisher={Frontiers Media SA}
}

@article{predhumeau2021pedestrian,
  title={Pedestrian behavior in shared spaces with autonomous vehicles: An integrated framework and review},
  author={Pr{\'e}dhumeau, Manon and Spalanzani, Anne and Dugdale, Julie},
  journal={IEEE Transactions on Intelligent Vehicles},
  volume={8},
  number={1},
  pages={438--457},
  year={2021},
  publisher={IEEE}
}

@misc{EUAIActWebsite2026,
  title        = {EU Artificial Intelligence Act},
  author       = {{Future of Life Institute}},
  year         = {2026},
  howpublished = {\url{https://artificialintelligenceact.eu/}},
  organisation = {Future of Life Institute},
  url          = {https://artificialintelligenceact.eu/}
}

@inproceedings{tran2025animal,
author = {Tran, Tram Thi Minh and Yu, Xinyan and Hoggenmueller, Marius and Parker, Callum and Schmitt, Paul and Berrio Perez, Julie Stephany and Worrall, Stewart and Tomitsch, Martin},
title = {Animal Interaction with Autonomous Mobility Systems: Designing for Multi-Species Coexistence},
year = {2025},
isbn = {9798400720130},
publisher = {Association for Computing Machinery},
address = {New York, NY, USA},
url = {https://doi.org/10.1145/3744333.3747834},
doi = {10.1145/3744333.3747834},
booktitle = {Proceedings of the 17th International Conference on Automotive User Interfaces and Interactive Vehicular Applications},
pages = {164–179},
numpages = {16},
location = {
},
series = {AutomotiveUI '25}
}

@inproceedings{tran2025adaptive,
author = {Tran, Tram Thi Minh and Kay, Judy and Worrall, Stewart and Hoggenmueller, Marius and Parker, Callum and Yu, Xinyan and Berr\'{\i}o Perez, Julie Stephany and Shan, Mao and Tomitsch, Martin},
title = {Towards Adaptive External Communication in Autonomous Vehicles: A Conceptual Design Framework},
year = {2025},
isbn = {9798400720147},
publisher = {Association for Computing Machinery},
address = {New York, NY, USA},
url = {https://doi.org/10.1145/3744335.3758494},
doi = {10.1145/3744335.3758494},
booktitle = {Adjunct Proceedings of the 17th International Conference on Automotive User Interfaces and Interactive Vehicular Applications},
pages = {131–137},
numpages = {7},
location = {
},
series = {AutomotiveUI Adjunct '25}
}

@inproceedings{dey2024multimodal,
author = {Dey, Debargha and Senan, Toros Ufuk and Hengeveld, Bart and Colley, Mark and Habibovic, Azra and Ju, Wendy},
title = {Multi-Modal eHMIs: The Relative Impact of Light and Sound in AV-Pedestrian Interaction},
year = {2024},
isbn = {9798400703300},
publisher = {Association for Computing Machinery},
address = {New York, NY, USA},
url = {https://doi.org/10.1145/3613904.3642031},
doi = {10.1145/3613904.3642031},
booktitle = {Proceedings of the 2024 CHI Conference on Human Factors in Computing Systems},
articleno = {91},
numpages = {16},
location = {Honolulu, HI, USA},
series = {CHI '24}
}

@article{deb2018pedestrians,
  title={Pedestrians’ receptivity toward fully automated vehicles: Research review and roadmap for future research},
  author={Deb, Shuchisnigdha and Rahman, Md Mahmudur and Strawderman, Lesley J and Garrison, Teena M},
  journal={IEEE Transactions on Human-Machine Systems},
  volume={48},
  number={3},
  pages={279--290},
  year={2018},
  publisher={IEEE}
}

@article{schieben2019designing,
  title={Designing the interaction of automated vehicles with other traffic participants: design considerations based on human needs and expectations},
  author={Schieben, Anna and Wilbrink, Marc and Kettwich, Carmen and Madigan, Ruth and Louw, Tyron and Merat, Natasha},
  journal={Cognition, Technology \& Work},
  volume={21},
  number={1},
  pages={69--85},
  year={2019},
  publisher={Springer}
}

@article{carmona2021ehmi,
  title={eHMI: Review and guidelines for deployment on autonomous vehicles},
  author={Carmona, Juan and Guindel, Carlos and Garcia, Fernando and De La Escalera, Arturo},
  journal={Sensors},
  volume={21},
  number={9},
  pages={2912},
  year={2021},
  publisher={MDPI}
}

@inproceedings{gao2022external,
  title={External HMI for Automated Vehicles: Adding a Communication Perspective for all Road Users},
  author={Gao, Ruolin and Martens, Marieke},
  booktitle={AHFE International Conference on Human Factors and Systems Interaction},
  year={2022},
  doi={10.54941/ahfe1001922},
  url={https://doi.org/10.54941/ahfe1001922},
publisher = {AHFE International},
address   = {USA},
numpages = {7}
}

@article{wilbrink2023principles,
  title={Principles for external human--machine interfaces},
  author={Wilbrink, Marc and Cieler, Stephan and Wei{\ss}, Sebastian L and Beggiato, Matthias and Joisten, Philip and Feierle, Alexander and Oehl, Michael},
  journal={Information},
  volume={14},
  number={8},
  pages={463},
  year={2023},
  publisher={MDPI}
}

@inproceedings{dey2022shape,
author = {Dey, Debargha and De Zeeuw, Coen and Bruns, Miguel and Martens, Marieke and Pfleging, Bastian},
title = {Shape-Changing Interfaces in the Automotive Context: A Taxonomy to Aid the Systematic Development of Intuitive Gesture-Based eHMIs},
year = {2022},
isbn = {9781450394154},
publisher = {Association for Computing Machinery},
address = {New York, NY, USA},
url = {https://doi.org/10.1145/3543174.3546085},
doi = {10.1145/3543174.3546085},
booktitle = {Proceedings of the 14th International Conference on Automotive User Interfaces and Interactive Vehicular Applications},
pages = {56–64},
numpages = {9},
location = {Seoul, Republic of Korea},
series = {AutomotiveUI '22}
}

@article{dey2020taming,
  title={Taming the eHMI jungle: A classification taxonomy to guide, compare, and assess the design principles of automated vehicles' external human-machine interfaces},
  author={Dey, Debargha and Habibovic, Azra and L{\"o}cken, Andreas and Wintersberger, Philipp and Pfleging, Bastian and Riener, Andreas and Martens, Marieke and Terken, Jacques},
  journal={Transportation Research Interdisciplinary Perspectives},
  volume={7},
  pages={100174},
  year={2020},
  publisher={Elsevier}
}

@article{wang2021can,
  title={How can autonomous vehicles convey emotions to pedestrians? A review of emotionally expressive non-humanoid robots},
  author={Wang, Yiyuan and Hespanhol, Luke and Tomitsch, Martin},
  journal={Multimodal Technologies and Interaction},
  volume={5},
  number={12},
  pages={84},
  year={2021},
  publisher={MDPI}
}

@inproceedings{bied2024autonomous,
  title={Autonomous vehicles as social agents: Vehicle to pedestrian communication from v2x, ehmi and hri perspectives},
  author={Bied, Manuel and Bruno, Barbara and Vinel, Alexey},
  booktitle={2024 20th International Conference on Wireless and Mobile Computing, Networking and Communications (WiMob)},
  pages={86--91},
  year={2024},
publisher = {IEEE},
address   = {Piscataway, NJ, USA}
}

@article{rasouli2019autonomous,
  title={Autonomous vehicles that interact with pedestrians: A survey of theory and practice},
  author={Rasouli, Amir and Tsotsos, John K},
  journal={IEEE transactions on intelligent transportation systems},
  volume={21},
  number={3},
  pages={900--918},
  year={2019},
  publisher={IEEE}
}

@article{sankeerthana2022strategic,
  title={A strategic review approach on adoption of autonomous vehicles and its risk perception by road users},
  author={Sankeerthana, Gone and Raghuram Kadali, B},
  journal={Innovative Infrastructure Solutions},
  volume={7},
  number={6},
  pages={351},
  year={2022},
  publisher={Springer}
}

@article{riegler2019systematic,
  title={A systematic review of augmented reality applications for automated driving: 2009--2020},
  author={Riegler, Andreas and Riener, Andreas and Holzmann, Clemens},
  journal={PRESENCE: Virtual and Augmented Reality},
  volume={28},
  pages={87--126},
  year={2019},
  publisher={MIT Press One Rogers Street, Cambridge, MA 02142-1209, USA journals-info~…}
}

@article{riegler2021augmented,
  title={Augmented reality for future mobility: insights from a literature review and HCI workshop},
  author={Riegler, Andreas and Riener, Andreas and Holzmann, Clemens},
  journal={i-com},
  volume={20},
  number={3},
  pages={295--318},
  year={2021},
  publisher={De Gruyter Oldenbourg}
}

@article{stefanidi2024augmented,
  title={Augmented Reality on the Move: A Systematic Literature Review for Vulnerable Road Users},
  author={Stefanidi, Helen and Tatzgern, Markus and Meschtscherjakov, Alexander},
  journal={Proceedings of the ACM on Human-Computer Interaction},
  volume={8},
  number={MHCI},
  pages={1--30},
  year={2024},
  publisher={ACM New York, NY, USA}
}

@ARTICLE{bastola2025driving,
  author={Bastola, Ashish and Wang, Hao and Boroujeni, Sayed Pedram Haeri and Brinkley, Julian and Moshayedi, Ata Jahangir and Razi, Abolfazl},
  journal={IEEE Access}, 
  title={Driving Toward Inclusion: A Systematic Review of AI-Powered Accessibility Enhancements for People With Disability in Autonomous Vehicles}, 
  year={2025},
  volume={13},
  number={},
  pages={61384-61415},
  doi={10.1109/ACCESS.2025.3555923},
publisher={IEEE}}

@article{omeiza2021explanations,
  title={Explanations in autonomous driving: A survey},
  author={Omeiza, Daniel and Webb, Helena and Jirotka, Marina and Kunze, Lars},
  journal={IEEE Transactions on Intelligent Transportation Systems},
  volume={23},
  number={8},
  pages={10142--10162},
  year={2021},
  publisher={IEEE}
}

@inproceedings{kumar2024human,
  title={Human-Centered AI for Autonomous Vehicles: A Review of Interaction Strategies and Technologies},
  author={Kumar, Abhishek and Rana, Kritika and Gupta, Rishaan and Gunta, Deenika and Mahida, Ankur and others},
  booktitle={2024 3rd Edition of IEEE Delhi Section Flagship Conference (DELCON)},
  pages={1--6},
  year={2024},
  publisher = {IEEE},
address   = {Piscataway, NJ, USA},
}

@inproceedings{gadermann2024increasing,
  title={Increasing Acceptance Through Design: Review of Evaluation Methods for Interaction Design in Mixed Traffic},
  author={Gadermann, Lars and Holder, Daniel and Maier, Thomas},
  booktitle={DS 130: Proceedings of NordDesign 2024},
  editor={Malmqvist, Johan and Candi, Marina and Saemundsson, Ragnheiður J. and Bystrom, Fredrik and Isaksson, Ola},
  pages={123--132},
  year={2024},
  publisher={Design Society},
  address={Glasgow, United Kingdom},
  series={NordDESIGN},
  doi={10.35199/NORDDESIGN2024.14},
  isbn={978-1-912254-21-7}
}

@inproceedings{zhang2021human,
author = {Zhang, Jiehuang and Shu, Ying and Yu, Han},
title = {Human-Machine Interaction for Autonomous Vehicles: A Review},
year = {2021},
isbn = {978-3-030-77625-1},
publisher = {Springer-Verlag},
address = {Berlin, Heidelberg},
url = {https://doi.org/10.1007/978-3-030-77626-8_13},
doi = {10.1007/978-3-030-77626-8_13},
booktitle = {Social Computing and Social Media: Experience Design and Social Network Analysis: 13th International Conference, SCSM 2021, Held as Part of the 23rd HCI International Conference, HCII 2021, Virtual Event, July 24–29, 2021, Proceedings, Part  I},
pages = {190–201},
numpages = {12}
}

@article{zhou2021factors,
  title={Factors affecting pedestrians’ trust in automated vehicles: Literature review and theoretical model},
  author={Zhou, Siyuan and Sun, Xu and Liu, Bingjian and Burnett, Gary},
  journal={IEEE Transactions on Human-Machine Systems},
  volume={52},
  number={3},
  pages={490--500},
  year={2021},
  publisher={IEEE}
}

@article{yuan2024driving,
  author    = {Yuan, Xiaofang and Deng, Fenghui and Yao, Xiaoyu and Wu, Yu},
  title     = {Driving Behaviour Research for Autonomous Vehicle Interaction Design: A Comprehensive Review and Future Directions},
  journal   = {International Journal of Human--Computer Interaction},
  volume    = {41},
  number    = {16},
  pages     = {10245--10265},
  year      = {2024},
  publisher = {Taylor \& Francis},
  doi       = {10.1080/10447318.2024.2432447},
  url       = {https://doi.org/10.1080/10447318.2024.2432447}
}

@inproceedings{deshmukh2023systematic,
  title={A systematic review of challenging scenarios involving automated vehicles and vulnerable road users},
  author={Deshmukh, Aditya and Wang, Zifei and Guo, Huizhong and Ammar, Dania and Sherony, Rini and Feng, Fred and Lin, Brian and Bao, Shan and Zhou, Feng},
  booktitle={Proceedings of the Human Factors and Ergonomics Society Annual Meeting},
  volume={67},
  pages={678--685},
  year={2023},
  publisher = {SAGE Publications},
address   = {Los Angeles, CA, USA}
}

@article{riegler2021systematic,
  title={A systematic review of virtual reality applications for automated driving: 2009--2020},
  author={Riegler, Andreas and Riener, Andreas and Holzmann, Clemens},
  journal={Frontiers in human dynamics},
  volume={3},
  pages={689856},
  year={2021},
  publisher={Frontiers Media SA}
}

@article{reyes2022vulnerable,
  title={Vulnerable road users and connected autonomous vehicles interaction: A survey},
  author={Reyes-Mu{\~n}oz, Ang{\'e}lica and Guerrero-Ib{\'a}{\~n}ez, Juan},
  journal={Sensors},
  volume={22},
  number={12},
  pages={4614},
  year={2022},
  publisher={MDPI}
}

@inproceedings{hollander2021taxonomy,
author = {Holl\"{a}nder, Kai and Colley, Mark and Rukzio, Enrico and Butz, Andreas},
title = {A Taxonomy of Vulnerable Road Users for HCI Based On A Systematic Literature Review},
year = {2021},
isbn = {9781450380966},
publisher = {Association for Computing Machinery},
address = {New York, NY, USA},
url = {https://doi.org/10.1145/3411764.3445480},
doi = {10.1145/3411764.3445480},
booktitle = {Proceedings of the 2021 CHI Conference on Human Factors in Computing Systems},
articleno = {158},
numpages = {13},
location = {Yokohama, Japan},
series = {CHI '21}
}

@misc{Singapore_AV,
  author       = {Singapore Land Transport Authority},
  title        = {Autonomous Vehicles},
  howpublished = {Land Transport Authority website: Industry \& Innovations – Technologies – Autonomous Vehicles},
  year         = {n.d.},
  url          = {https://www.lta.gov.sg/content/ltagov/en/industry_innovations/technologies/autonomous_vehicles.html}
}

@misc{SAE_J3134_2019,
  title        = {Automated Driving System (ADS) Marker Lamp},
  author  = {SAE International},
  number       = {J3134\_201905},
  type         = {SAE Recommended Practice},
  year         = {2019},
  month        = {May},
  doi          = {10.4271/J3134_201905},
  url          = {https://doi.org/10.4271/J3134_201905},
  note         = {Issued 31 May 2019}
}

@misc{ISO_TR23049_2018,
  author       = {{International Organization for Standardization (ISO)}},
  title        = {Road Vehicles — Ergonomic aspects of external visual communication from automated vehicles to other road users},
  type         = {Technical Report},
  number       = {ISO/TR 23049:2018},
  institution  = {ISO},
  year         = {2018},
  month        = {September},
  edition      = {1},
  pages        = {7},
  url          = {https://www.iso.org/standard/74397.html}
}

@misc{ETSI_TR102638,
  author       = {{European Telecommunications Standards Institute (ETSI)}},
  title        = {Intelligent Transport Systems (ITS); Vehicular Communications; Basic Set of Applications; Release 2},
  type         = {Technical Report},
  number       = {ETSI TR 102 638 V2.1.1},
  institution  = {ETSI},
  year         = {2024},
  month        = {April},
  day          = {16},
  edition      = {2.1.1},
  pages        = {114},
  url          = {https://www.etsi.org/deliver/etsi_tr/102600_102699/102638/02.01.01_60/tr_102638v020101p.pdf}
}

@misc{tuv2023china,
title        = {China: Draft National Standard on Light Signaling Devices and Systems for Vehicles and Their Trailers},
author       = {{TÜV Rheinland}},
year         = 2023,
url          = {https://www.tuv.com/regulations-and-standards/en/china-draft-national-standard-on-light-signaling-devices-and-systems-for-vehicles-and-their-trailers.html}
}

@misc{GTB_GlobalLightingGroup,
  author       = {{The International Automotive Lighting and Light Signalling Expert Group (GTB)}},
  title        = {GTB – The International Automotive Lighting and Light Signalling Expert Group},
  howpublished = {GTB website},
  year         = {n.d.},
  url          = {https://gtb-lighting.org/}
}

@misc{UNECE_GRE,
  author       = {{United Nations Economic Commission for Europe (UNECE)}},
  title        = {Working Party on Lighting and Light-Signalling — Introduction},
  howpublished = {UNECE website},
  year         = {n.d.},
  url          = {https://unece.org/transportvehicle-regulations/working-party-lighting-and-light-signalling-introduction}
}

@misc{UNECE_GRVA,
  author       = {{United Nations Economic Commission for Europe (UNECE)}},
  title        = {Working Party on Automated/Autonomous and Connected Vehicles — Introduction},
  howpublished = {UNECE website},
  year         = {2018},
  url          = {https://unece.org/transport/road-transport/working-party-automatedautonomous-and-connected-vehicles-introduction}
}

@misc{ISO_TR23735,
  author       = {{International Organization for Standardization (ISO)}},
  title        = {Road vehicles — Ergonomic design guidance for external visual communication from automated vehicles to other road users},
  type         = {Technical Report (NP)},
  number       = {ISO/NP TR 23735},
  institution  = {ISO},
  year         = {2019},
}

@misc{ISO_18682,
  author       = {{International Organization for Standardization (ISO)}},
  title        = {Intelligent transport systems — External hazard detection and notification systems — Basic requirements},
  number       = {ISO 18682:2016},
  type         = {International Standard},
  institution  = {ISO},
  year         = {2016},
  month        = {October},
  edition      = {1},
  pages        = {19},
  url          = {https://www.iso.org/standard/63250.html}
}

@misc{ISO_TR4448,
  author       = {{International Organization for Standardization (ISO)}},
  title        = {Intelligent transport systems — Public-area mobile robots (PMR) — Part 1: Overview of paradigm},
  type         = {Technical Report},
  number       = {ISO/TR 4448-1:2024},
  institution  = {ISO},
  year         = {2024},
  month        = {August},
  edition      = {1},
  pages        = {24},
  url          = {https://www.iso.org/standard/81068.html}
}

@misc{ISO_TR12204,
  author       = {{International Organization for Standardization (ISO)}},
  title        = {Road vehicles — Ergonomic aspects of transport information and control systems — Introduction to integrating safety critical and time critical warning signals},
  type         = {Technical Report},
  number       = {ISO/TR 12204:2012},
  institution  = {ISO},
  year         = {2012},
  month        = {November},
  edition      = {1},
  pages        = {50},
  url          = {https://www.iso.org/standard/51275.html}
}

@misc{ISO_TR23720,
  author       = {{International Organization for Standardization (ISO)}},
  title        = {Road Vehicles — Methods for evaluating other road user behavior in the presence of automated vehicle external communication},
  type         = {Technical Report (AWI)},
  number       = {ISO/AWI TR 23720},
  institution  = {ISO},
  year         = {2019},
}

@article{tekkesinoglu2025advancing,
  title={Advancing explainable autonomous vehicle systems: A comprehensive review and research roadmap},
  author={Tekkesinoglu, Sule and Habibovic, Azra and Kunze, Lars},
  journal={ACM Transactions on Human-Robot Interaction},
  volume={14},
  number={3},
  pages={1--46},
  year={2025},
  publisher={ACM New York, NY}
}

@misc{ISO_9241_303,
  author       = {{International Organization for Standardization (ISO)}},
  title        = {Ergonomics of human-system interaction — Part 303: Requirements for electronic visual displays},
  type         = {Standard},
  number       = {ISO 9241-303:2011},
  institution  = {ISO},
  year         = {2011},
  month        = {April},
  edition      = {1},
  pages        = {107},
  url          = {https://www.iso.org/standard/57992.html}
}

@misc{SAE_J578_2020,
  title        = {Chromaticity Requirements for Ground Vehicle Lamps and Lighting Equipment},
  author  = {SAE International},
  number       = {J578\_202004},
  type         = {SAE Standard},
  year         = {2020},
  month        = {April},
  doi          = {10.4271/J578\_202004},
  url          = {https://doi.org/10.4271/J578_202004},
}

@misc{UNECE_R149,
  author       = {{United Nations Economic Commission for Europe (UNECE)}},
  title        = {United Nations Regulation No.\ 149: Uniform provisions concerning the approval of road illumination devices (lamps) and systems for power-driven vehicles},
  howpublished = {UNECE, published 15 November 2019},
  year         = {2019},
  url          = {https://unece.org/sites/default/files/2021-05/R149e.pdf}
}

@misc{lyft2020notification,
  title        = {Autonomous vehicle notification system},
  author       = {{Lyft, Inc.}},
  year         = {2020},
  number       = {US10607491B2},
  url          = {https://patents.google.com/patent/US10607491B2/en},
  note         = {Filed 2018, issued 2020}
}

@misc{waymo2015pedestrian,
  title        = {Pedestrian notifications},
  author       = {{Google LLC}},
  year         = {2015},
  number       = {US8954252B1},
  url          = {https://patents.google.com/patent/US8954252B1/en},
  note         = {Filed 2013, issued 2015}
}

@misc{uber2018lightoutput,
  title        = {Light output system for a self-driving vehicle},
  author       = {{Uber Technologies, Inc.}},
  year         = {2018},
  number       = {US10160378B2},
  url          = {https://patents.google.com/patent/US10160378B2/},
  note         = {Filed 2017, issued 2018}
}

@misc{gmc2024identification,
  title        = {Autonomous vehicle identification},
  author       = {{GM Cruise Holdings LLC}},
  year         = {2024},
  number       = {US12038289B2},
  url          = {https://patents.google.com/patent/US12038289B2/},
  note         = {Filed 2021, issued 2024}
}

@misc{nissan2024notification,
  title        = {Autonomous vehicle notification system and method},
  author       = {{Nissan North America Inc.}},
  year         = {2024},
  number       = {EP3665564B1},
  url          = {https://patents.google.com/patent/EP3665564B1},
  note         = {Filed 2017, issued 2024}
}

@misc{apple2023onewayfilter,
  title        = {System with one-way filter over light-emitting elements},
  author       = {{Apple Inc.}},
  year         = {2023},
  number       = {WO2023141369A1},
  url          = {https://patents.google.com/patent/WO2023141369A1/},
  note         = {Filed 2022, published 2023}
}

@misc{gm2023responses,
  author       = {GM Global Technology Operations LLC},
  title        = {Responses to Vulnerable Road User's Adversarial Behavior},
  year         = {2023},
  number       = {US20240383501A1},
  url          = {https://image-ppubs.uspto.gov/dirsearch-public/print/downloadPdf/20240383501}
}

@online{Rinspeed2017,
  title = {Oasis Press Book},
  author = {Rinspeed},
  year = {2017},
  url = {https://www.rinspeed.eu/upload/conceptfiles/rinspeed-oasis-partnerbook2017.pdf},
  urldate = {2025-09-09},
  howpublished = {Online},
}

@online{Daimler2017,
  title = {Autonomous concept car smart vision EQ fortwo},
  author = {Daimler},
  year = {2017},
  url = {https://media.mbusa.com/releases/release-80848dccd3f3680a764667ad530987e9-autonomous-concept-car-smart-vision-eq-fortwo},
  urldate = {2025-09-09},
  howpublished = {Online},
}

@online{JaguarLandRover2019,
title = {Jaguar Land Rover Lights Up the Road Ahead for Self-Driving Vehicles of the Future},  
author = {Jaguar Land Rover},
  year = {2019},
  url = {https://media.jaguarlandrover.com/news/2019/01/jaguar-land-rover-lights-road-ahead-self-driving-vehicles-future},
  urldate = {2025-09-09},
  howpublished = {Online},
}

@online{JaguarLandRover2018,
title = {Jaguar Land Rover’s Virtual Eyes Look at Trust in Self-Driving Cars},
author = {Jaguar Land Rover},
  year = {2018},
  url = {https://www.jlr.com/news/2018/08/jaguar-land-rovers-virtual-eyes-look-trust-self-driving-cars},
  urldate = {2025-09-09},
  howpublished = {Online},
}

@online{Kia2018,
  title = {Kia’s Niro EV concept is a battery-powered 201bhp SUV},
  author = {Kia},
  year = {2018},
  url = {https://www.topgear.com/car-news/consumer-electronics-show/kias-niro-ev-concept-battery-powered-201bhp-suv},
  urldate = {2025-09-09},
  howpublished = {Online},
}

@online{MercedesBenz2019-2,
  title = {Self-Driving Mercedes-Benz S-Class Puts on a Turquoise Light Show},
  author = {Mercedes-Benz},
  year = {2019},
  url = {https://www.autoevolution.com/news/self-driving-mercedes-benz-s-class-puts-on-a-turquoise-light-show-132111.html},
  urldate = {2025-09-09},
  howpublished = {Online},
}

@online{NissanMotorCorporation2015,
  title = {Nissan IDS Concept: Nissan’s vision for the future of EVs and autonomous driving},
  author = {Nissan Motor Corporation},
  year = {2015},
  url = {https://global.nissannews.com/en/releases/151028-01-e},
  urldate = {2025-09-09},
  howpublished = {Online},
}

@online{Semcon2016,
  title = {The Smiling Car},
  author = {Semcon},
  year = {2016},
  url = {https://www.semcon.com/article/the-smiling-car},
  urldate = {2025-09-09},
  howpublished = {Online},
}

@online{Audi2015,
  title = {Audi will use headlight lasers to protect pedestrians},
  author = {Audi},
  year = {2015},
  url = {https://www.techradar.com/news/car-tech/audi-will-use-headlight-lasers-to-protect-pedestrians-1310899/2},
  urldate = {2025-09-09},
  howpublished = {Online},
}

@online{BMW2016,
  title = {BMW Celebrates 100 Years with Vision Next 100 Concept},
  author = {BMW},
  year = {2016},
  url = {https://blog.bestride.com/news/technology/bmw-vision-next-100/},
  urldate = {2025-09-09},
  howpublished = {Online},
}

@online{Daimler2015,
  title = {Mercedes-Benz F015 Luxury In Motion revealed},
  author = {Daimler},
  year = {2015},
  url = {https://www.drive.com.au/news/mercedesbenz-f015-luxury-in-motion-revealed-20150106-12ix8u/},
  urldate = {2025-09-09},
  howpublished = {Online},
}

@online{Hyundai2023,
  title = {Hyundai's New Headlights Can Project Virtual Pedestrian Crossing On Dark Roads},
  author = {Hyundai},
  year = {2023},
  url = {https://carbuzz.com/news/hyundais-new-headlights-can-project-virtual-pedestrian-crossing-on-dark-roads/},
  urldate = {2025-09-09},
  howpublished = {Online},
}

@online{Navya2017,
  title = {Navya Factory Visit},
  author = {Navya},
  year = {2017},
  url = {https://thelastdriverlicenseholder.com/2018/06/29/navya-factory-visit/},
  urldate = {2025-09-09},
  howpublished = {Online},
}

@online{Kiwibot2024,
  title = {Kiwibot acquires an ad startup to turn its delivery robots into mobile billboards},
  author = {Kiwibot},
  year = {2024},
  url = {https://techcrunch.com/2024/09/19/kiwibot-acquires-ad-startup-to-turn-its-delivery-robots-into-mobile-billboards/},
  urldate = {2025-09-09},
  howpublished = {Online},
}

@online{Ottobots2023,
  title = {Multiple Screens for Enhanced User Experience and Branding},
  author = {Ottobots},
  year = {2023},
  url = {https://ottonomy.io/smart-cities/},
  urldate = {2025-09-09},
  howpublished = {Online},
}

@online{SonyHonda2023,
  title = {Light bars will communicate with the outside world},
  author = {Sony Honda},
  year = {2023},
  url = {https://www.autonews.com/technology/new-auto-tech-media-bar-external-digital-display/},
  urldate = {2025-09-09},
  howpublished = {Online},
}

@online{ToyotaMotorCorporation2018,
  title = {Autonomous Electric Toyota to Provide Mobility for Olympic Athletes},
  author = {Toyota Motor Corporation},
  year = {2018},
  url = {https://www.mandurahtoyota.com.au/news/autonomous-electric-toyota-to-provide-mobility-for-olympic-athletes/},
  urldate = {2025-09-09},
  howpublished = {Online},
}

@online{VolvoCars2018,
title = {Volvo 360c Concept Gives Autonomous Cars a Purpose and a Voice},
author = {Volvo Cars},
  year = {2018},
  url = {https://www.slashgear.com/volvo-360c-autonomous-car-concept-flying-alternative-pedestrian-communication-05544441/},
  urldate = {2025-09-09},
  howpublished = {Online},
}

@online{Coco2021,
  title = {Coco 1 remotely piloted delivery robot is headed for the streets of LA},
  author = {Coco},
  year = {2021},
  url = {https://newatlas.com/robotics/coco-1-delivery-robot/},
  urldate = {2025-09-09},
  howpublished = {Online},
}

@online{Kiwibot2022,
  title = {Kiwibot announces rebranding: we deliver emotions!},
  author = {Kiwibot},
  year = {2022},
  url = {https://www.kiwibot.com/articles/kiwibot-announces-rebranding-we-deliver-emotions},
  urldate = {2025-09-09},
  howpublished = {Online},
}

@online{ServeRobotics2024,
  title = {SEC Filling - Serve Robotics},
  author = {Serve Robotics},
  year = {2024},
  url = {https://www.youtube.com/watch?v=n9MJorTjpxU&t=58s},
  urldate = {2025-09-09},
  howpublished = {Online},
}

@online{BaidusApollo2021,
  title = {Chinese tech giant Baidu begins publicly testing Apollo Go robotaxis in Shanghai},
  author = {Baidu’s Apollo},
  year = {2021},
  url = {https://techcrunch.com/2021/09/13/chinese-tech-giant-baidu-begins-publicly-testing-apollo-go-robotaxis-in-shanghai/},
  urldate = {2025-09-09},
  howpublished = {Online},
}

@online{PonyAI2019,
  title = {chinese-self-driving-contender-pony-ai-strikes-pact-with-toyota},
  author = {Pony.AI},
  year = {2019},
  url = {https://www.bloomberg.com/news/articles/2019-08-26/chinese-self-driving-contender-pony-ai-strikes-pact-with-toyota},
  urldate = {2025-09-09},
  howpublished = {Online},
}

@online{Intel2018,
  title = {Intel study fine-tunes the friendliness in self-driving cars},
  author = {Intel},
  year = {2018},
  url = {https://www.cnet.com/roadshow/news/intel-studies-how-to-make-self-driving-cars-friendly/},
  urldate = {2025-09-09},
  howpublished = {Online},
}

@online{WeRideChina2020,
  title = {Autonomous driving tech in China},
  author = {WeRide},
  year = {2020},
  url = {https://tier4.jp/en/media/detail/?sys_id=5fFXVAxgyEqTycnNCFQFXs&category=BLOG},
  urldate = {2025-09-09},
  howpublished = {Online},
}

@online{Driveai2016,
  title = {Drive.ai wants to help autonomous cars talk with the people around them},
  author = {Drive.ai},
  year = {2016},
  url = {https://www.theverge.com/2016/8/30/12700290/drive-ai-autonomous-car-human-robot-interface},
  urldate = {2025-09-09},
  howpublished = {Online},
}

@online{GoogleWaymo2022,
  title = {New Waymo One features inspired by DOT's Inclusive Design Challenge},
  author = {Google Waymo},
  year = {2022},
  url = {https://waymo.com/blog/2022/08/new-waymo-one-features-inspired-by-dots},
  urldate = {2025-09-09},
  howpublished = {Online},
}

@online{GoogleWaymo2025,
  title = {Residents say Waymo robotaxis are driving them mad. Can AI and humans coexist?},
  author = {Google Waymo},
  year = {2025},
  url = {https://edition.cnn.com/2025/06/25/us/santa-monica-waymo-battles},
  urldate = {2025-09-09},
  howpublished = {Online},
}

@online{GoogleWaymo2020,
  title = {Designing the 5th-generation Waymo Driver},
  author = {Google Waymo},
  year = {2020},
  url = {https://waymo.com/blog/2020/03/designing-5th-generation-waymo-driver},
  urldate = {2025-09-09},
  howpublished = {Online},
}

@online{GoogleWaymo2023,
  title = {How will driverless cars ‘talk’ to pedestrians? Waymo has a few ideas},
  author = {Google Waymo},
  year = {2023},
  url = {https://www.theverge.com/2023/10/13/23913251/waymo-roof-dome-communicate-intent-pedestrian-driver},
  urldate = {2025-09-09},
  howpublished = {Online},
}

@online{Zoox2021,
  title = {How self-driving cars will 'talk' to pedestrians},
  author = {Zoox},
  year = {2021},
  url = {https://mashable.com/article/auotnomous-vehicles-talking-to-pedestrians},
  urldate = {2025-09-09},
  howpublished = {Online},
}

@inproceedings{altaie2025scalable,
author = {Al-Taie, Ammar and Freeman, Euan and Pollick, Frank and Brewster, Stephen Anthony},
title = {evARything, evARywhere, all at once: Exploring Scalable Holistic Autonomous Vehicle-Cyclist Interfaces},
year = {2025},
isbn = {9798400713941},
publisher = {Association for Computing Machinery},
address = {New York, NY, USA},
url = {https://doi.org/10.1145/3706598.3713412},
doi = {10.1145/3706598.3713412},
booktitle = {Proceedings of the 2025 CHI Conference on Human Factors in Computing Systems},
articleno = {82},
numpages = {18},
location = {
},
series = {CHI '25}
}

@inproceedings{cumbal2025visualising,
author = {Cumbal, Ronald and Calvo-Barajas, Natalia and Escobar-Planas, Marina and Rouchitsas, Alexandros and Castellano, Ginevra},
title = {Visualizing Confidence in Delivery Robots: Insights from Two Online Studies},
year = {2025},
isbn = {9798400713958},
publisher = {Association for Computing Machinery},
address = {New York, NY, USA},
url = {https://doi.org/10.1145/3706599.3719695},
doi = {10.1145/3706599.3719695},
booktitle = {Proceedings of the Extended Abstracts of the CHI Conference on Human Factors in Computing Systems},
articleno = {578},
numpages = {7},
location = {
},
series = {CHI EA '25}
}

@inproceedings{colley2025optimisation,
author = {Colley, Mark and Jansen, Pascal and Keskar, Mugdha and Rukzio, Enrico},
title = {Improving External Communication of Automated Vehicles Using Bayesian Optimization},
year = {2025},
isbn = {9798400713941},
publisher = {Association for Computing Machinery},
address = {New York, NY, USA},
url = {https://doi.org/10.1145/3706598.3714187},
doi = {10.1145/3706598.3714187},
booktitle = {Proceedings of the 2025 CHI Conference on Human Factors in Computing Systems},
articleno = {80},
numpages = {16},
location = {
},
series = {CHI '25}
}

@inproceedings{altaie2025cultures,
author = {Al-Taie, Ammar and Matviienko, Andrii and O'Hagan, Joseph and Pollick, Frank and Brewster, Stephen Anthony},
title = {Around the World in 60 Cyclists: Evaluating Autonomous Vehicle-Cyclist Interfaces Across Cultures},
year = {2025},
isbn = {9798400713941},
publisher = {Association for Computing Machinery},
address = {New York, NY, USA},
url = {https://doi.org/10.1145/3706598.3713407},
doi = {10.1145/3706598.3713407},
booktitle = {Proceedings of the 2025 CHI Conference on Human Factors in Computing Systems},
articleno = {217},
numpages = {18},
location = {
},
series = {CHI '25}
}

@inproceedings{rouchitsas2025framework,
  title={A Framework for Disentangling Efficiency from Effectiveness in External HMI Evaluation Procedures for Automated Vehicles},
  author={Rouchitsas, Anastasios},
  booktitle={Proceedings of the 11th International Conference on Vehicle Technology and Intelligent Transport Systems (VEHITS 2025)},
  pages={616--621},
  year={2025},
  publisher={SCITEPRESS -- Science and Technology Publications},
  address={Set{\'u}bal, Portugal},
  doi={10.5220/0013435900003941},
  isbn={978-989-758-745-0},
  issn={2184-495X}
}

@article{lee2025mitigating,
author = {Jieun Lee and Tatsuru Daimon},
title = {Not Always Good: Mitigating Pedestrians’ Less Careful Crossing Behavior by External Human-Machine Interfaces on Automated Vehicles},
journal = {International Journal of Human–Computer Interaction},
volume = {41},
number = {8},
pages = {4528--4540},
year = {2025},
publisher = {Taylor \& Francis},
doi = {10.1080/10447318.2024.2352212},
}

@Article{yan2025comparing,
AUTHOR = {Yan, Ming and Rampino, Lucia and Caruso, Giandomenico},
TITLE = {Comparing User Acceptance in Human–Machine Interfaces Assessments of Shared Autonomous Vehicles: A Standardized Test Procedure},
JOURNAL = {Applied Sciences},
VOLUME = {15},
YEAR = {2025},
NUMBER = {1},
ARTICLE-NUMBER = {45},
URL = {https://www.mdpi.com/2076-3417/15/1/45},
ISSN = {2076-3417},
DOI = {10.3390/app15010045},
numpages = {25}
}

@inproceedings{gadermann2025codecharts,
author = {Gadermann, Lars and Holder, Daniel and Maier, Thomas},
title = {CodeCharts and AI as Alternatives to On-site Eyetracking for Vehicle Design Evaluation},
year = {2025},
isbn = {978-3-031-92688-4},
publisher = {Springer-Verlag},
address = {Berlin, Heidelberg},
url = {https://doi.org/10.1007/978-3-031-92689-1_11},
doi = {10.1007/978-3-031-92689-1_11},
booktitle = {HCI in Mobility, Transport, and Automotive Systems: 7th International Conference, MobiTAS 2025, Held as Part of the 27th HCI International Conference, HCII 2025, Gothenburg, Sweden, June 22–27, 2025, Proceedings, Part I},
pages = {179–193},
numpages = {15},
location = {Gothenburg, Sweden}
}

@ARTICLE{liu2025preinstruction,
  author={Liu, Hailong and Hirayama, Takatsugu},
  journal={IEEE Transactions on Intelligent Transportation Systems}, 
  title={Pre-Instruction for Pedestrians Interacting Autonomous Vehicles With eHMI: Effects on Their Psychology and Walking Behavior}, 
  year={2025},
  volume={26},
  number={8},
  pages={11313-11324},
  doi={10.1109/TITS.2025.3560621}}

@article{yi2024impact,
title = {The impact of nighttime car body lighting on pedestrians’ distraction: A virtual reality simulation based on bottom-up attention mechanism},
journal = {Safety Science},
volume = {180},
pages = {106633},
year = {2024},
issn = {0925-7535},
doi = {https://doi.org/10.1016/j.ssci.2024.106633},
url = {https://www.sciencedirect.com/science/article/pii/S0925753524002236},
author = {Xiangwei Yi and Rui Zhao and Yandan Lin},
}

@inproceedings{gao2024role,
author = {Gao, Ruolin and Verstegen, Rutger and Dong, Haoyu and Bazilinskyy, Pavlo and Martens, Marieke},
title = {Incorporating Multiple Users' Perspectives in HMI Design for Automated Vehicles: Exploration of a Role-Switching Approach},
year = {2024},
isbn = {9798400705205},
publisher = {Association for Computing Machinery},
address = {New York, NY, USA},
url = {https://doi.org/10.1145/3641308.3685047},
doi = {10.1145/3641308.3685047},
booktitle = {Adjunct Proceedings of the 16th International Conference on Automotive User Interfaces and Interactive Vehicular Applications},
pages = {197–202},
numpages = {6},
location = {Stanford, CA, USA},
series = {AutomotiveUI '24 Adjunct}
}

@inproceedings{dong2024insideout,
author = {Dong, Jiayuan and Gowda, Nikhil and Wang, Yiyuan and Choe, Mungyeong and Alsaid, Areen and Alvarez, Ignacio and Krome, Sven and Jeon, Myounghoon},
title = {Inside Out: Emotion GaRage Vol. V},
year = {2024},
isbn = {9798400705205},
publisher = {Association for Computing Machinery},
address = {New York, NY, USA},
url = {https://doi.org/10.1145/3641308.3677403},
doi = {10.1145/3641308.3677403},
booktitle = {Adjunct Proceedings of the 16th International Conference on Automotive User Interfaces and Interactive Vehicular Applications},
pages = {260–263},
numpages = {4},
location = {Stanford, CA, USA},
series = {AutomotiveUI '24 Adjunct}
}

@inproceedings{tran2024mapping,
author = {Tran, Tram Thi Minh and Yu, Xinyan and Wang, Yiyuan and Parker, Callum and Tomitsch, Martin},
title = {Mapping Pedestrian-to-Driver Gestures: Implications for Autonomous Vehicle Bidirectional Interaction},
year = {2024},
isbn = {9798400705205},
publisher = {Association for Computing Machinery},
address = {New York, NY, USA},
url = {https://doi.org/10.1145/3641308.3685014},
doi = {10.1145/3641308.3685014},
booktitle = {Adjunct Proceedings of the 16th International Conference on Automotive User Interfaces and Interactive Vehicular Applications},
pages = {1–7},
numpages = {7},
location = {Stanford, CA, USA},
series = {AutomotiveUI '24 Adjunct}
}

@inproceedings{moore2020defense,
author = {Moore, Dylan and Currano, Rebecca and Shanks, Michael and Sirkin, David},
title = {Defense Against the Dark Cars: Design Principles for Griefing of Autonomous Vehicles},
year = {2020},
isbn = {9781450367462},
publisher = {Association for Computing Machinery},
address = {New York, NY, USA},
url = {https://doi.org/10.1145/3319502.3374796},
doi = {10.1145/3319502.3374796},
booktitle = {Proceedings of the 2020 ACM/IEEE International Conference on Human-Robot Interaction},
pages = {201–209},
numpages = {9},
location = {Cambridge, United Kingdom},
series = {HRI '20}
}

@inproceedings{passero2024honkable,
author = {Passero, Sergio and Pelikan, Hannah Rm and Broth, Mathias and Brown, Barry},
title = {Honkable Gestalts: Why Autonomous Vehicles Get Honked At},
year = {2024},
isbn = {9798400705106},
publisher = {Association for Computing Machinery},
address = {New York, NY, USA},
url = {https://doi.org/10.1145/3640792.3675732},
doi = {10.1145/3640792.3675732},
booktitle = {Proceedings of the 16th International Conference on Automotive User Interfaces and Interactive Vehicular Applications},
pages = {317–328},
numpages = {12},
location = {Stanford, CA, USA},
series = {AutomotiveUI '24}
}

@inproceedings{bazilinskyy2024multiagent,
author = {Bazilinskyy, Pavlo and Ebel, Patrick and Walker, Francesco and Dey, Debargha and Tran, Tram Thi Minh},
title = {It Is Not Always Just One Road User: Workshop on Multi-Agent Automotive Research},
year = {2024},
isbn = {9798400705205},
publisher = {Association for Computing Machinery},
address = {New York, NY, USA},
url = {https://doi.org/10.1145/3641308.3677400},
doi = {10.1145/3641308.3677400},
booktitle = {Adjunct Proceedings of the 16th International Conference on Automotive User Interfaces and Interactive Vehicular Applications},
pages = {268–272},
numpages = {5},
location = {Stanford, CA, USA},
series = {AutomotiveUI '24 Adjunct}
}

@inproceedings{wang2024physio,
author = {Wang, Yiyuan and Tran, Tram Thi Minh and Tomitsch, Martin},
title = {Physiological Measurements in Automated Vehicle-Pedestrian Research: Review and Future Opportunities},
year = {2024},
isbn = {9798400705205},
publisher = {Association for Computing Machinery},
address = {New York, NY, USA},
url = {https://doi.org/10.1145/3641308.3685043},
doi = {10.1145/3641308.3685043},
booktitle = {Adjunct Proceedings of the 16th International Conference on Automotive User Interfaces and Interactive Vehicular Applications},
pages = {172–177},
numpages = {6},
location = {Stanford, CA, USA},
series = {AutomotiveUI '24 Adjunct}
}

@article{harkin2024vulnerable,
title = {How do vulnerable road users evaluate automated vehicles in urban traffic? A focus group study with pedestrians, cyclists, e-scooter riders, older adults, and people with walking disabilities},
journal = {Transportation Research Part F: Traffic Psychology and Behaviour},
volume = {104},
pages = {59-71},
year = {2024},
issn = {1369-8478},
doi = {https://doi.org/10.1016/j.trf.2024.05.017},
url = {https://www.sciencedirect.com/science/article/pii/S1369847824001165},
author = {Kevin A. Harkin and A. Marie Harkin and Christina Gögel and Jens Schade and Tibor Petzoldt},
}

@inproceedings{yu2024agent,
author = {Yu, Xinyan and Hoggenm\"{u}ller, Marius and Tomitsch, Martin},
title = {From Agent Autonomy to Casual Collaboration: A Design Investigation on Help-Seeking Urban Robots},
year = {2024},
isbn = {9798400703300},
publisher = {Association for Computing Machinery},
address = {New York, NY, USA},
url = {https://doi.org/10.1145/3613904.3642389},
doi = {10.1145/3613904.3642389},
booktitle = {Proceedings of the 2024 CHI Conference on Human Factors in Computing Systems},
articleno = {587},
numpages = {14},
location = {Honolulu, HI, USA},
series = {CHI '24}
}

@inproceedings{tran2024interconnected,
author = {Tran, Tram Thi Minh and Parker, Callum and Hoggenm\"{u}ller, Marius and Wang, Yiyuan and Tomitsch, Martin},
title = {Exploring the Impact of Interconnected External Interfaces in Autonomous Vehicles on Pedestrian Safety and Experience},
year = {2024},
isbn = {9798400703300},
publisher = {Association for Computing Machinery},
address = {New York, NY, USA},
url = {https://doi.org/10.1145/3613904.3642118},
doi = {10.1145/3613904.3642118},
booktitle = {Proceedings of the 2024 CHI Conference on Human Factors in Computing Systems},
articleno = {89},
numpages = {17},
location = {Honolulu, HI, USA},
series = {CHI '24}
}

@Article{almeida2024reward,
AUTHOR = {Almeida, Raul and Sousa, Emanuel and Machado, Dário and Pereira, Frederico and Faria, Susana and Freitas, Elisabete},
TITLE = {Analysis of the Interaction between Humans and Autonomous Vehicles Equipped with External Human–Machine Interfaces: The Effect of an Experimental Reward Mechanism on Pedestrian Crossing Behavior in a Virtual Environment},
JOURNAL = {Sustainability},
VOLUME = {16},
YEAR = {2024},
NUMBER = {8},
numpages = {23},
ARTICLE-NUMBER = {3236},
URL = {https://www.mdpi.com/2071-1050/16/8/3236},
ISSN = {2071-1050},
DOI = {10.3390/su16083236}
}

@article{ammar2024identifying,
  author  = {Ammar, Dania and Wu, Yi and Guo, Huizhong and Misra, Aditi and Jia, Bochen and Bao, Shan},
  title   = {Identifying User Needs and Current Challenges of External Interface Design for AV--VRU Communications: Insights from an Expert Survey Data Analysis},
  journal = {Transportation Research Record},
  volume  = {2678},
  number  = {4},
  pages   = {228--242},
  year    = {2024},
  doi     = {10.1177/03611981231184239},
  url     = {https://doi.org/10.1177/03611981231184239}
}

@misc{milford2024integrating,
  title={Integrating Multi-Stakeholder Communication into Autonomous Vehicle Trolley Problem},
  author={Milford, Stephen R. and Laxton, Victoria and Vinel, Alexey},
  year={2024},
  month={Jan},
  doi={10.1109/WiMob61911.2024.10770323},
  url={https://hdl.handle.net/20.500.14716/137719},
  publisher={IEEE Computer Society},
}

@article{hochman2024pedestrians,
author = {Michal Hochman and Yisrael Parmet and Tal Oron-Gilad},
title = {Pedestrians’ intent to cross in a fully autonomous vehicle environment; looking at crossing opportunity, eHMI message and wait time},
journal = {Traffic Injury Prevention},
volume = {25},
number = {sup1},
pages = {S126--S136},
year = {2024},
publisher = {Taylor \& Francis},
doi = {10.1080/15389588.2024.2372801},
}

@INPROCEEDINGS{iwamoto2024skin,
  author={Iwamoto, Yuya and Kawanaka, Haruki and Oguri, Koji},
  booktitle={2024 IEEE 13th Global Conference on Consumer Electronics (GCCE)}, 
  title={Estimation of Human Anxiety Based on Skin Conductance Response and its Application to eHMI Performance Evaluation}, 
  year={2024},
  volume={},
  number={},
  pages={1012-1014},
  doi={10.1109/GCCE62371.2024.10760472},
publisher = {IEEE},
address   = {Piscataway, NJ, USA}}

@ARTICLE{zhang2024shared,
AUTHOR = {Zhang, Xiaochen and Song, Ziyang and Huang, Qianbo and Pan, Ziyi and Li, Wujing and Gong, Ruining and Zhao, Bi},
TITLE = {Shared eHMI: Bridging Human–Machine Understanding in Autonomous Wheelchair Navigation},
JOURNAL = {Applied Sciences},
VOLUME = {14},
YEAR = {2024},
numpages ={25},
NUMBER = {1},
ARTICLE-NUMBER = {463},
URL = {https://www.mdpi.com/2076-3417/14/1/463},
ISSN = {2076-3417},
DOI = {10.3390/app14010463}
}

@article{zhao2023invisible,
title = {The ‘invisible gorilla’ during pedestrian-AV interaction: Effects of secondary tasks on pedestrians’ reaction to eHMIs},
journal = {Accident Analysis \& Prevention},
volume = {192},
pages = {107246},
year = {2023},
issn = {0001-4575},
doi = {https://doi.org/10.1016/j.aap.2023.107246},
url = {https://www.sciencedirect.com/science/article/pii/S0001457523002932},
author = {Xiaoyuan Zhao and Xiaomeng Li and Andry Rakotonirainy and Samira {Bourgeois- Bougrine} and Dominique Gruyer and Patricia Delhomme},
}

@article{fu2023adopting,
title = {Adopting an HMI for overtaking assistance - Impact of distance display, advice, and guidance information on driver gaze and performance},
journal = {Accident Analysis \& Prevention},
volume = {191},
pages = {107204},
year = {2023},
issn = {0001-4575},
doi = {https://doi.org/10.1016/j.aap.2023.107204},
url = {https://www.sciencedirect.com/science/article/pii/S0001457523002518},
author = {Rui Fu and Wenxiao Liu and Hailun Zhang and Xue Liu and Wei Yuan},
}

@article{bluhm2023frontal,
title = {Effects of a frontal brake light on (automated) vehicles on children’s willingness to cross the road},
journal = {Transportation Research Part F: Traffic Psychology and Behaviour},
volume = {98},
pages = {269-279},
year = {2023},
issn = {1369-8478},
doi = {https://doi.org/10.1016/j.trf.2023.09.014},
url = {https://www.sciencedirect.com/science/article/pii/S1369847823001973},
author = {Luka-Franziska Bluhm and Daniel Eisele and Wolfgang Schubert and Rainer Banse},
}

@inproceedings{tran2023scoping,
author = {Tran, Tram Thi Minh and Parker, Callum and Tomitsch, Martin},
title = {Scoping Out the Scalability Issues of Autonomous Vehicle-Pedestrian Interaction},
year = {2023},
isbn = {9798400701054},
publisher = {Association for Computing Machinery},
address = {New York, NY, USA},
url = {https://doi.org/10.1145/3580585.3607167},
doi = {10.1145/3580585.3607167},
booktitle = {Proceedings of the 15th International Conference on Automotive User Interfaces and Interactive Vehicular Applications},
pages = {167–177},
numpages = {11},
location = {Ingolstadt, Germany},
series = {AutomotiveUI '23}
}

@inproceedings{joshi2023mapping,
author = {Joshi, Swapna and Block, Avram and Schmitt, Paul},
title = {Autonomous Vehicle and External Road User Interfaces: Mapping of Standards Gaps and Opportunities},
year = {2023},
isbn = {9798400701122},
publisher = {Association for Computing Machinery},
address = {New York, NY, USA},
url = {https://doi.org/10.1145/3581961.3609898},
doi = {10.1145/3581961.3609898},
booktitle = {Adjunct Proceedings of the 15th International Conference on Automotive User Interfaces and Interactive Vehicular Applications},
pages = {30–35},
numpages = {6},
location = {Ingolstadt, Germany},
series = {AutomotiveUI '23 Adjunct}
}

@inproceedings{dong2023holistic,
author = {Dong, Haoyu and Tran, Tram Thi Minh and Bazilinskyy, Pavlo and Hoggenm\"{u}ller, Marius and Dey, Debargha and Cazacu, Silvia and Franssen, Mervyn and Gao, Ruolin},
title = {Holistic HMI Design for Automated Vehicles: Bridging In-Vehicle and External Communication},
year = {2023},
isbn = {9798400701122},
publisher = {Association for Computing Machinery},
address = {New York, NY, USA},
url = {https://doi.org/10.1145/3581961.3609837},
doi = {10.1145/3581961.3609837},
booktitle = {Adjunct Proceedings of the 15th International Conference on Automotive User Interfaces and Interactive Vehicular Applications},
pages = {237–240},
numpages = {4},
location = {Ingolstadt, Germany},
series = {AutomotiveUI '23 Adjunct}
}

@inproceedings{winkelmann2023node,
author = {Winkelmann, Sven and B\"{u}ttner, Max and Deivasihamani, Dharani and von Hoffmann, Alexander and Flohr, Fabian},
title = {Using Node-RED as a Low-Code Approach to Model Interaction Logic of Machine-Learning-Supported eHMIs for the Virtual Driving Simulator Carla},
year = {2023},
isbn = {9798400701122},
publisher = {Association for Computing Machinery},
address = {New York, NY, USA},
url = {https://doi.org/10.1145/3581961.3609844},
doi = {10.1145/3581961.3609844},
booktitle = {Adjunct Proceedings of the 15th International Conference on Automotive User Interfaces and Interactive Vehicular Applications},
pages = {323–326},
numpages = {4},
location = {Ingolstadt, Germany},
series = {AutomotiveUI '23 Adjunct}
}

@inproceedings{lee2023IntVRsection,
author = {Lee, Jeannie Su Ann and Yeo, Adriel and Kwok, Benjamin Wei Jie and Wong, Zi Feng and Poh, Emran and Chan, Raymond},
title = {IntVRsection: Virtual Reality Environment for Evaluating Signalized and Unsignalized Intersection Scenarios},
year = {2023},
isbn = {9798400701122},
publisher = {Association for Computing Machinery},
address = {New York, NY, USA},
url = {https://doi.org/10.1145/3581961.3610373},
doi = {10.1145/3581961.3610373},
booktitle = {Adjunct Proceedings of the 15th International Conference on Automotive User Interfaces and Interactive Vehicular Applications},
pages = {300–303},
numpages = {4},
location = {Ingolstadt, Germany},
series = {AutomotiveUI '23 Adjunct}
}

@Article{dongas2023virtual,
AUTHOR = {Dongas, Robert and Grace, Kazjon and Gillespie, Samuel and Hoggenmueller, Marius and Tomitsch, Martin and Worrall, Stewart},
TITLE = {Virtual Urban Field Studies: Evaluating Urban Interaction Design Using Context-Based Interface Prototypes},
JOURNAL = {Multimodal Technologies and Interaction},
VOLUME = {7},
YEAR = {2023},
NUMBER = {8},
numpages = {27},
ARTICLE-NUMBER = {82},
URL = {https://www.mdpi.com/2414-4088/7/8/82},
ISSN = {2414-4088},
DOI = {10.3390/mti7080082}
}

@article{colley2023scalability,
author = {Colley, Mark and Britten, Julian and Rukzio, Enrico},
title = {Scalability in External Communication of Automated Vehicles: Evaluation and Recommendations},
year = {2023},
issue_date = {June 2023},
publisher = {Association for Computing Machinery},
address = {New York, NY, USA},
volume = {7},
number = {2},
url = {https://doi.org/10.1145/3596248},
doi = {10.1145/3596248},
journal = {Proc. ACM Interact. Mob. Wearable Ubiquitous Technol.},
month = jun,
articleno = {51},
numpages = {26}
}

@inproceedings{altaie2023keep,
author = {Al-Taie, Ammar and Abdrabou, Yasmeen and Macdonald, Shaun Alexander and Pollick, Frank and Brewster, Stephen Anthony},
title = {Keep it Real: Investigating Driver-Cyclist Interaction in Real-World Traffic},
year = {2023},
isbn = {9781450394215},
publisher = {Association for Computing Machinery},
address = {New York, NY, USA},
url = {https://doi.org/10.1145/3544548.3581049},
doi = {10.1145/3544548.3581049},
booktitle = {Proceedings of the 2023 CHI Conference on Human Factors in Computing Systems},
articleno = {769},
numpages = {15},
location = {Hamburg, Germany},
series = {CHI '23}
}

@inproceedings{lanzer2023interaction,
author = {Lanzer, Mirjam and Koniakowsky, Ina and Colley, Mark and Baumann, Martin},
title = {Interaction Effects of Pedestrian Behavior, Smartphone Distraction and External Communication of Automated Vehicles on Crossing and Gaze Behavior},
year = {2023},
isbn = {9781450394215},
publisher = {Association for Computing Machinery},
address = {New York, NY, USA},
url = {https://doi.org/10.1145/3544548.3581303},
doi = {10.1145/3544548.3581303},
booktitle = {Proceedings of the 2023 CHI Conference on Human Factors in Computing Systems},
articleno = {768},
numpages = {18},
location = {Hamburg, Germany},
series = {CHI '23}
}

@Article{lim2023visibility,
AUTHOR = {Lim, Dokshin and Kwon, Yongwhee},
TITLE = {How to Design the eHMI of AVs for Urgent Warning to Other Drivers with Limited Visibility?},
JOURNAL = {Sensors},
VOLUME = {23},
YEAR = {2023},
NUMBER = {7},
numpages = {17},
ARTICLE-NUMBER = {3721},
URL = {https://www.mdpi.com/1424-8220/23/7/3721},
PubMedID = {37050781},
ISSN = {1424-8220},
DOI = {10.3390/s23073721}
}

@inproceedings{yu2023way,
author = {Yu, Xinyan and Hoggenm\"{u}ller, Marius and Tomitsch, Martin},
title = {Your Way Or My Way: Improving Human-Robot Co-Navigation Through Robot Intent and Pedestrian Prediction Visualisations},
year = {2023},
isbn = {9781450399647},
publisher = {Association for Computing Machinery},
address = {New York, NY, USA},
url = {https://doi.org/10.1145/3568162.3576992},
doi = {10.1145/3568162.3576992},
booktitle = {Proceedings of the 2023 ACM/IEEE International Conference on Human-Robot Interaction},
pages = {211–221},
numpages = {11},
location = {Stockholm, Sweden},
series = {HRI '23}
}

@inproceedings{lee2023safe,
author = {Lee, Seong Hee and Patil, Vaidehi and Britten, Nicholas and Block, Avram and Pandya, Aryaman and Jung, Malte F. and Schmitt, Paul},
title = {Safe to Approach: Insights on Autonomous Vehicle Interaction Protocols with First Responders},
year = {2023},
isbn = {9781450399708},
publisher = {Association for Computing Machinery},
address = {New York, NY, USA},
url = {https://doi.org/10.1145/3568294.3580114},
doi = {10.1145/3568294.3580114},
booktitle = {Companion of the 2023 ACM/IEEE International Conference on Human-Robot Interaction},
pages = {399–402},
numpages = {4},
location = {Stockholm, Sweden},
series = {HRI '23}
}

@Article{rouchitsas2023smiles,
AUTHOR = {Rouchitsas, Alexandros and Alm, Håkan},
TITLE = {Smiles and Angry Faces vs. Nods and Head Shakes: Facial Expressions at the Service of Autonomous Vehicles},
JOURNAL = {Multimodal Technologies and Interaction},
VOLUME = {7},
YEAR = {2023},
NUMBER = {2},
numpages = {20},
ARTICLE-NUMBER = {10},
URL = {https://www.mdpi.com/2414-4088/7/2/10},
ISSN = {2414-4088},
DOI = {10.3390/mti7020010}
}

@inproceedings{zheng2023literature,
author = {Zheng, Yahua and Wu, Kangrui and Shi, Ruisi and Zhu, Xiaopeng and Zhang, Jingyu},
title = {A Literature Review of Current Practices to Evaluate the Usability of External Human Machine Interface},
year = {2023},
isbn = {978-3-031-35388-8},
publisher = {Springer-Verlag},
address = {Berlin, Heidelberg},
url = {https://doi.org/10.1007/978-3-031-35389-5_40},
doi = {10.1007/978-3-031-35389-5_40},
booktitle = {Engineering Psychology and Cognitive Ergonomics: 20th International Conference, EPCE 2023, Held as Part of the 25th HCI International Conference, HCII 2023, Copenhagen, Denmark, July 23–28, 2023, Proceedings, Part II},
pages = {573–586},
numpages = {14},
location = {Copenhagen, Denmark}
}

@inproceedings{serrano2023digital,
  author    = {Serrano, S. Mart{\'\i}n and Izquierdo, R. and Daza, I. Garc{\'\i}a and Sotelo, M. A. and Llorca, D. Fern{\'a}ndez},
  title     = {Digital twin in virtual reality for human-vehicle interactions in the context of autonomous driving},
  booktitle = {Proceedings of the 2023 IEEE 26th International Conference on Intelligent Transportation Systems (ITSC)},
  year      = {2023},
  pages     = {590--595},
  publisher = {IEEE},
  address   = {New York, NY, USA},
  doi       = {10.1109/ITSC57777.2023.10421914}
}

@inproceedings{shmueli2023ehmi,
  author    = {Shmueli, Y. and Degani, A.},
  title     = {eHMI Design: Theoretical Foundations and Methodological Process},
  booktitle = {Proceedings of the 18th International Joint Conference on Computer Vision, Imaging and Computer Graphics Theory and Applications (VISIGRAPP 2023) -- HUCAPP},
  year      = {2023},
  pages     = {201--212},
  publisher = {SciTePress},
  organization = {INSTICC},
  address   = {Set{\'u}bal, Portugal},
  doi       = {10.5220/0011686600003417},
  isbn      = {978-989-758-634-7},
  issn      = {2184-4321}
}

@inproceedings{sun2023onroad,
  author    = {Sun, Xiaodong and Shah, Ali Hassan and Ao, Jinlong and Miao, Wenqing and Lin, Yandan and Chen, Wenfang},
  title     = {On-road projection symbols for future vehicles: A survey study for Chinese roads},
  booktitle = {Proceedings of the 2022 19th China International Forum on Solid State Lighting and the 2022 8th International Forum on Wide Bandgap Semiconductors},
  year      = {2023},
  pages     = {231--240},
  publisher = {IEEE},
  address   = {New York, NY, USA},
  doi       = {10.1109/SSLChinaIFWS57942.2023.10071038}
}

@INPROCEEDINGS{yeo2023cycling,
  author={Yeo, Adriel and Kwok, Benjamin W.J. and Wong, Zi-Feng and Koh, Guo-Xun and Tan, Ryan W. and Yamin, Krystal and Liew, Yeni and Yan, Derwin H. and Lee, Jeannie S.},
  booktitle={2023 IEEE International Conference on Service Operations and Logistics, and Informatics (SOLI)}, 
  title={Cycling Simulation in Virtual Reality for Autonomous Vehicle Traffic Scenarios}, 
  year={2023},
  volume={},
  number={},
  pages={1-6},
publisher = {IEEE},
  address   = {New York, NY, USA},
  doi={10.1109/SOLI60636.2023.10425141}}

@article{Brown2023designing,
author = {Brown, Barry and Laurier, Eric and Vinkhuyzen, Erik},
title = {Designing Motion: Lessons for Self-driving and Robotic Motion from Human Traffic Interaction},
year = {2022},
issue_date = {January 2023},
publisher = {Association for Computing Machinery},
address = {New York, NY, USA},
volume = {7},
number = {GROUP},
url = {https://doi.org/10.1145/3567555},
doi = {10.1145/3567555},
journal = {Proc. ACM Hum.-Comput. Interact.},
month = dec,
articleno = {5},
numpages = {21}
}

@inproceedings{locken2022accessible,
author = {L\"{o}cken, Andreas and Matviienko, Andrii and Colley, Mark and Dey, Debargha and Habibovic, Azra and Lee, Yee Mun and Riener, Andreas},
title = {Accessible Automated Automotive Workshop Series (A3WS): International Perspective on Inclusive External Human-Machine Interfaces},
year = {2022},
isbn = {9781450394284},
publisher = {Association for Computing Machinery},
address = {New York, NY, USA},
url = {https://doi.org/10.1145/3544999.3551347},
doi = {10.1145/3544999.3551347},
booktitle = {Adjunct Proceedings of the 14th International Conference on Automotive User Interfaces and Interactive Vehicular Applications},
pages = {192–195},
numpages = {4},
location = {Seoul, Republic of Korea},
series = {AutomotiveUI '22}
}

@article{kaleefathullah2022misleading,
  title={External Human--Machine Interfaces Can Be Misleading: An Examination of Trust Development and Misuse in a CAVE-Based Pedestrian Simulation Environment},
  author={Kaleefathullah, A. A. and Merat, Natasha and Lee, Y. M. and Eisma, Y. B. and Madigan, Ruth and Garcia, J. and de Winter, Joost},
  journal={Human Factors: The Journal of the Human Factors and Ergonomics Society},
  volume={64},
  number={6},
  pages={1070--1085},
  year={2022},
  doi={10.1177/0018720820970751},
  publisher={SAGE Publications},
  address={Los Angeles, CA, USA}
}

@article{loew2022goahead,
  author  = {Loew, Alexandra and Graefe, Julia and Heil, Lukas and Guthardt, Anne and Boos, Annika and Dietrich, Andr{\'e} and Bengler, Klaus},
  title   = {Go Ahead, Please!—Evaluation of External Human--Machine Interfaces in a Real-World Crossing Scenario},
  journal = {Frontiers in Computer Science},
  volume  = {4},
  numpages ={14},
  year    = {2022},
  doi     = {10.3389/fcomp.2022.863072},
  url     = {https://www.frontiersin.org/articles/10.3389/fcomp.2022.863072}
}

@inproceedings{colley2022time,
author = {Colley, Mark and Bajrovic, Elvedin and Rukzio, Enrico},
title = {Effects of Pedestrian Behavior, Time Pressure, and Repeated Exposure on Crossing Decisions in Front of Automated Vehicles Equipped with External Communication},
year = {2022},
isbn = {9781450391573},
publisher = {Association for Computing Machinery},
address = {New York, NY, USA},
url = {https://doi.org/10.1145/3491102.3517571},
doi = {10.1145/3491102.3517571},
booktitle = {Proceedings of the 2022 CHI Conference on Human Factors in Computing Systems},
articleno = {367},
numpages = {11},
location = {New Orleans, LA, USA},
series = {CHI '22}
}

@ARTICLE{hollander2022take,
    
AUTHOR={Holländer, Kai  and Hoggenmüller, Marius  and Gruber, Romy  and Völkel, Sarah Theres  and Butz, Andreas },
           
TITLE={Take It to the Curb: Scalable Communication Between Autonomous Cars and Vulnerable Road Users Through Curbstone Displays},
          
JOURNAL={Frontiers in Computer Science},
          
VOLUME={Volume 4 - 2022},
  
YEAR={2022},
  
URL={https://www.frontiersin.org/journals/computer-science/articles/10.3389/fcomp.2022.844245},
  
DOI={10.3389/fcomp.2022.844245},
  
ISSN={2624-9898},
numpages = {18}}

@article{singer2022display,
author = {Singer, Timo and Kobbert, Jonas and Zandi, Babak and Khanh, Tran Quoc},
title = {Displaying the Driving State of Automated Vehicles to Other Road Users: An International, Virtual Reality-Based Study as a First Step for the Harmonized Regulations of Novel Signaling Devices},
year = {2022},
issue_date = {April 2022},
publisher = {IEEE Press},
volume = {23},
number = {4},
issn = {1524-9050},
url = {https://doi.org/10.1109/TITS.2020.3032777},
doi = {10.1109/TITS.2020.3032777},
journal = {Trans. Intell. Transport. Sys.},
month = apr,
pages = {2904–2918},
numpages = {15}
}

@inproceedings{vonsawitzky2022hazard,
author = {von Sawitzky, Tamara and Grauschopf, Thomas and Riener, Andreas},
title = {Hazard Notifications for Cyclists: Comparison of Awareness Message Modalities in a Mixed Reality Study},
year = {2022},
isbn = {9781450391443},
publisher = {Association for Computing Machinery},
address = {New York, NY, USA},
url = {https://doi.org/10.1145/3490099.3511127},
doi = {10.1145/3490099.3511127},
booktitle = {Proceedings of the 27th International Conference on Intelligent User Interfaces},
pages = {310–322},
numpages = {13},
location = {Helsinki, Finland},
series = {IUI '22}
}

@article{berge2022cyclist,
title = {Do cyclists need HMIs in future automated traffic? An interview study},
journal = {Transportation Research Part F: Traffic Psychology and Behaviour},
volume = {84},
pages = {33-52},
year = {2022},
issn = {1369-8478},
doi = {https://doi.org/10.1016/j.trf.2021.11.013},
url = {https://www.sciencedirect.com/science/article/pii/S1369847821002631},
author = {Siri Hegna Berge and Marjan Hagenzieker and Haneen Farah and Joost {de Winter}},
}

@article{colley2022truck,
  author  = {Colley, Mark and Mytilineos, Stefanos and Walch, Marcel and Gugenheimer, Jan and Rukzio, Enrico},
  title   = {Requirements for the Interaction with Highly Automated Construction Site Delivery Trucks},
  journal = {Frontiers in Human Dynamics},
  volume  = {4},
  numpages = {11},
  year    = {2022},
  doi     = {10.3389/fhumd.2022.794890},
  url     = {https://www.frontiersin.org/articles/10.3389/fhumd.2022.794890}
}

@inproceedings{hensch2022effects,
author="Hensch, Ann-Christin
and Krei{\ss}ig, Isabel
and Beggiato, Matthias
and Krems, Josef F.",
editor="Ahram, Tareq
and Taiar, Redha",
title="The Effects of eHMI Failures on Elderly Participants' Assessment of Automated Vehicle Communication Signals",
booktitle="Human Interaction, Emerging Technologies and Future Systems V",
year="2022",
publisher="Springer International Publishing",
address="Cham",
pages="355--363",
isbn="978-3-030-85540-6"
}

@article{orlicky2021assessment,
  title={Assessment of External Interface of Autonomous Vehicles},
  author={Orlick{\'y}, Adam and Mashko, Alina and M{\'\i}k, Josef},
  journal={Acta Polytechnica},
  volume={61},
  number={6},
  pages={733--739},
  year={2021},
  doi={10.14311/AP.2021.61.0733}
}

@Article{riener2021shuttle,
AUTHOR = {Riener, Andreas and Schlackl, Dominik and Malsam, Julia and Huber, Josef and Homm, Benjamin and Kaczmar, Marion and Kleitsch, Iris and Megos, Alina and Park, Eunji and Sanverdi, Gülsüm and Schmidt, Sabrina and Bracaci, Daniel and Anees, Esha},
TITLE = {Improving the UX for Users of Automated Shuttle Buses in Public Transport: Investigating Aspects of Exterior Communication and Interior Design},
JOURNAL = {Multimodal Technologies and Interaction},
VOLUME = {5},
numpages = {34},
YEAR = {2021},
NUMBER = {10},
ARTICLE-NUMBER = {61},
URL = {https://www.mdpi.com/2414-4088/5/10/61},
ISSN = {2414-4088},
DOI = {10.3390/mti5100061}
}

@inproceedings{sahin2021prosocial,
author = {Sahin, Hatice and Mueller, Heiko and Sadeghian, Shadan and Dey, Debargha and L\"{o}cken, Andreas and Matviienko, Andrii and Colley, Mark and Habibovic, Azra and Wintersberger, Philipp},
title = {Workshop on Prosocial Behavior in Future Mixed Traffic},
year = {2021},
isbn = {9781450386418},
publisher = {Association for Computing Machinery},
address = {New York, NY, USA},
url = {https://doi.org/10.1145/3473682.3477438},
doi = {10.1145/3473682.3477438},
booktitle = {13th International Conference on Automotive User Interfaces and Interactive Vehicular Applications},
pages = {167–170},
numpages = {4},
location = {Leeds, United Kingdom},
series = {AutomotiveUI '21 Adjunct}
}

@inproceedings{lau2021investigating,
author = {Lau, Merle and Jipp, Meike and Oehl, Michael},
title = {Investigating the Interplay between eHMI and dHMI for Automated Buses: How Do Contradictory Signals Influence a Pedestrian's Willingness to Cross?},
year = {2021},
isbn = {9781450386418},
publisher = {Association for Computing Machinery},
address = {New York, NY, USA},
url = {https://doi.org/10.1145/3473682.3480284},
doi = {10.1145/3473682.3480284},
booktitle = {13th International Conference on Automotive User Interfaces and Interactive Vehicular Applications},
pages = {152–155},
numpages = {4},
location = {Leeds, United Kingdom},
series = {AutomotiveUI '21 Adjunct}
}

@inproceedings{tabone2021towards,
author = {Tabone, Wilbert and Lee, Yee Mun and Merat, Natasha and Happee, Riender and de Winter, Joost},
title = {Towards future pedestrian-vehicle interactions: Introducing theoretically-supported AR prototypes},
year = {2021},
isbn = {9781450380638},
publisher = {Association for Computing Machinery},
address = {New York, NY, USA},
url = {https://doi.org/10.1145/3409118.3475149},
doi = {10.1145/3409118.3475149},
booktitle = {13th International Conference on Automotive User Interfaces and Interactive Vehicular Applications},
pages = {209–218},
numpages = {10},
location = {Leeds, United Kingdom},
series = {AutomotiveUI '21}
}

@inproceedings{mirnig2021stop,
author = {Mirnig, Alexander G. and G\"{a}rtner, Magdalena and Wallner, Vivien and Gafert, Michael and Braun, Hanna and Fr\"{o}hlich, Peter and Suette, Stefan and Sypniewski, Jakub and Meschtscherjakov, Alexander and Tscheligi, Manfred},
title = {Stop or Go? Let me Know! A Field Study on Visual External Communication for Automated Shuttles},
year = {2021},
isbn = {9781450380638},
publisher = {Association for Computing Machinery},
address = {New York, NY, USA},
url = {https://doi.org/10.1145/3409118.3475131},
doi = {10.1145/3409118.3475131},
booktitle = {13th International Conference on Automotive User Interfaces and Interactive Vehicular Applications},
pages = {287–295},
numpages = {9},
location = {Leeds, United Kingdom},
series = {AutomotiveUI '21}
}

@inproceedings{drechsler2021mixed,
author = {Funk Drechsler, Maikol and Peintner, Jakob Benedikt and Seifert, Georg and Huber, Werner and Riener, Andreas},
title = {Mixed Reality Environment for Testing Automated Vehicle and Pedestrian Interaction},
year = {2021},
isbn = {9781450386418},
publisher = {Association for Computing Machinery},
address = {New York, NY, USA},
url = {https://doi.org/10.1145/3473682.3481878},
doi = {10.1145/3473682.3481878},
booktitle = {13th International Conference on Automotive User Interfaces and Interactive Vehicular Applications},
pages = {229–232},
numpages = {4},
location = {Leeds, United Kingdom},
series = {AutomotiveUI '21 Adjunct}
}

@inproceedings{dey2021flow,
author = {Dey, Debargha and Temmink, Brent and Sonnemans, Daan and Den Teuling, Karijn and van Berkel, Lotte and Pfleging, Bastian},
title = {FlowMotion: Exploring the Intuitiveness of Fluid Motion Based Communication in eHMI Design for Vehicle-Pedestrian Communication},
year = {2021},
isbn = {9781450386418},
publisher = {Association for Computing Machinery},
address = {New York, NY, USA},
url = {https://doi.org/10.1145/3473682.3480279},
doi = {10.1145/3473682.3480279},
booktitle = {13th International Conference on Automotive User Interfaces and Interactive Vehicular Applications},
pages = {128–131},
numpages = {4},
location = {Leeds, United Kingdom},
series = {AutomotiveUI '21 Adjunct}
}

@article{hu2021review,
author = {Hu, Lin and Zhou, Xiqin and Zhang, Xin and Wang, Fang and Li, Qiqi and Wu, Wenguang},
title = {A review on key challenges in intelligent vehicles: Safety and driver-oriented features},
journal = {IET Intelligent Transport Systems},
volume = {15},
number = {9},
pages = {1093-1105},
doi = {https://doi.org/10.1049/itr2.12088},
url = {https://ietresearch.onlinelibrary.wiley.com/doi/abs/10.1049/itr2.12088},
year = {2021}
}

@INPROCEEDINGS{liu2021instruction,
  author={Liu, Hailong and Hirayama, Takatsugu and Watanabe, Masaya},
  booktitle={2021 IEEE Intelligent Vehicles Symposium (IV)}, 
  title={Importance of Instruction for Pedestrian-Automated Driving Vehicle Interaction with an External Human Machine Interface: Effects on Pedestrians' Situation Awareness, Trust, Perceived Risks and Decision Making}, 
  year={2021},
  volume={},
  number={},
  pages={748-754},
  doi={10.1109/IV48863.2021.9575246},
publisher = {IEEE},
address   = {Piscataway, NJ, USA}}

@article{oudshoorn2021bio,
title = {Bio-inspired intent communication for automated vehicles},
journal = {Transportation Research Part F: Traffic Psychology and Behaviour},
volume = {80},
pages = {127-140},
year = {2021},
issn = {1369-8478},
doi = {https://doi.org/10.1016/j.trf.2021.03.021},
url = {https://www.sciencedirect.com/science/article/pii/S1369847821000759},
author = {Max Oudshoorn and Joost {de Winter} and Pavlo Bazilinskyy and Dimitra Dodou},
}

@inproceedings{lee2021discovering,
author = {Lee, Jaemyung and park, wonyoung and Lee, Sangsu},
title = {Discovering the Design Challenges of Autonomous Vehicles through Exploring Scenarios via an Immersive Design Workshop},
year = {2021},
isbn = {9781450384766},
publisher = {Association for Computing Machinery},
address = {New York, NY, USA},
url = {https://doi.org/10.1145/3461778.3462119},
doi = {10.1145/3461778.3462119},
booktitle = {Proceedings of the 2021 ACM Designing Interactive Systems Conference},
pages = {322–338},
numpages = {17},
location = {Virtual Event, USA},
series = {DIS '21}
}

@inproceedings{asha2021wheelchairs,
author = {Asha, Ashratuz Zavin and Smith, Christopher and Freeman, Georgina and Crump, Sean and Somanath, Sowmya and Oehlberg, Lora and Sharlin, Ehud},
title = {Co-Designing Interactions between Pedestrians in Wheelchairs and Autonomous Vehicles},
year = {2021},
isbn = {9781450384766},
publisher = {Association for Computing Machinery},
address = {New York, NY, USA},
url = {https://doi.org/10.1145/3461778.3462068},
doi = {10.1145/3461778.3462068},
booktitle = {Proceedings of the 2021 ACM Designing Interactive Systems Conference},
pages = {339–351},
numpages = {13},
location = {Virtual Event, USA},
series = {DIS '21}
}

@INPROCEEDINGS{orlicky2021microsimulation,
  author={Orlický, Adam and Mashko, Alina and Mík, Josef},
  booktitle={2021 Smart City Symposium Prague (SCSP)}, 
  title={Microsimulation model for assessment of eHMI of autonomous vehicles}, 
  year={2021},
  volume={},
  number={},
  pages={1-5},
  doi={10.1109/SCSP52043.2021.9447377},
publisher = {IEEE},
address   = {Piscataway, NJ, USA}
}

@inproceedings{hoggenmueller2021context,
author = {Hoggenm\"{u}ller, Marius and Tomitsch, Martin and Hespanhol, Luke and Tran, Tram Thi Minh and Worrall, Stewart and Nebot, Eduardo},
title = {Context-Based Interface Prototyping: Understanding the Effect of Prototype Representation on User Feedback},
year = {2021},
isbn = {9781450380966},
publisher = {Association for Computing Machinery},
address = {New York, NY, USA},
url = {https://doi.org/10.1145/3411764.3445159},
doi = {10.1145/3411764.3445159},
booktitle = {Proceedings of the 2021 CHI Conference on Human Factors in Computing Systems},
articleno = {370},
numpages = {14},
location = {Yokohama, Japan},
series = {CHI '21}
}

@inproceedings{faas2021calibrating,
author = {M. Faas, Stefanie and Kraus, Johannes and Schoenhals, Alexander and Baumann, Martin},
title = {Calibrating Pedestrians' Trust in Automated Vehicles: Does an Intent Display in an External HMI Support Trust Calibration and Safe Crossing Behavior?},
year = {2021},
isbn = {9781450380966},
publisher = {Association for Computing Machinery},
address = {New York, NY, USA},
url = {https://doi.org/10.1145/3411764.3445738},
doi = {10.1145/3411764.3445738},
booktitle = {Proceedings of the 2021 CHI Conference on Human Factors in Computing Systems},
articleno = {157},
numpages = {17},
location = {Yokohama, Japan},
series = {CHI '21}
}

@article{lee2021road,
author = {Lee, Yee Mun and Madigan, Ruth and Giles, Oscar and Garach-Morcillo, Laura and Markkula, Gustav and Fox, Charles and Camara, Fanta and Rothmueller, Markus and Vendelbo-Larsen, Signe Alexandra and Rasmussen, Pernille Holm and Dietrich, Andre and Nathanael, Dimitris and Portouli, Villy and Schieben, Anna and Merat, Natasha},
title = {Road users rarely use explicit communication when interacting in today’s traffic: implications for automated vehicles},
year = {2021},
issue_date = {May 2021},
publisher = {Springer-Verlag},
address = {Berlin, Heidelberg},
volume = {23},
number = {2},
issn = {1435-5558},
url = {https://doi.org/10.1007/s10111-020-00635-y},
doi = {10.1007/s10111-020-00635-y},
journal = {Cogn. Technol. Work},
month = may,
pages = {367–380},
numpages = {14}
}

@article{eisma2021perspective,
title = {External human-machine interfaces: Effects of message perspective},
journal = {Transportation Research Part F: Traffic Psychology and Behaviour},
volume = {78},
pages = {30-41},
year = {2021},
issn = {1369-8478},
doi = {https://doi.org/10.1016/j.trf.2021.01.013},
url = {https://www.sciencedirect.com/science/article/pii/S1369847821000206},
author = {Y.B. Eisma and A. Reiff and L. Kooijman and D. Dodou and J.C.F. {de Winter}},
}

@ARTICLE{tabone2021vulnerable,
title = {Vulnerable road users and the coming wave of automated vehicles: Expert perspectives},
journal = {Transportation Research Interdisciplinary Perspectives},
volume = {9},
pages = {100293},
year = {2021},
issn = {2590-1982},
doi = {https://doi.org/10.1016/j.trip.2020.100293},
url = {https://www.sciencedirect.com/science/article/pii/S2590198220302049},
author = {Wilbert Tabone and Joost {de Winter} and Claudia Ackermann and Jonas Bärgman and Martin Baumann and Shuchisnigdha Deb and Colleen Emmenegger and Azra Habibovic and Marjan Hagenzieker and P.A. Hancock and Riender Happee and Josef Krems and John D. Lee and Marieke Martens and Natasha Merat and Don Norman and Thomas B. Sheridan and Neville A. Stanton},
}

@InProceedings{lim2021ux,
author="Lim, Dokshin
and Hwangbo, Hwan",
editor="Ahram, Tareq Z.
and Falc{\~a}o, Christianne S.",
title="UX Design for Holistic User Journey of Future Robotaxi",
booktitle="Advances in Usability, User Experience, Wearable and Assistive Technology",
year="2021",
publisher="Springer International Publishing",
address="Cham",
pages="976--984",
isbn="978-3-030-80091-8"
}

@article{schrauth2021acceptance,
  title   = {The acceptance of conditionally automated cars from the perspective of different road user groups},
  author  = {Schrauth, Bernhard and Funk, Walter and Maier, Sarah and Kraetsch, Clemens},
  journal = {European Journal of Transport and Infrastructure Research},
  volume  = {21},
  number  = {4},
  pages   = {81--103},
  year    = {2021},
  doi     = {10.18757/ejtir.2021.21.4.5466},
  url     = {https://journals.open.tudelft.nl/ejtir/article/view/5466}
}

@article{deWinter2021visualattention,
author = {Joost de Winter and Pavlo Bazilinskyy and Dale Wesdorp and Valerie de Vlam and Belle Hopmans and Just Visscher and Dimitra Dodou},
title = {How do pedestrians distribute their visual attention when walking through a parking garage? An eye-tracking study},
journal = {Ergonomics},
volume = {64},
number = {6},
pages = {793--805},
year = {2021},
publisher = {Taylor \& Francis},
doi = {10.1080/00140139.2020.1862310},
}

@inproceedings{hoggenmuller2020tangible,
author = {Hoggenm\"{u}ller, Marius and Tomitsch, Martin and Parker, Callum and Nguyen, Trung Thanh and Zhou, Dawei and Worrall, Stewart and Nebot, Eduardo},
title = {A Tangible Multi-Display Toolkit to Support the Collaborative Design Exploration of AV-Pedestrian Interfaces},
year = {2021},
isbn = {9781450389754},
publisher = {Association for Computing Machinery},
address = {New York, NY, USA},
url = {https://doi.org/10.1145/3441000.3441031},
doi = {10.1145/3441000.3441031},
booktitle = {Proceedings of the 32nd Australian Conference on Human-Computer Interaction},
pages = {25–35},
numpages = {11},
location = {Sydney, NSW, Australia},
series = {OzCHI '20}
}

@inproceedings{hollander2020save,
author = {Holl\"{a}nder, Kai and Kr\"{u}ger, Andy and Butz, Andreas},
title = {Save the Smombies: App-Assisted Street Crossing},
year = {2020},
isbn = {9781450375160},
publisher = {Association for Computing Machinery},
address = {New York, NY, USA},
url = {https://doi.org/10.1145/3379503.3403547},
doi = {10.1145/3379503.3403547},
booktitle = {22nd International Conference on Human-Computer Interaction with Mobile Devices and Services},
articleno = {22},
numpages = {11},
location = {Oldenburg, Germany},
series = {MobileHCI '20}
}

@inproceedings{chang2020gender,
author = {Chang, Chia-Ming},
title = {A Gender Study of Communication Interfaces between an Autonomous Car and a Pedestrian},
year = {2020},
isbn = {9781450380669},
publisher = {Association for Computing Machinery},
address = {New York, NY, USA},
url = {https://doi.org/10.1145/3409251.3411719},
doi = {10.1145/3409251.3411719},
booktitle = {12th International Conference on Automotive User Interfaces and Interactive Vehicular Applications},
pages = {42–45},
numpages = {4},
location = {Virtual Event, DC, USA},
series = {AutomotiveUI '20}
}

@inproceedings{colley2020designspace,
author = {Colley, Mark and Rukzio, Enrico},
title = {A Design Space for External Communication of Autonomous Vehicles},
year = {2020},
isbn = {9781450380652},
publisher = {Association for Computing Machinery},
address = {New York, NY, USA},
url = {https://doi.org/10.1145/3409120.3410646},
doi = {10.1145/3409120.3410646},
booktitle = {12th International Conference on Automotive User Interfaces and Interactive Vehicular Applications},
pages = {212–222},
numpages = {11},
location = {Virtual Event, DC, USA},
series = {AutomotiveUI '20}
}

@inproceedings{dalipi2020benchmark,
author = {Dalipi, Ana Fiona and Liu, Dongfang and Guo, Xiaolei and Chen, Yingjie and Mousas, Christos},
title = {VR-PAVIB: The Virtual Reality Pedestrian-Autonomous Vehicle Interaction Benchmark},
year = {2020},
isbn = {9781450380669},
publisher = {Association for Computing Machinery},
address = {New York, NY, USA},
url = {https://doi.org/10.1145/3409251.3411718},
doi = {10.1145/3409251.3411718},
booktitle = {12th International Conference on Automotive User Interfaces and Interactive Vehicular Applications},
pages = {38–41},
numpages = {4},
location = {Virtual Event, DC, USA},
series = {AutomotiveUI '20}
}

@Article{faas2020efficient,
AUTHOR = {Faas, Stefanie M. and Mattes, Stefan and Kao, Andrea C. and Baumann, Martin},
TITLE = {Efficient Paradigm to Measure Street-Crossing Onset Time of Pedestrians in Video-Based Interactions with Vehicles},
JOURNAL = {Information},
VOLUME = {11},
YEAR = {2020},
NUMBER = {7},
numpages = {21},
ARTICLE-NUMBER = {360},
URL = {https://www.mdpi.com/2078-2489/11/7/360},
ISSN = {2078-2489},
DOI = {10.3390/info11070360}
}

@inproceedings{locken2020wecare,
author = {L\"{o}cken, Andreas and Colley, Mark and Matviienko, Andrii and Holl\"{a}nder, Kai and Dey, Debargha and Habibovic, Azra and Kun, Andrew L and Boll, Susanne and Riener, Andreas},
title = {WeCARe: Workshop on Inclusive Communication between Automated Vehicles and Vulnerable Road Users},
year = {2021},
isbn = {9781450380522},
publisher = {Association for Computing Machinery},
address = {New York, NY, USA},
url = {https://doi.org/10.1145/3406324.3424587},
doi = {10.1145/3406324.3424587},
booktitle = {22nd International Conference on Human-Computer Interaction with Mobile Devices and Services},
articleno = {43},
numpages = {5},
location = {Oldenburg, Germany},
series = {MobileHCI '20}
}

@Article{feierle2020multi,
AUTHOR = {Feierle, Alexander and Rettenmaier, Michael and Zeitlmeir, Florian and Bengler, Klaus},
TITLE = {Multi-Vehicle Simulation in Urban Automated Driving: Technical Implementation and Added Benefit},
JOURNAL = {Information},
VOLUME = {11},
YEAR = {2020},
NUMBER = {5},
numpages = {21},
ARTICLE-NUMBER = {272},
URL = {https://www.mdpi.com/2078-2489/11/5/272},
ISSN = {2078-2489},
DOI = {10.3390/info11050272}
}

@inproceedings{asha2020wheelchair,
author = {Asha, Ashratuz Zavin and Smith, Christopher and Oehlberg, Lora and Somanath, Sowmya and Sharlin, Ehud},
title = {Views from the Wheelchair: Understanding Interaction between Autonomous Vehicle and Pedestrians with Reduced Mobility},
year = {2020},
isbn = {9781450368193},
publisher = {Association for Computing Machinery},
address = {New York, NY, USA},
url = {https://doi.org/10.1145/3334480.3383041},
doi = {10.1145/3334480.3383041},
booktitle = {Extended Abstracts of the 2020 CHI Conference on Human Factors in Computing Systems},
pages = {1–8},
numpages = {8},
location = {Honolulu, HI, USA},
series = {CHI EA '20}
}

@inproceedings{ackermans2020effects,
author = {Ackermans, Sander and Dey, Debargha and Ruijten, Peter and Cuijpers, Raymond H. and Pfleging, Bastian},
title = {The Effects of Explicit Intention Communication, Conspicuous Sensors, and Pedestrian Attitude in Interactions with Automated Vehicles},
year = {2020},
isbn = {9781450367080},
publisher = {Association for Computing Machinery},
address = {New York, NY, USA},
url = {https://doi.org/10.1145/3313831.3376197},
doi = {10.1145/3313831.3376197},
booktitle = {Proceedings of the 2020 CHI Conference on Human Factors in Computing Systems},
pages = {1–14},
numpages = {14},
location = {Honolulu, HI, USA},
series = {CHI '20}
}

@inproceedings{faas2020longitudinal,
author = {Faas, Stefanie M. and Kao, Andrea C. and Baumann, Martin},
title = {A Longitudinal Video Study on Communicating Status and Intent for Self-Driving Vehicle Pedestrian Interaction},
year = {2020},
isbn = {9781450367080},
publisher = {Association for Computing Machinery},
address = {New York, NY, USA},
url = {https://doi.org/10.1145/3313831.3376484},
doi = {10.1145/3313831.3376484},
booktitle = {Proceedings of the 2020 CHI Conference on Human Factors in Computing Systems},
pages = {1–14},
numpages = {14},
location = {Honolulu, HI, USA},
series = {CHI '20}
}

@inproceedings{colley2020inclusive,
author = {Colley, Mark and Walch, Marcel and Gugenheimer, Jan and Askari, Ali and Rukzio, Enrico},
title = {Towards Inclusive External Communication of Autonomous Vehicles for Pedestrians with Vision Impairments},
year = {2020},
isbn = {9781450367080},
publisher = {Association for Computing Machinery},
address = {New York, NY, USA},
url = {https://doi.org/10.1145/3313831.3376472},
doi = {10.1145/3313831.3376472},
booktitle = {Proceedings of the 2020 CHI Conference on Human Factors in Computing Systems},
pages = {1–14},
numpages = {14},
location = {Honolulu, HI, USA},
series = {CHI '20}
}

@inproceedings{dey2020color,
author = {Dey, Debargha and Habibovic, Azra and Pfleging, Bastian and Martens, Marieke and Terken, Jacques},
title = {Color and Animation Preferences for a Light Band eHMI in Interactions Between Automated Vehicles and Pedestrians},
year = {2020},
isbn = {9781450367080},
publisher = {Association for Computing Machinery},
address = {New York, NY, USA},
url = {https://doi.org/10.1145/3313831.3376325},
doi = {10.1145/3313831.3376325},
booktitle = {Proceedings of the 2020 CHI Conference on Human Factors in Computing Systems},
pages = {1–13},
numpages = {13},
location = {Honolulu, HI, USA},
series = {CHI '20}
}

@Article{kaß2020standardised,
AUTHOR = {Kaß, Christina and Schoch, Stefanie and Naujoks, Frederik and Hergeth, Sebastian and Keinath, Andreas and Neukum, Alexandra},
TITLE = {Standardized Test Procedure for External Human–Machine Interfaces of Automated Vehicles},
JOURNAL = {Information},
VOLUME = {11},
YEAR = {2020},
NUMBER = {3},
numpages = {19},
ARTICLE-NUMBER = {173},
URL = {https://www.mdpi.com/2078-2489/11/3/173},
ISSN = {2078-2489},
DOI = {10.3390/info11030173}
}

@Article{Bengler2020,
AUTHOR = {Bengler, Klaus and Rettenmaier, Michael and Fritz, Nicole and Feierle, Alexander},
TITLE = {From HMI to HMIs: Towards an HMI Framework for Automated Driving},
JOURNAL = {Information},
VOLUME = {11},
YEAR = {2020},
NUMBER = {2},
numpages = {17},
ARTICLE-NUMBER = {61},
URL = {https://www.mdpi.com/2078-2489/11/2/61},
ISSN = {2078-2489},
DOI = {10.3390/info11020061}
}

@article{hagenzieker2020interactions,
author = {Marjan P. Hagenzieker and Sander van der Kint and Luuk Vissers and Ingrid N. L. G van Schagen and Jonathan de Bruin and Paul van Gent and Jacques J. F. Commandeur},
title = {Interactions between cyclists and automated vehicles: Results of a photo experiment*},
journal = {Journal of Transportation Safety \& Security},
volume = {12},
number = {1},
pages = {94--115},
year = {2020},
publisher = {Taylor \& Francis},
doi = {10.1080/19439962.2019.1591556},
}

@inproceedings{kaß2020methodological,
author = {Ka\ss{}, Christina and Schoch, Stefanie and Naujoks, Frederik and Hergeth, Sebastian and Keinath, Andreas and Neukum, Alexandra},
title = {A Methodological Approach to Determine the Benefits of External HMI During Interactions Between Cyclists and Automated Vehicles: A Bicycle Simulator Study},
year = {2020},
isbn = {978-3-030-50536-3},
publisher = {Springer-Verlag},
address = {Berlin, Heidelberg},
url = {https://doi.org/10.1007/978-3-030-50537-0_16},
doi = {10.1007/978-3-030-50537-0_16},
booktitle = {HCI in Mobility, Transport, and Automotive Systems. Driving Behavior, Urban and Smart Mobility: Second International Conference, MobiTAS 2020, Held as Part of the 22nd HCI International Conference, HCII 2020, Copenhagen, Denmark, July 19–24, 2020, Proceedings, Part II},
pages = {211–227},
numpages = {17},
location = {Copenhagen, Denmark}
}

@ARTICLE{Leveque2020where,
  author={Lévêque, Lucie and Ranchet, Maud and Deniel, Jonathan and Bornard, Jean-Charles and Bellet, Thierry},
  journal={IEEE Access}, 
  title={Where Do Pedestrians Look When Crossing? A State of the Art of the Eye-Tracking Studies}, 
  year={2020},
  volume={8},
  number={},
  pages={164833-164843},
  doi={10.1109/ACCESS.2020.3021208},
publisher = {IEEE},
address   = {Piscataway, NJ, USA}}

@article{faas2020external,
title = {External HMI for self-driving vehicles: Which information shall be displayed?},
journal = {Transportation Research Part F: Traffic Psychology and Behaviour},
volume = {68},
pages = {171-186},
year = {2020},
issn = {1369-8478},
doi = {https://doi.org/10.1016/j.trf.2019.12.009},
url = {https://www.sciencedirect.com/science/article/pii/S1369847819305479},
author = {Stefanie M. Faas and Lesley-Ann Mathis and Martin Baumann},
}

@inproceedings{wesseling2020exploring,
author = {Wesseling, Anne-Marie Julie Barthe and Mugge, Ruth and van Grondelle, Elmer and Othersen, Ina},
title = {Exploring Universal and Cultural Preferences for Different Concepts of Autonomous Vehicles’ External Communication in China, USA and Germany},
year = {2020},
isbn = {978-3-030-49787-3},
publisher = {Springer-Verlag},
address = {Berlin, Heidelberg},
url = {https://doi.org/10.1007/978-3-030-49788-0_49},
doi = {10.1007/978-3-030-49788-0_49},
booktitle = {Cross-Cultural Design. User Experience of Products, Services, and Intelligent Environments: 12th International Conference, CCD 2020, Held as Part of the 22nd HCI International Conference, HCII 2020, Copenhagen, Denmark, July 19–24, 2020, Proceedings, Part I},
pages = {641–660},
numpages = {20},
location = {Copenhagen, Denmark}
}

@ARTICLE{domeyer2020vehicle,
  author={Domeyer, Joshua E. and Lee, John D. and Toyoda, Heishiro},
  journal={IEEE Access}, 
  title={Vehicle Automation–Other Road User Communication and Coordination: Theory and Mechanisms}, 
  year={2020},
  volume={8},
  number={},
  pages={19860-19872},
  doi={10.1109/ACCESS.2020.2969233}}

@Article{kooijman2019ehmis,
AUTHOR = {Kooijman, Lars and Happee, Riender and de Winter, Joost C. F.},
TITLE = {How Do eHMIs Affect Pedestrians’ Crossing Behavior? A Study Using a Head-Mounted Display Combined with a Motion Suit},
JOURNAL = {Information},
VOLUME = {10},
YEAR = {2019},
NUMBER = {12},
numpages = {18},
ARTICLE-NUMBER = {386},
URL = {https://www.mdpi.com/2078-2489/10/12/386},
ISSN = {2078-2489},
DOI = {10.3390/info10120386}
}

@incollection{yan2022inclusive,
  author    = {Yan, M. and Rampino, L. and Zhao, H. and Caruso, A.},
  title     = {Implications of Human-Machine Interface for Inclusive Shared Autonomous Vehicles},
  booktitle = {Human Factors in Transportation},
  editor    = {Plant, Katie and Praetorius, Gesa},
  series    = {AHFE Open Access},
  volume    = {60},
  year      = {2022},
  publisher = {AHFE International},
  address   = {USA},
  doi       = {10.54941/ahfe1002488}
}

\appendix

\section{Workflow and Open Data Resources}
\label{sec:workflow}

This appendix outlines the end-to-end workflow used to construct the dataset, chart study variables, and perform all analyses. The figure distils the process steps, tools, and verification measures to ensure methodological transparency. All underlying materials required for reproducibility, including screening logs, charted variables, and the supporting Python scripts, are available on OSF: \url{https://osf.io/7esmq}

\begin{figure*}[htbp]
    \centering
    \includegraphics[width=1\linewidth]{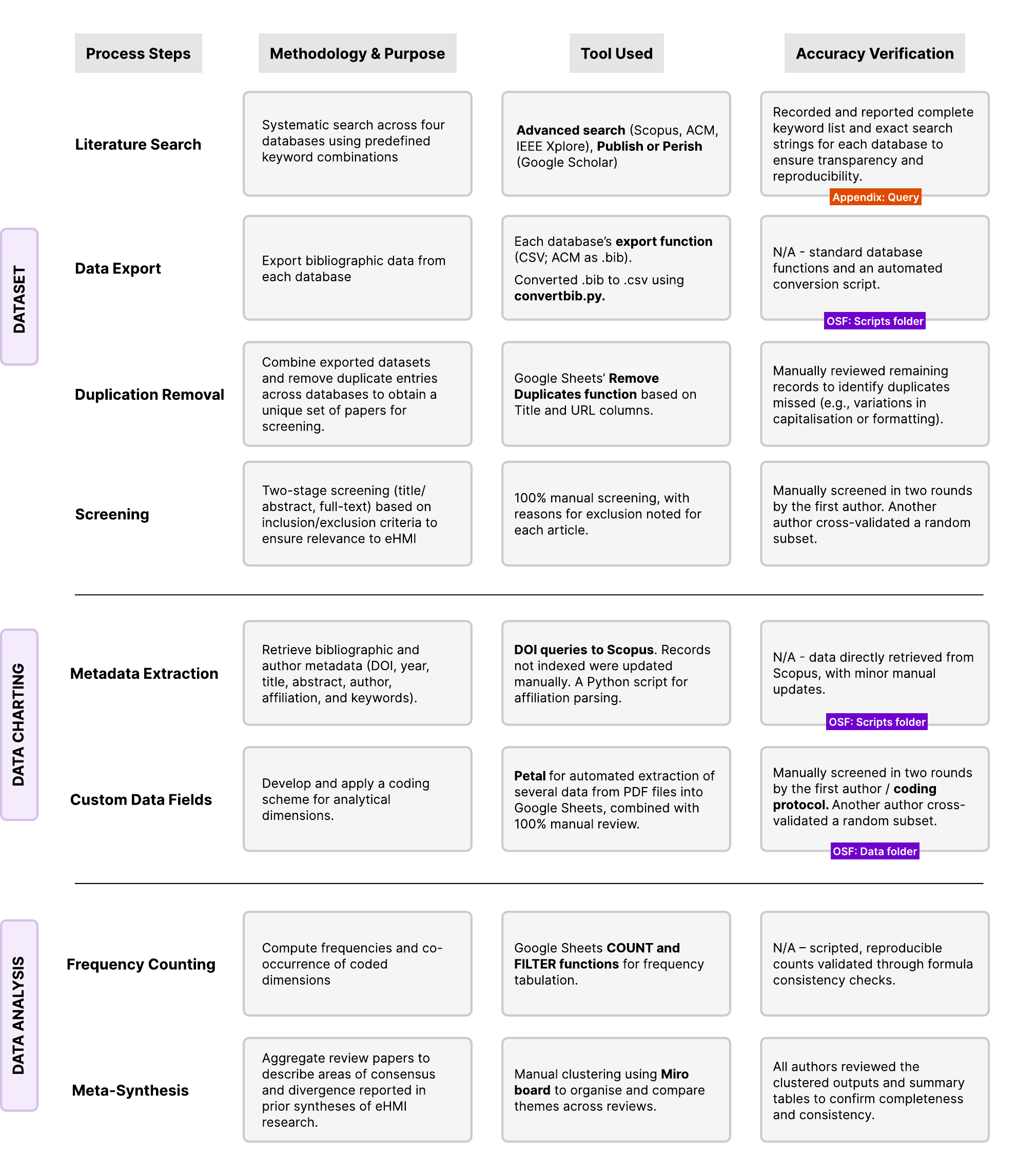}
 \caption{Workflow for dataset construction, charting, and analysis of eHMI studies.}
    \label{fig:workflow}
    \Description{.}
\end{figure*}

\section{AI Declaration of Use}

Apart from Petal, ChatGPT was employed to create Python scripts for the following purposes:

\begin{itemize}
    \item \textbf{Bibliography and metadata conversion}: Transforming bibliographic data exported from Scopus, ACM, IEEE, and Google Scholar into a structured \texttt{.csv} format, with back-conversion into \texttt{.bib} files for consistency with citation management in \LaTeX.
        
    \item \textbf{Coding support}: Flagging papers with industry or corporate authorship by scanning author affiliations, and extracting the country of the first author as well as countries of all authors from their affiliations.
\end{itemize}

All scripts were used strictly to support data organisation and preprocessing. Analytical reasoning, coding decisions, and interpretation of the results were conducted by the authors.

\begin{table*}[t]
\caption{Search queries and results across academic databases}
\label{tab:queries}
\small
\renewcommand{\arraystretch}{1.15}
\begin{tabular}{p{0.15\linewidth} p{0.75\linewidth}}
\toprule
\textbf{Source} & \textbf{Query, Date, and Results} \\
\midrule

\textbf{Scopus} &
\textbf{Date:} 9 July 2025\par
\textbf{Query:}\par
{\ttfamily
TITLE-ABS-KEY ( ( ``autonomous'' OR ``automated'' OR ``self-driving'' OR ``driverless'' )
AND ( ``car'' OR ``vehicle'' )
AND ( ``eHMI'' OR ``external communication'' OR
``external human--machine interface'' OR ``external display'' ) )
}\par
\textbf{Results:} 411
\\

\midrule

\textbf{ACM Digital Library} &
\textbf{Date:} 9 July 2025\par
\textbf{Query A:}\par
{\ttfamily
AllField:(( ``autonomous'' OR ``automated'' OR ``self-driving'' OR ``driverless'' )
AND ( ``car'' OR ``vehicle'' )
AND ( ``eHMI'' OR ``external communication'' OR
``external human-machine interface'' OR ``external display'' ))
}\par
\textbf{Results:} 418\par
\textbf{After cleaning (excluding Proceedings):} 290
\\

\midrule

\textbf{IEEE Xplore} &
\textbf{Date:} 9 July 2025\par
\textbf{Query A:}\par
{\ttfamily
(``Full Text \& Metadata'':``autonomous'' OR ``automated'' OR
``self-driving'' OR ``driverless'')
AND (``car'' OR ``vehicle'')
AND (``eHMI'' OR ``external communication'' OR
``external human-machine interface'' OR ``external display'')
}\par
\textbf{Results:} 780\par
\textbf{Filters applied:} Conferences, Journals, Early Access Articles
\\

\midrule

\textbf{Google Scholar} &
\textbf{Date:} 9 July 2025\par
\textbf{Results:} 919 (via Publish or Perish; query term: ``eHMI'')
\\

\bottomrule
\end{tabular}
\end{table*}

\begin{table*}
\centering
\caption{List of companies used for manual search of industry eHMI concepts and deployments.}
\label{tab:companies}
\small
\renewcommand{\arraystretch}{1.2}
\begin{tabular}{p{0.4\linewidth} p{0.5\linewidth}}
\toprule
\textbf{Category} & \textbf{Companies} \\ 
\midrule
Major AV OEMs / service operators & Waymo, Cruise, Zoox, Motional, Aurora Innovation, Mobileye, Baidu Apollo, AutoX, Pony.ai, WeRide, May Mobility, Oxbotica, EasyMile, Navya \\
Delivery robot \& sidewalk AV companies & Nuro, Starship Technologies, Amazon Scout, Serve Robotics, Kiwibot, Coco, Ottonomy, Uber, Yandex, Postmates \\
Automotive OEMs with AV/eHMI concept work & Ford, General Motors, Toyota, Honda, Nissan, Hyundai, Kia, Mercedes-Benz, BMW, Audi, Volkswagen, Volvo Cars, Volvo Group, Jaguar Land Rover, Stellantis, BYD, Geely, SAIC Motor \\
Tech companies with AV platforms & Apple, Google, Intel, Nvidia, Qualcomm \\
\bottomrule
\end{tabular}
\end{table*}

\begin{table*}[t]
\centering
\caption{Exact Petal prompt and a representative output example.}
\label{tab:petal}
\small
\renewcommand{\arraystretch}{1.4}
\begin{tabular}{p{0.4\linewidth} p{0.5\linewidth}}
\toprule
\textbf{Exact Prompt} & \textbf{Petal Output Example} (Paper~\cite{rouchitsas2023smiles}) \\
\midrule
Extract details on the study’s evaluation. Provide your answer as a structured list, even if some fields are missing. Include: 
\newline \textbf{Evaluation Method}: (e.g., video-based survey, VR simulation, field study, Wizard-of-Oz). 
\newline \textbf{Prototype Details}: (e.g., type of eHMI tested, display modality, vehicle setup). 
\newline \textbf{Vehicle Type}: (e.g., passenger car AV, shuttle, delivery robot). 
\newline \textbf{Target Road Users}: (e.g., pedestrians, cyclists, manual drivers, other VRUs). 
\newline \textbf{Sample Size}: (number of participants if reported). 
\newline \textbf{Participant Demographics}: (age groups, gender, special populations such as older adults, children, or accessibility groups). 
\newline \textbf{Study Location / Country}: (company/university/country participants were recruited or where the study was conducted).
&
\textbf{Evaluation Method}: Laboratory-based video-based survey
\newline \textbf{Prototype Details}: eHMI concept with full-sized virtual human controller (VHC), presented on a monitor
\newline \textbf{Vehicle Type}: Autonomous vehicle (AV)
\newline \textbf{Target Road Users}: Pedestrians
\newline \textbf{Sample Size}: 45 participants
\newline \textbf{Participant Demographics}: Mean age = 25.8 years, 30 male, 15 female, diverse nationalities, 44 participants assumed the role of a pedestrian on a daily basis
\newline \textbf{Study Location / Country}: Umeå, Sweden \\ 
\bottomrule
\end{tabular}
\end{table*}

\end{document}